\documentclass[twocolumn]{aastex63}

\usepackage{amsmath}

\newcommand{\msun}{\hbox{M$_{\odot}$}}
\newcommand{\rsun}{\hbox{R$_{\odot}$}}

\newcommand{\kms}{\,km\,s$^{-1}$}

\usepackage{xcolor}

\graphicspath{{./}{figures/}}

\shorttitle{SN 2022an}
\shortauthors{Chen Ping}

\begin{document}

\title{Detection of persistent helium absorption in the 91bg-like type Ia Supernova 2022an}

\correspondingauthor{Ping Chen}
\email{ping.chen@zju.edu.cn}

\author[0000-0003-0853-6427]{Ping Chen}\affil{Institute for Advanced Study in Physics, Zhejiang University, Hangzhou 310058, China}\affil{Institute for Astronomy, School of Physics, Zhejiang University, Hangzhou 310058, China}\affil{Department of Particle Physics and Astrophysics, Weizmann Institute of Science, 234 Herzl St, 7610001 Rehovot, Israel}

\author[0000-0002-3653-5598]{Avishay Gal-Yam}\affil{Department of Particle Physics and Astrophysics, Weizmann Institute of Science, 234 Herzl St, 7610001 Rehovot, Israel}

\author[0000-0002-1027-0990]{Subo Dong}\affil{Department of Astronomy, School of Physics, Peking University, 5 Yiheyuan Road, Haidian District, Beijing 100871, China}\affil{Kavli Institute of Astronomy and Astrophysics, Peking University,  5 Yiheyuan Road, Haidian District, Beijing 100871, China}\affil{National Astronomical Observatories, Chinese Academy of Sciences, 20A Datun Road, Chaoyang District, Beijing 100101, China}

\author[0000-0001-8531-9536]{Renyue Cen}\affil{Center for Cosmology and Computational Astrophysics, Institute for Advanced Study in Physics, Zhejiang University, Hangzhou 310058, China}\affil{Institute for Astronomy, School of Physics, Zhejiang University, Hangzhou 310058, China}

\author[0000-0003-0584-2920]{Boaz Katz}\affil{Department of Particle Physics and Astrophysics, Weizmann Institute of Science, 234 Herzl St, 7610001 Rehovot, Israel}

\author[0000-0002-9770-3508]{Kate Maguire}\affil{School of Physics, Trinity College Dublin, The University of Dublin, Dublin 2, Ireland} %

\author[0000-0001-6797-1889]{Steve Schulze}
\affil{Department of Particle Physics and Astrophysics, Weizmann Institute of Science, 234 Herzl St, 7610001 Rehovot, Israel}

\author[0000-0003-1546-6615]{Jesper Sollerman}\affil{The Oskar Klein Centre, Department of Astronomy, Stockholm University, Albanova University Center, 106 91 Stockholm, Sweden}

\author[0000-0003-0227-3451]{Joseph P Anderson}\affil{European Southern Observatory, Alonso de C\'ordova 3107, Vitacura, Casilla 19001, Santiago, Chile} %

\author[0000-0002-1066-6098]{Ting-Wan Chen}\affil{Graduate Institute of Astronomy, National Central University, 300 Jhongda Road, 32001 Jhongli, Taiwan}  %

\author[0000-0002-1296-6887]{L. Galbany}
\affil{Institute of Space Sciences (ICE-CSIC), Campus UAB, Carrer de Can Magrans, s/n, E-08193 Barcelona, Spain}
\affil{Institut d'Estudis Espacials de Catalunya (IEEC), 08860 Castelldefels (Barcelona), Spain}

\author[0000-0002-1650-1518]{Mariusz Gromadzki}\affil{Astronomical Observatory, University of Warsaw, Al. Ujazdowskie 4, 00-478 Warszawa, Poland} %

\author[0000-0002-7866-4531]{Chang Liu}
\affil{Department of Physics and Astronomy, Northwestern University, 2145 Sheridan Rd, Evanston, IL 60208, USA}
\affil{Center for Interdisciplinary Exploration and Research in Astrophysics (CIERA), Northwestern University, 1800 Sherman Ave, Evanston, IL 60201, USA}
\affil{NSF-Simons AI Institute for the Sky (SkAI), 172 E. Chestnut St., Chicago, IL 60611, USA}

\author[0000-0001-9515-478X]{Adam~A.~Miller}
\affil{Department of Physics and Astronomy, Northwestern University, 2145 Sheridan Rd, Evanston, IL 60208, USA}
\affil{Center for Interdisciplinary Exploration and Research in Astrophysics (CIERA), Northwestern University, 1800 Sherman Ave, Evanston, IL 60201, USA}
\affil{NSF-Simons AI Institute for the Sky (SkAI), 172 E. Chestnut St., Chicago, IL 60611, USA}

\author[0000-0003-3939-7167]{Tomás E. Müller-Bravo}\affil{School of Physics, Trinity College Dublin, The University of Dublin, Dublin 2, Ireland}\affil{Instituto de Ciencias Exactas y Naturales (ICEN), Universidad Arturo Prat, Chile}

\author[0000-0003-4743-1679]{Tanja Petrushevska}
\affil{Center for Astrophysics and Cosmology, University of Nova Gorica, Vipavska 11c, 5270 Ajdov\v{s}\v{c}ina, Slovenia}

\author[0000-0003-0006-0188]{Giuliano Pignata}
\affil{Instituto de Alta Investigaci\'on, Universidad de Tarapac\'a, Casilla 7D, Arica, Chile} 

\begin{abstract}
We present optical and near-infrared (NIR) observations of the fast-declining Type Ia supernova (SN\,Ia) 2022an. The photometric and spectroscopic properties identify it as a standard 91bg-like event; however, our data reveal a relatively narrow absorption feature with a full width at half-maximum (FWHM) of 75 \AA, near $1.037\,\mu$m in the rest frame of the observed spectra that persists from around 30 days to nearly 90 days after maximum light. We attribute this feature to the \ion{He}{1}~$1.083\,\mu$m line with a blueshifted velocity of $1.3\times10^{4}$\kms\, and a FWHM of $2.1\times10^{3}$\kms, supported by the detection of multiple optical \ion{He}{1} transitions at earlier epochs with a higher velocity of around $1.5\times10^{4}$\kms. The high velocity of the helium cannot be explained by helium external to the progenitor at the time of explosion, such as stripped surface helium from a companion star. The properties of the helium absorption in the spectra of SN\,2022an instead point to unburned material in the outer ejecta, thus providing the most compelling evidence to date for helium-bearing ejecta in a 91bg-like SN~Ia. Such helium has been predicted in sub-Chandrasekhar-mass double-detonation explosions involving a surface helium shell. No theoretical calculations of modern helium-shell double detonations have been performed at epochs similar to those observed for SN\,2022an to study the effects of helium on their spectra, revealing a gap between observations and theoretical calculations in our understanding of how helium manifests in SNe Ia. Nevertheless, the discovery of persistent helium absorption in SN\,2022an demonstrates the diagnostic power of NIR spectroscopy for understanding thermonuclear supernova explosions by probing the abundance and structure of their ejecta. 

\end{abstract}

\keywords{supernovae: individual (SN 2022an) --- supernovae: Type Ia --- supernovae: 91bg-like --- helium lines --- near-infrared spectroscopy}

\section{Introduction}
Type Ia supernovae (SNe Ia) play a central role in astrophysics as cosmological distance indicators and as the outcomes of close-binary evolution \citep[see, e.g.,][for reviews.]{Goobar2011, Liu2023RAA}. Although it is now well established on both theoretical and observational grounds that SNe Ia result from the thermonuclear explosions of white dwarfs (WDs), the nature of their progenitor systems and the details of the explosion mechanism remain challenging open questions. 

The diversity within the SN Ia population remains an active area of study \citep[see, e.g.,][]{Taubenberger2017, Jha2019, Dimitriadis2025, Alburai2026} and reflects a variety of candidate explosion channels and progenitor configurations \citep[see, e.g.,][]{Hillebrandt2000, Maoz2014, Livio2018, Liu2023RAA}. Among the most well-studied photometric properties of SNe Ia is their distribution in the parameter space of peak luminosity and decline rate, from which the width–luminosity relation \citep[WLR; see, e.g.,][]{Phillips2017_WLR} was established. The WLR and the SNe Ia that follow it (hereafter WLR-sequence SNe Ia) serve as the cornerstone for the standardization of these ``standardizable candles.'' Some peculiar SN Ia subtypes, such as 02cx-like SN Ia (also known as SN Iax), 02es-like SN Ia, and 09dc-like SN Ia, clearly do not obey the WLR \citep{Taubenberger2017, Chen2019}. Two other peculiar groups are 91T-like SNe Ia with SN\,1991T as the prototype \citep{Filippenko1992a_91T, Phillips1992_91T}, and 91bg-like SNe Ia with SN\,1991bg as the prototype \citep{Filippenko1992b_1991bg, Leibundgut1993_1991bg}, which represent the luminous, slow-declining end and the dim, fast-declining ends of the SN Ia population, respectively. While 91T-like SNe Ia have commonly been treated as WLR-sequence SNe Ia \citep{Phillips2024}, it remains a matter of debate whether 91bg-like events represent the low-luminosity, fast-declining end of the WLR \citep{Phillips2026}. This is an essential matter to address in understanding the progenitor systems and explosion mechanisms of SNe Ia, particularly if we believe that all WLR-sequence SNe Ia share the same explosion model \citep{Sharon2022}. Specifically, it is crucial to determine whether 91bg-like events represent a distinct explosion channel or arise from the same channel only with extreme parameters. 

The introduction of the color-stretch parameter $s_{BV}$ demonstrated that 91bg-like SNe~Ia also follow the WLR \citep{Burns2014, Burns2018, Chen2023, Graur2024, Phillips2026}, making them an indispensable component of any unified physical framework aimed at explaining the full width--luminosity sequence of SNe~Ia. In this context, the WLR-sequence SNe~Ia include 91T-like SNe~Ia \citep{Phillips2024}, normal-luminosity SNe~Ia, which largely overlap with the core-normal objects in the Branch diagram \citep{Branch2006}, transitional SNe~Ia \citep{Hsiao2015}, and 91bg-like SNe~Ia. This continuity suggests that the apparent diversity among these subclasses may arise from a continuous distribution of explosion properties within a common physical framework.  These subgroups should therefore not be treated as isolated or fundamentally distinct classes, but rather as closely related manifestations of a continuous SN~Ia population governed by similar underlying explosion physics.

From a physical perspective, the defining observational properties of 91bg-like SNe~Ia, including rapid photometric evolution, low peak luminosities, and strong low-ionization features such as enhanced \ion{Ti}{2} absorption, collectively point to ejecta with low mass, reduced radioactive heating, and relatively cool thermodynamic conditions \citep{Mazzali1997}. These characteristics are naturally produced in sub-Chandrasekhar-mass (hereafter sub-$M\mathrm{_{Ch}}$) explosions, which synthesize less $^{56}$Ni and have shorter diffusion timescales \citep{Blondin2017, Goldstein2018, Shen2021a}, thereby providing a physically motivated explanation for the observed properties of this subclass. Beyond their photometric and spectroscopic properties, 91bg-like SNe~Ia preferentially occur in early-type host galaxies, indicating old progenitor populations and long delay times \citep{Howell2001a, Barkhudaryan2019, Panther2019}. This is compatible with sub-$M\mathrm{_{Ch}}$ detonations, although such environments do not uniquely distinguish among different explosion channels.

The sub-$M\mathrm{_{Ch}}$ detonation of a low-mass WD with mass around 0.85--0.9 \msun\, matches the observed properties of 91bg-like SNe Ia \citep{Sim2010, Ruiter2013, Blondin2017, Blondin2018, Shen2021a, Shen2021b}. Specifically, \citet{Blondin2018} performed radiative-transfer simulations for a sub-$M\mathrm{_{Ch}}$ SCH2p0 \footnote{The \textsc{SCH} series denotes sub-Chandrasekhar-mass models resulting from pure central detonations of C--O WDs \citep{Blondin2017}. The suffix \textsc{2p0} identifies a specific model with a total ejecta mass of $0.90\,M_\odot$ synthesizing $0.12\,M_\odot$ of $^{56}$Ni.} model, resulting from the pure central detonation of a C--O WD, and compared the results to those of an $M\mathrm{_{Ch}}$ delayed-detonation model DDC25, which produces the same $^{56}$Ni yield of 0.12 \msun. They found unprecedented agreement between the sub-$M\mathrm{_{Ch}}$ model and the optical and near-infrared (NIR) observations of SN\,1999by. It is important to note that the sub-$M\mathrm{_{Ch}}$ models considered by \citet{Blondin2018} were constructed to explore the radiative-transfer consequences of centrally detonated low-mass WDs, without specifying the physical mechanism responsible for initiating the detonation. In physically motivated scenarios, such a central detonation may be triggered by the ignition of an accreted surface helium layer,  providing a natural pathway to sub-$M\mathrm{_{Ch}}$ explosions \citep{Nomoto1982a, Nomoto1982b, Woosley1986, Bildsten2007, Shen2009, Fink2010}.

A long-standing theoretical prediction of helium-detonation-triggered sub-$M_\mathrm{{Ch}}$ explosions is the presence of unburned helium at high velocities \citep{Woosley1994, Fink2010, Kromer2010, Shen2014}. Non-local thermodynamic equilibrium (NLTE) radiative-transfer calculations including \ion{He}{1} excitation by nonthermal electrons consistently predict observable absorption features in early-time spectra, particularly through the strong \ion{He}{1} 1.083 and 2.058 $\mu$m transitions \citep{Dessart2015, Boyle2017, Collins2023, Callan2025}. Detecting such features would provide direct evidence for a helium-rich shell and, therefore, strong support for sub-$M_\mathrm{{Ch}}$ detonations.

Despite these theoretical expectations, observational evidence for helium in SNe Ia has remained sparse and ambiguous. The NIR ``W-shaped'' absorption feature near 1.05 $\mu$m, commonly observed before maximum light in fast-declining SNe Ia has been variously attributed to \ion{Mg}{2} 1.0927 $\mu$m, \ion{C}{1} 1.0693 $\mu$m, or \ion{He}{1} 1.083 $\mu$m \citep{Hsiao2015, Wyatt2021, Li2022_2012ij, Lu2023_CSP_SNIa_NIR, Pearson2024}. The degeneracy among these species, combined with severe NLTE effects, limited wavelength coverage, and rapid line evolution, has prevented a definitive identification. Clear helium signatures have, however, been detected in some interacting SNe Ia \citep[e.g., SN\,2020eyj;][]{Kool2023_2020eyj}. The \ion{He}{1} emission features in SN\,2020eyj arise from the interaction between the supernova ejecta and the helium-rich circumstellar material (CSM). The CSM has been proposed to come either from a thick wind driven by mass transfer from a helium-star companion \citep{Kool2023_2020eyj} or from a helium-rich common envelope ejected before explosion \citep{Soker2023, Meng2026}.

In this context, SN 2022an presents a rare opportunity. Although its photometric and spectroscopic properties firmly place it among the 91bg-like SNe Ia, our optical and NIR follow-up observations reveal an absorption feature near 1.037 $\mu$m that persists until nearly +90 days after maximum light, which is much later than the typical disappearance of the 1.05 $\mu$m complex in other SNe Ia \citep{Lu2023_CSP_SNIa_NIR}. This feature is consistently reproduced by \ion{He}{1} 1.083 $\mu$m absorption with a blueshift of $\sim 1.2-1.3\times 10^4$\kms, and its identification is independently supported by the presence of weaker optical \ion{He}{1} transitions at matching velocities. Such persistent multiline helium absorption has never been previously confirmed in a 91bg-like SN Ia.

In this paper, we present multiepoch optical and NIR observations of SN 2022an and report the detection of persistent \ion{He}{1} absorption features in both wavelength regimes. Section~\ref{sec:obs_and_reduction} summarizes the observations and data reduction. Section~\ref{sec:properties} establishes SN\,2022an as a typical 91bg-like SN Ia. Section~\ref{sec:helium_iden} presents the identification and temporal evolution of the helium features. Section~\ref{sec:discussion} discusses the origin and implications of the detected helium in the context of SN Ia explosion models and progenitor channels.

\section{Observations and Data Reduction}
\label{sec:obs_and_reduction}
SN 2022an (R.A. = 12$^\mathrm{h}$43$^\mathrm{m}$32\fs240, decl. = $-$41\degr21\arcmin57\farcs56) was discovered by the All-Sky Automated Survey for SuperNovae \citep[ASAS-SN;][]{Shappee2014_asassn} on 2022 January 4 and reported to the Transient Name Server (TNS) as ASASSN-22ac on 2022 January 5 \citep{Stanek2022_discovery}. It was also detected by ATLAS and Gaia, with the internal names ATLAS22bhv and Gaia22ajh, respectively. It was classified as a 91bg-like SN Ia on January 8 \citep {Jacobson-Galan2022_classification_2022an}, after which we began follow-up observations in both optical and NIR wavelengths.  Details of the imaging observations and photometry are given in Appendix A. The photometric data of SN\,2022an are presented in Table~\ref{tab:phot_BVgri} for the $BVgri$ bands and in Table~\ref{tab:phot_JHKs} for the $JHK_s$ bands. We also obtained the ASAS-SN $g$-band light curve around maximum light \citep{Kochanek2017_asassn}, which is presented in Table~\ref{tab:phot_asassn}. The $BV$ and $JHKs$ magnitudes are given in the Vega system, and the $gri$ magnitudes are given in the AB system. A summary of the light curves of SN\,2022an is shown in Figure ~\ref{fig:lcs}. Details of the spectroscopic observations and data reduction are given in Appendix B. The spectral sequence of SN\,2022an is shown in Figure ~\ref{fig:specs}.

\begin{figure*}
\centerline{\includegraphics[width=19cm]{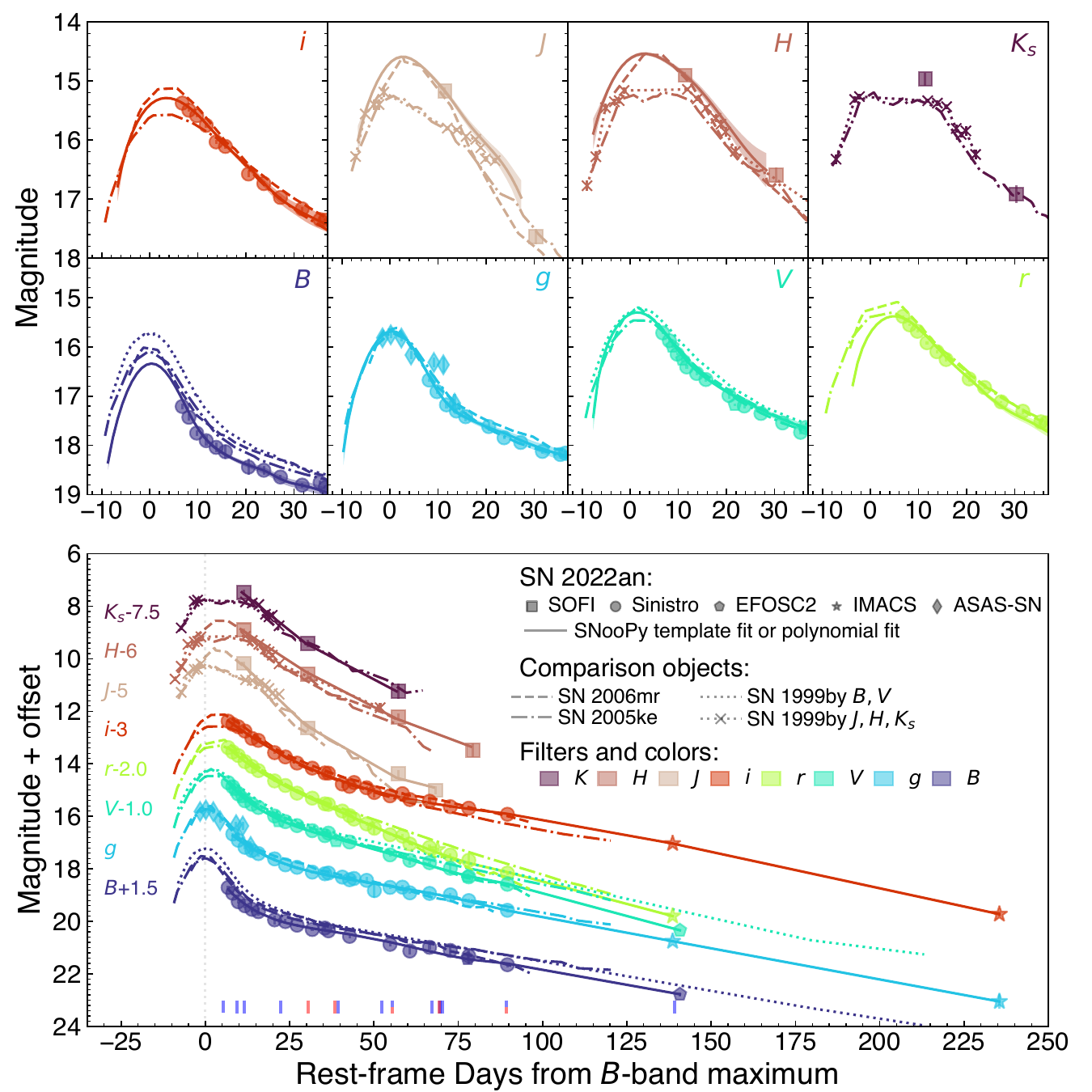}}
\caption{Multiband light curves of SN\,2022an and comparisons with other 91bg-like SNe Ia. The bottom panel shows the full range of the light curves, while the top panels show zoomed-in views around the maximum light. All panels share the same color, symbol, and line-style scheme shown in the legend of the bottom panel: colors distinguish different filters, symbols indicate the instruments used for the SN\,2022an photometry, and line styles distinguish different objects. In the top panels, the solid lines show the template-fitting results for the SN\,2022an light curves obtained using the template-matching method implemented in SNooPy \citep{Burns2011, Burns2014}. In the bottom panel, the solid lines show polynomial fits to the SN\,2022an light curves. Three low-luminosity 91bg-like SNe Ia, SN\,1999by \citep{Hoflich2002_1999by, Garnavich2004_1999by}, SN\,2005ke \citep{Krisciunas2017_CSP_dr3}, and SN\,2006mr \citep{Krisciunas2017_CSP_dr3}, are shown for comparison. The magnitudes of the reference objects have been shifted to match the distance of SN\,2022an. The adopted distance moduli are: $\mu_ {\rm SN\,1999by}=30.75$ mag, $\mu_{\rm SN\,2005ke}=31.60$ mag, and $\mu_{\rm SN\,2006mr}=31.25$ mag, respectively. For SN\,2006mr, an additional shift of $-1$ mag has been applied in all filters after correcting for the distance difference. Galactic extinction has been corrected for all objects; host-galaxy extinction has not. The epochs of the spectra listed in Table \ref{tab:spec_log} are indicated as vertical lines along the bottom of the panel. Epochs of the optical spectra are shown in blue, while those of the NIR spectra are shown in red.
} 
\label{fig:lcs}
\end{figure*}

\begin{figure*}
\centerline{\includegraphics[width=20cm]{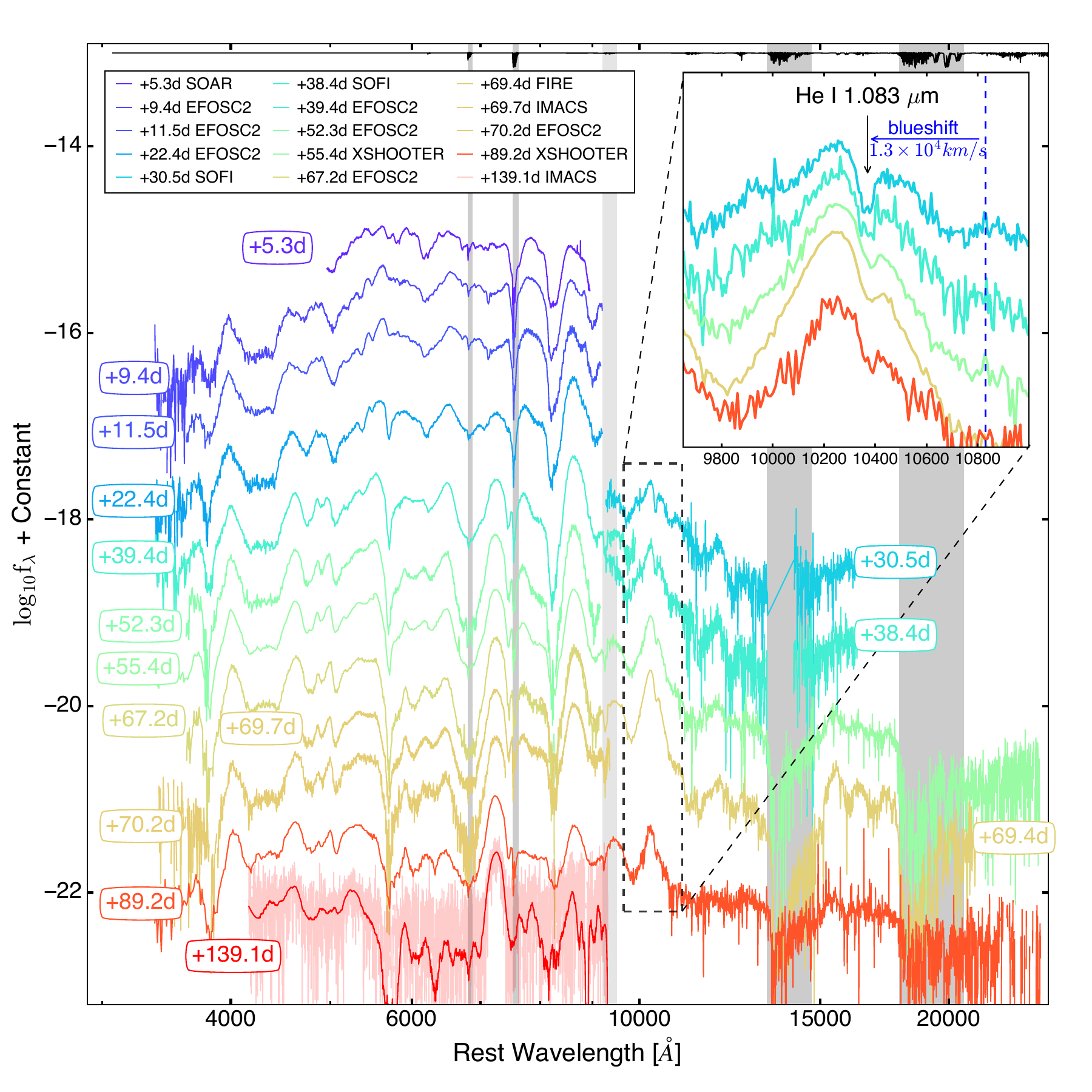}}
\caption{Optical and NIR spectra of SN\,2022an. The phases of the spectra are given with respect to the time of $B$-band maximum, JD=2459582. A telluric spectrum is shown at the top of the plot, and regions strongly affected by telluric absorption are marked with vertical gray bands. The inset panel shows a zoom-in view of the detected absorption feature indicated by the black arrow, which, as discussed in the following sections, we attribute to the blueshifted \ion{He}{1} line from its rest-frame wavelength of 1.083 $\mu m$ (vertical dashed line). }
\label{fig:specs}
\end{figure*}

\section{SN\,2022an as a typical 91bg-like SN Ia}
\label{sec:properties}
\subsection{Host galaxy and extinction}
SN\,2022an is located 17\farcs8 away from the center of NGC\,4645B at a redshift of $z=0.0091$ ($z =0.009070\pm0.000133$ from \citealt{Wegner2003} and $z =0.00910\pm0.00006$ from \citealt{Ogando2008})\footnote{Note that the fiducial value of $z=0.007397$ for NGC 4645B in the NASA/IPAC Extragalactic Data \citep[NED;][]{ned} is incorrect. The fiducial redshift for this galaxy is based on \ion{H}{1} 21-cm observations, which measure an \ion{H}{1} line profile that may include emission not only from the target galaxy but also anything else in the 14\arcmin (FWHM) beam, resulting in a significantly different redshift (centroid) than measured with optical telescopes having much higher resolution centered on the stars and gas bound to the galaxy. }. NGC\,4645B is a lenticular galaxy with a weak bar structure (SAB0), consistent with the preference of 91bg-like SNe Ia to occur in early-type hosts with old stellar populations and low star-formation rates. The luminosity distance to NGC\,4645B derived from the redshift is $d_L=37.6$ Mpc (distance modulus $\mu=32.88$ mag), adopting a standard Lambda cold dark matter ($\Lambda$CDM) cosmology with $H_0=73$\,km\,s$^{-1}$\,Mpc$^{-1}$, $\Omega_m=0.27$ and $\Omega_\Lambda=0.73$. NGC\,4645B belongs to the Centaurus Cluster of galaxies, and it is close to the so-called ``Great Attractor'' \citep{Tonry2000}. This implies that the peculiar galaxy velocity relative to the Hubble flow constitutes a significant fraction of its observed total radial velocity, thereby introducing a large uncertainty into the Hubble-flow distance measurement. Using the surface-brightness fluctuation method, \cite{Tonry2000} measured a distance to the Centaurus Cluster of $33.8 \pm 1.4$ Mpc ($\mu = 32.64 \pm 0.08$ mag), while \cite{Mieske2003} measured a distance of $41.3 \pm 2.1$ Mpc ($\mu = 33.08 \pm 0.11$ mag). We adopt $\mu=32.88$ mag with an uncertainty of 0.3 mag (corresponding to a relative distance uncertainty of 14\%) for the remainder of our analysis.
The Galactic extinction along the line of sight to SN\,2022an is E($B-V$) = 0.166 mag \citep{Schlafly2011}. We apply the reddening law from \cite{Fitzpatrick1999} with $R_V=3.1$ to correct for the Galactic extinction. We measure a \ion{Na}{1} D1+D2 equivalent width (EW) of $<0.1$\,\AA\, for the sodium absorption at the redshift of NGC\,4645B, indicating that the extinction from the host galaxy is probably negligible. Unless otherwise stated, we therefore assume zero host-galaxy extinction in our analysis.

\subsection{Photometric properties}
\label{sec:phot_property}
Our multiband follow-up observations for SN\,2022an begin after maximum light, with only the ASAS-SN $g$-band light curve covering the phase around peak brightness. We used a fifth-order polynomial to fit the first 20 days of the $g$-band light curve and obtained an estimated $g$-band maximum at JD= $2459581.4\pm0.3$. Since the time difference between the $B$-band and $g$-band maxima is small for SNe Ia \citep{Folatelli2010}, we treat JD = 2459581.4 as the epoch of $B$-band maximum. A conservative uncertainty of 1 day is assigned to this estimate, and this epoch is used as the reference time throughout the paper. 

\begin{figure}[t]
\centering
\includegraphics[width=\columnwidth]{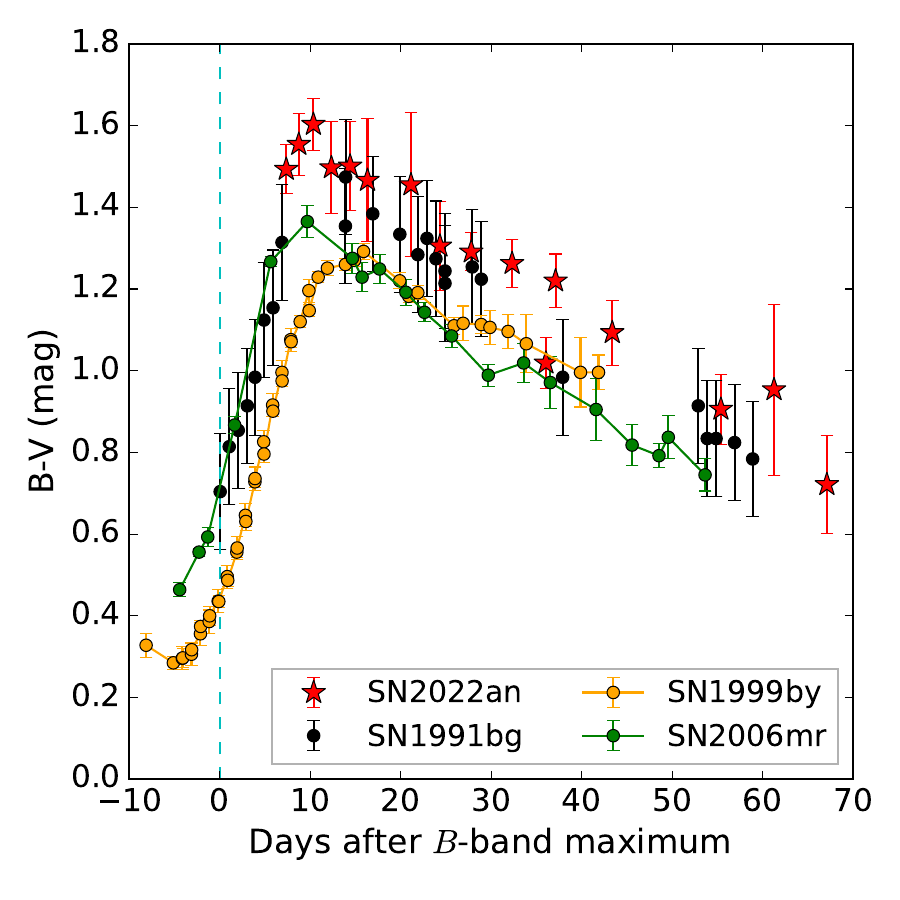}
\caption{$B-V$ color curves of SN\,2022an and several 91bg-like SNe Ia, including SN\,1991bg \citep{Filippenko1992b_91bg, Leibundgut1993}, SN\,1999by \citep{Hoflich2002_1999by, Garnavich2004_1999by} and SN\,2006mr\citep{Krisciunas2017_CSP_dr3}. Galactic extinction has been corrected for all objects.}
\label{fig:BVcolor}
\end{figure}

The multiband light curves of SN\,2022an (Figure ~\ref{fig:lcs}) show a good resemblance to those of 91bg-like SNe Ia, such as SN\,1999by \citep{Hoflich2002_1999by, Garnavich2004_1999by}, SN\,2005ke \citep{Krisciunas2017_CSP_dr3}, and SN\,2006mr \citep{Krisciunas2017_CSP_dr3}. The $B-V$ color curve of SN\,2022an is shown in Figure ~\ref{fig:BVcolor}. Some 91bg-like SNe Ia, including SN\,1991bg \citep{Filippenko1992b_91bg, Leibundgut1993}, SN\,1999by \citep{Hoflich2002_1999by, Garnavich2004_1999by} and SN\,2006mr \citep{Krisciunas2017_CSP_dr3}, are shown for comparison. SN\,1991bg is the prototype object of its class. SN\,1999by, SN\,2005ke, and SN\,2006mr are among the best-observed 91bg-like SNe Ia with optical and NIR spectrophotometric data available in the literature, and we therefore use them as comparison objects throughout this work. We measure a color-stretch parameter of $s_{BV} = 0.30\pm0.08$ for SN\,2022an using the method described in \citealt{Chen2023}.

We also performed multiband light-curve fitting using the template-matching method adopted by SNooPy \citep{Burns2011, Burns2014}. We used the \textit{max\_model} to generate the template light curves and used the color-stretch parameter $s_{BV}$ for the light-curve parameter. The best-fit model light curves are shown in the top panels of Figure~\ref{fig:lcs}. The resulting best-fit parameters are a B-band maximum time of $T_{\rm max} (B) = 2459581.6\pm0.7$ and $s_{BV} = 0.32\pm0.04$. Both values are consistent with those measured directly from the light curves without template matching.

Our photometric data cover three distinct wavelength regimes. Around maximum light, only $g$-band photometry is available (\textit{Phase~1}). From approximately 10 to 80 days after $B$-band maximum, both optical and NIR photometry were obtained (\textit{Phase~2}). At later epochs, only optical photometry is available (\textit{Phase~3}). For \textit{Phase~1} and \textit{Phase~2}, we constructed a pseudobolometric light curve of SN\,2022an by adopting a method similar to that described in \citet{Chen2024}, in which spectral templates are calibrated to the photometry and integrated over a selected wavelength range. For each epoch listed in Table~\ref{tab:phot_BVgri} during \textit{Phase~2}, we constructed a spectral template using either the closest observed spectrum in phase or a weighted average of the two nearest spectra. The resulting templates cover the wavelength range 3750--25,000\,\AA. For epochs lacking NIR spectroscopy, or for spectra that do not extend to 25,000\,\AA, we interpolated the broadband NIR spectral energy distribution using the available NIR photometry and concatenated it with the observed spectrum. Each spectral template was then scaled to match the observed optical photometry and interpolated NIR photometry (see the solid curves in Figure~\ref{fig:lcs}). For \textit{Phase~1}, we used the template-fit light curves (solid lines in the top panels of Figure~\ref{fig:lcs}) together with the SCH2p0 model spectra of \cite{Blondin2018} as the photometric and spectral template, respectively. The calibrated spectra were integrated over 3750--25,000\,\AA\ to derive the pseudobolometric luminosity, which we hereafter refer to as the optical--NIR luminosity, $L_{\mathrm{onir}}$.

The ultraviolet contribution to the bolometric luminosity of 91bg-like SNe\,Ia has not been explicitly quantified in the literature. Model spectra from \citet{Blondin2018} suggest that emission in the wavelength range 2500--3750\,\AA\ contributes $\lesssim5\%$ of the total luminosity from approximately 1 to 230 days after explosion. We therefore neglect the ultraviolet contribution in our pseudobolometric estimates.

SN\,2022an exhibits a markedly different evolution of the NIR contribution to the total bolometric luminosity compared to other SN\,Ia subtypes. The NIR fraction ($\lambda>9150$\AA) is around $\sim30\%$ near maximum light, rises to $\sim40\%$ about 1 week after $B$-band peak, and then declines steadily to $\lesssim20\%$ by roughly 2 months after maximum. In contrast, other SN\,Ia subclasses show a relatively small NIR contribution near peak brightness, followed by an increase to a maximum around the epoch of the NIR secondary maximum before declining. This behavior has been observed in normal SNe\,Ia \citep[e.g., Figure~23 of][]{Wang2009}, transitional SNe\,Ia \citep[e.g., Figure~11 of][]{Gall2018}, and 09dc-like peculiar SNe\,Ia \citep[e.g., Figure~9 of][]{Chen2019}. 

The evolution observed in SN\,2022an is consistent with the relatively bright NIR peak luminosities of 91bg-like SNe Ia compared to the optical peak luminosities, as well as with the rapid postmaximum decline of their NIR light curves, which lack prominent secondary maxima. These results underscore the importance of NIR photometry for accurately determining the luminosity evolution of 91bg-like SNe\,Ia.

\begin{figure}[t]
\centering
\includegraphics[width=\columnwidth]{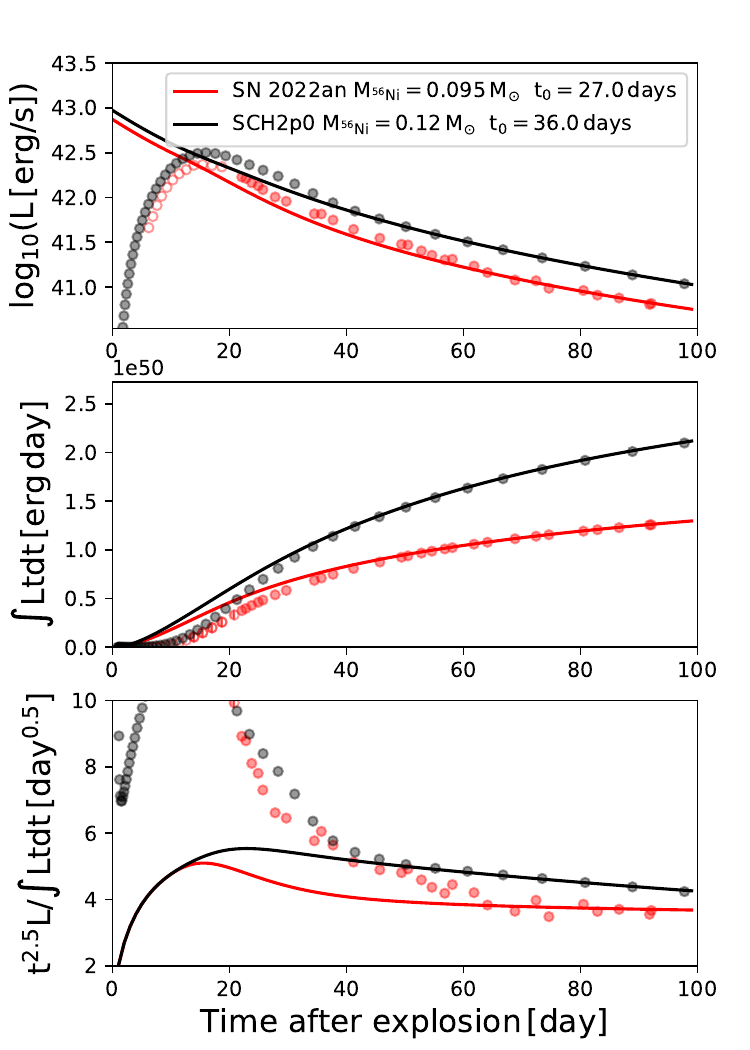}
\caption{The pseudobolometric light curve of SN\,2022an (red) fit with the radioactive $\mathrm{^{56}Ni}$ decay model obtained using the method outlined in \citet{Wygoda2019}. The synthetic light curve of the SCH2p0 model (black) from \citet{Blondin2018} is shown for comparison. The three panels show the pseudobolometric luminosity (top), time-weighted integrated luminosity (middle), and the $\mathrm{^{56}Ni}$-mass-independent quantity defined as the ratio of $t^{2.5}L/\int Ltdt$ (bottom).
The \textit{Phase~1} data for SN\,2022an is shown with unfilled symbols.}
\label{fig:t0_Ni56_fit}
\end{figure}

We use the derived bolometric light curve of SN\,2022an to estimate its $\mathrm{^{56}Ni}$ mass and $\gamma$-ray escape time scale $t_0$ by fitting it to a radioactive $\mathrm{^{56}Ni}$ decay model \citep{Jeffery1999, Katz2013, Wygoda2019}. We adopt the method outlined in \citet{Wygoda2019}. The best-fit results, constrained by data obtained more than 50 days after explosion, are shown with the red lines in Figure~\ref{fig:t0_Ni56_fit} and give $\mathrm{M_{^{56}Ni}=0.095\, M_\odot}$ and $\mathrm{t_0=27.0}$\,days. A rise time of $14\pm2$ days from explosion to the $B$-band peak is adopted for SN\,2022an \citep{Ganeshalingam2011}.  For comparison, $\mathrm{M_{^{56}Ni}=0.12\, M_\odot}$ and $\mathrm{t_0=36.0}$\,days are derived from the synthetic bolometric light curve of the SCH2p0 model of \citet{Blondin2018}.  

\begin{figure}[t]
\centering
\includegraphics[width=\columnwidth]{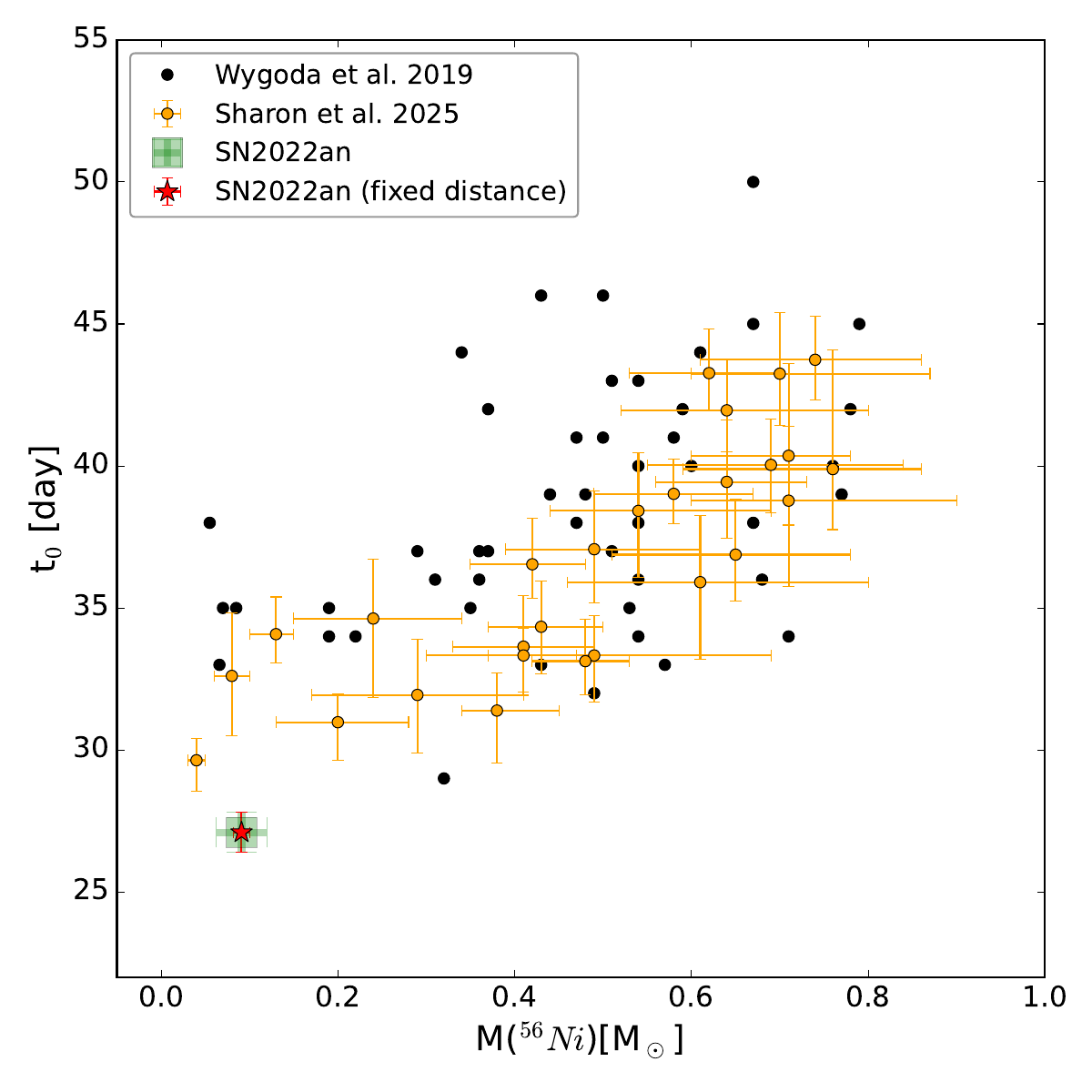}
\caption{The $\gamma$-ray escape time $t_0$ and $^{56}$Ni mass of SN\,2022an in the context of the SN Ia population. The comparison samples are taken from \citet{Wygoda2019} and \citet{Sharon2025}. The asterisk symbol indicates the result obtained without accounting for the error budget due to the uncertain distance to SN\,2022an.}
\label{fig:t0_Ni56}
\end{figure}

We compare the $\mathrm{M_{^{56}Ni}}$ and $t_0$ of SN\,2022an with those of other SNe Ia derived from observational data \citep{Wygoda2019, Sharon2025}, as shown in Figure~\ref{fig:t0_Ni56}. We note that SN\,2022an has a smaller $t_0$ than the other SNe Ia, reflecting a lower ejecta column density that allows $\gamma$-rays to escape at earlier epochs and causes the bolometric light curve to decline more rapidly at late times.

\subsection{Spectroscopic properties}
\label{sec:spec_property}
\begin{figure}[t]
\centering  \includegraphics[width=\columnwidth]{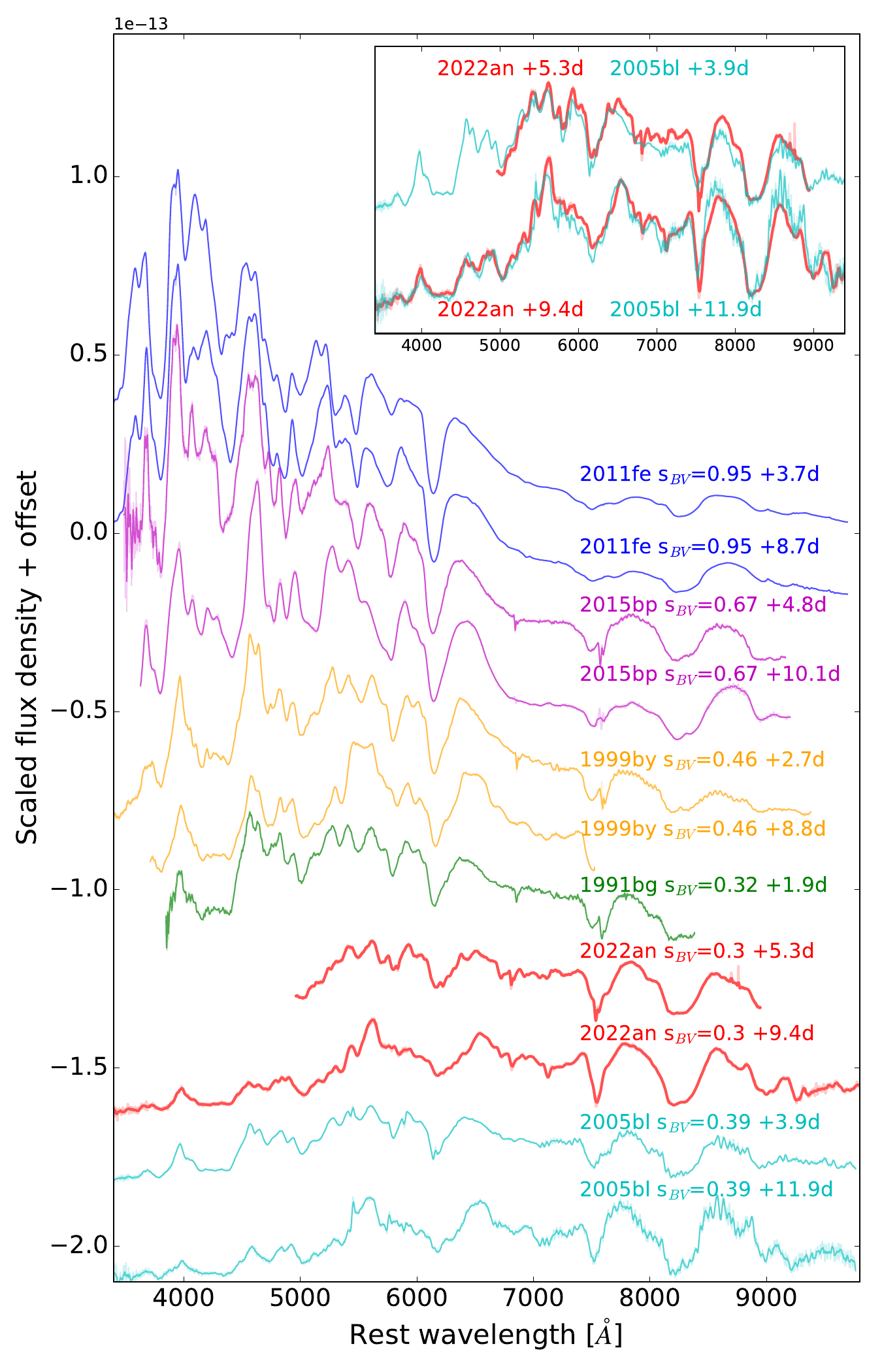} \\
\caption{Spectra of SN\,2022an compared with those of the normal SN Ia SN\,2011fe \citep{Pereira2013_2011fe}, the transitional Ia SN\,2015bp \citep{Wyatt2021_2015bp}, SN\,1991bg \citep{Filippenko1992b_1991bg, Leibundgut1993_1991bg}, and the 91bg-like SN\,1999by \citep{Hoflich2002_1999by, Garnavich2004_1999by} and SN\,2005bl \citep{Taubenberger2008_2005bl, Hachinger2009_2005bl}. The color-stretch parameter $s_{BV}$ of each supernova and the phase of each spectrum relative to the $B$-band maximum are listed.}
\label{fig:compare_optspec_peak}
\end{figure}

\begin{figure}[t]
\centering  \includegraphics[width=\columnwidth]{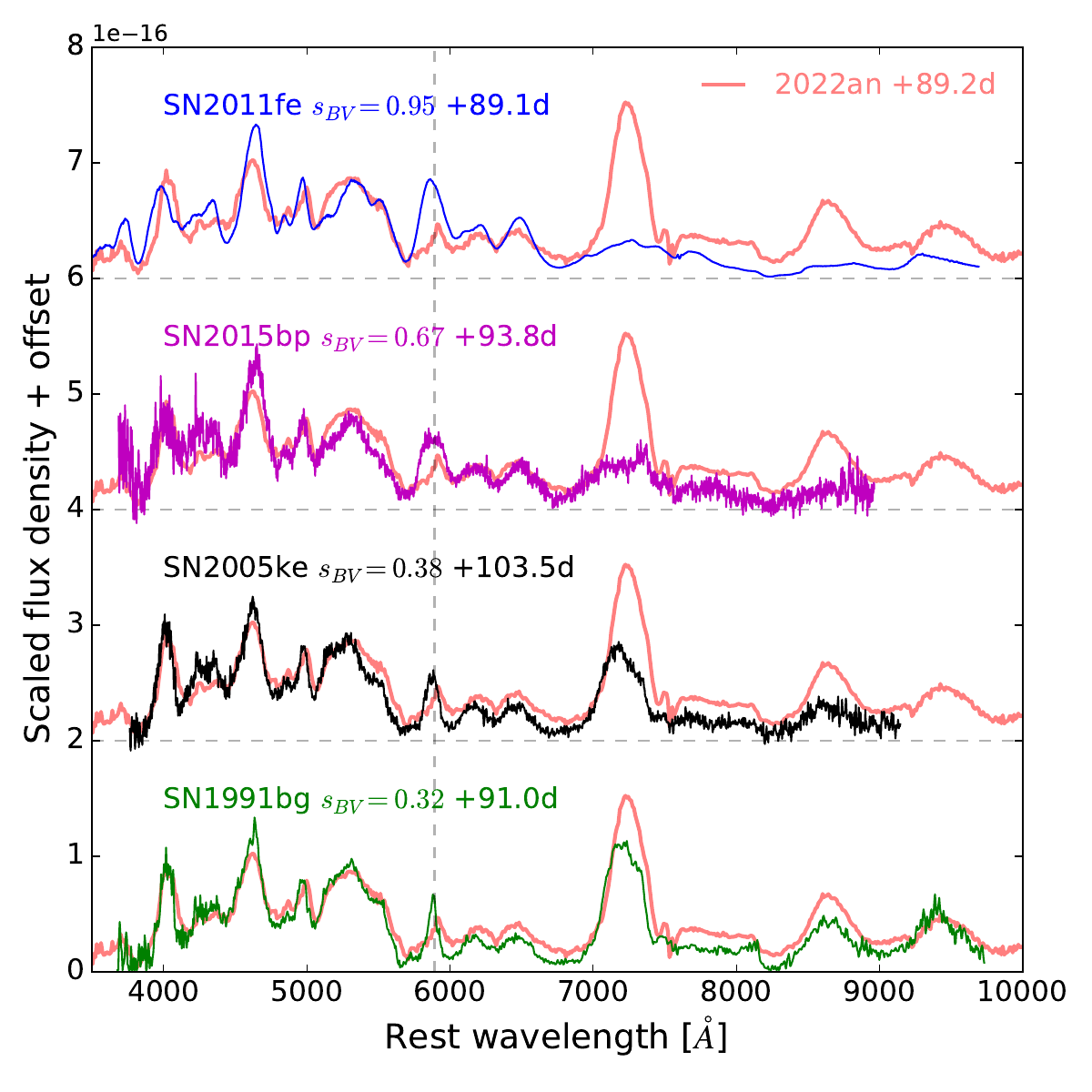} \\
\caption{Spectra of SN\,2022an compared with those of the core-normal Ia SN\,2011fe \citep{Pereira2013_2011fe}, the transitional Ia SN\,2015bp \citep{Wyatt2021_2015bp}, SN\,1991bg \citep{Filippenko1992b_1991bg, Leibundgut1993_1991bg}, and the 91bg-like SN\,2005ke \citep{Folatelli2013_CSPspec_dr1}. The color-stretch parameter $s_{BV}$ of each supernova and the phase of each spectrum after the $B$-band maximum are given after the supernova name. The comparison spectra have been scaled to match the integrated flux of SN\,2022an over the wavelength range 4000-5500\,\AA. The comparison pairs have been shifted for clarity, with the applied offsets indicated by the horizontal dashed lines. The vertical dashed line marks the rest-frame wavelength of [\ion{Co}{3}] $\lambda$5892 \citep{Dong2018}. }
\label{fig:compare_optspec_nebular}
\end{figure}

We compare the spectra of  SN\,2022an obtained at +5.3 and +9.4 days post peak with those of other WLR-sequence SNe Ia at similar phases in Figure~\ref{fig:compare_optspec_peak}, including the core-normal SN\,2011fe \citep{Pereira2013_2011fe}, the transitional SN\,2015bp \citep{Wyatt2021_2015bp}, SN\,1991bg \citep{Filippenko1992b_1991bg, Leibundgut1993_1991bg}, and two other low-luminosity 91bg-like SNe Ia, SN\,1999by \citep{Hoflich2002_1999by, Garnavich2004_1999by} and SN\,2005bl \citep{Taubenberger2008_2005bl, Hachinger2009_2005bl}. SN\,2022an shows excellent agreement with the spectra of 91bg-like SNe Ia, as highlighted by the comparison to SN\,2005bl in the inset panel of Figure~\ref{fig:compare_optspec_peak}. We measure a \ion{Si}{2} $\lambda$6355 velocity of 9400\kms ~from the +5.3d spectrum of SN\,2022an, consistent with the velocities measured in 91bg-like SNe Ia at similar phases (see, e.g., Figure~19 of \citealt{Taubenberger2008_2005bl}).
A comparison of the early-nebular-phase spectra of SN\,2022an with those of other SNe Ia is shown in Figure~\ref{fig:compare_optspec_nebular}. The comparison sample includes the core-normal SN\,2011fe \citep{Pereira2013_2011fe}, the transitional SN\,2015bp \citep{Wyatt2021_2015bp}, SN\,1991bg \citep{Filippenko1992b_1991bg, Leibundgut1993_1991bg}, and the 91bg-like SN\,2005ke \citep{Folatelli2013_CSPspec_dr1}. Once again, SN\,2022an matches well the 91bg-like events, especially in its strong [\ion{Ca}{2}] $\lambda\lambda$7291,7323 and \ion{Ca}{2} NIR triplet emission lines, which are weak or not clearly seen in core-normal and transitional SNe Ia. The [\ion{Co}{3}] $\lambda$5892 line is redshifted by $\approx 1200$\kms~relative to the rest-frame wavelength ( vertical dashed line in Figure~\ref{fig:compare_optspec_nebular}), implying an aspheric distribution of $^{56}$Ni in the ejecta \citep{Dong2018}.

\begin{figure}[t]
\centering
\includegraphics[width=\columnwidth]{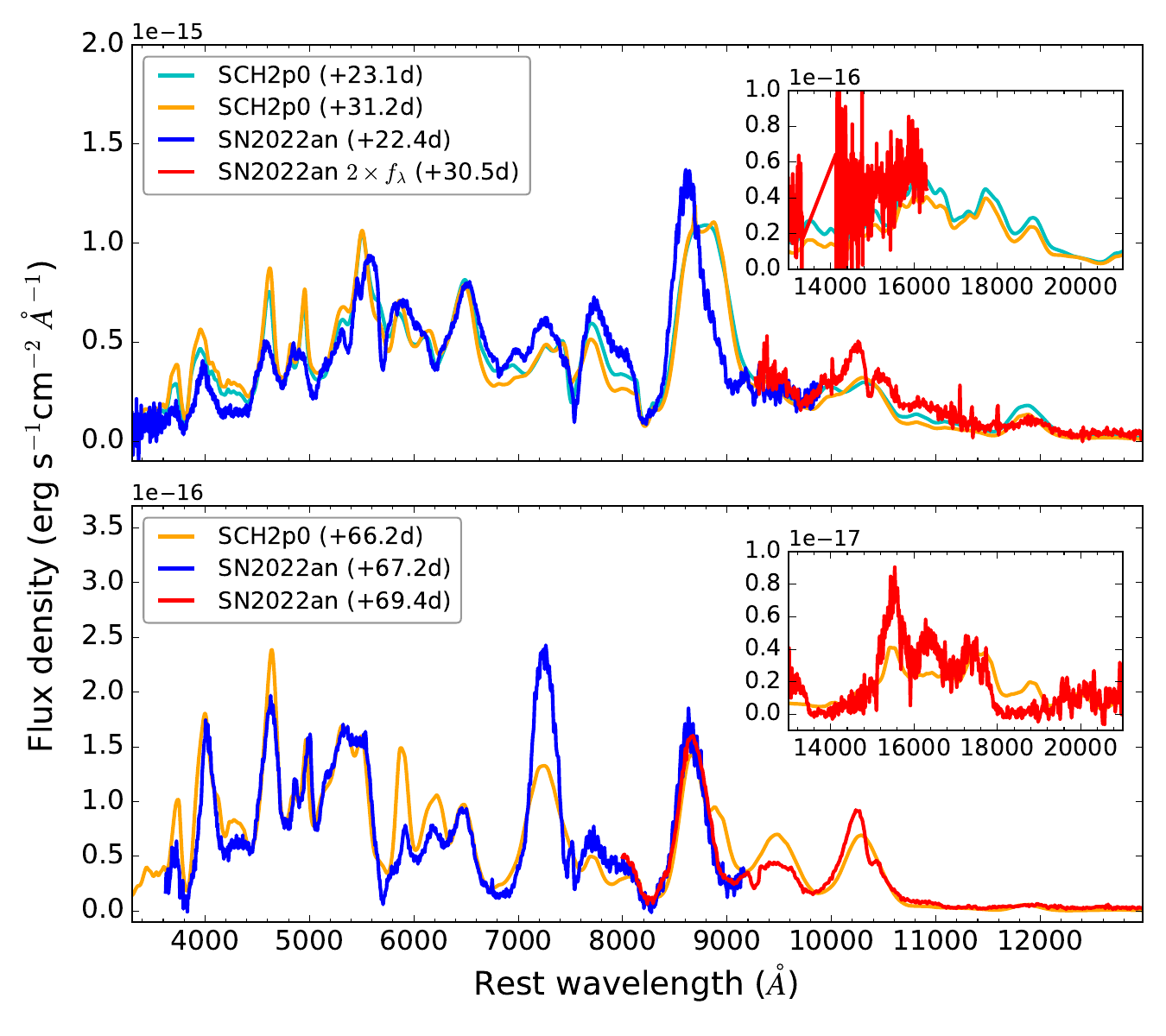} \\
\caption{Spectra of SN\,2022an compared with the SCH2p0 model spectra of \cite{Blondin2018} at similar phases. The model spectra have been scaled to match the integrated flux over the wavelength range 4000 -- 8000 \AA.}
\label{fig:compare_model}
\end{figure}

We did not find any NIR spectra of 91bg-like SNe Ia at phases similar to those of SN 2022an in the literature for direct comparison. Instead, we compare two NIR spectra of SN 2022an, taken approximately 1 and 2 months after $B$-band maximum, along with their corresponding optical spectra at similar phases, to the model spectra of \cite{Blondin2018} in Figure~\ref{fig:compare_model}. These model spectra have been demonstrated to match those of SN 1999by closely. A general resemblance is also observed between the model spectra and those of SN 2022an.

Overall, SN\,2022an exploded in an old environment and shares spectrophotometric properties with 91bg-like SNe Ia, i.e., subluminous peak magnitudes, rapidly declining light curves, and cool, low-ionization spectra with strong \ion{Ti}{2} absorption in the photospheric spectra and conspicuous \ion{Ca}{2} emission during the nebular phases. These properties establish SN\,2022an as a typical 91bg-like SN Ia.

\section{Detection of helium absorption}
\label{sec:helium_iden}
The NLTE radiative-transfer calculations of modern helium-shell double-detonation models predict detectable helium absorption in their early-phase spectra \citep[e.g.,][see Section~\ref{sec:model_spec} for a discussion on these model spectra]{Dessart2015, Boyle2017, Collins2023, Callan2025}. These absorption features arise from unburned helium surviving in the outer ejecta shell \citep[e.g.,][]{Kromer2010, Shen2010, Woosley2011}. However, the detection of such helium absorption features in SN Ia spectra has long been hindered by limited wavelength coverage and low signal-to-noise ratios \citep{Liu2023_2020jgb}. The tentative \ion{He}{1} identifications reported in the literature generally lack cross-validation across multiple transitions \citep[e.g.,][]{Collins2023, Liu2023_2020jgb, Callan2025}. 

\begin{figure*}
\centerline{\includegraphics[width=18cm]{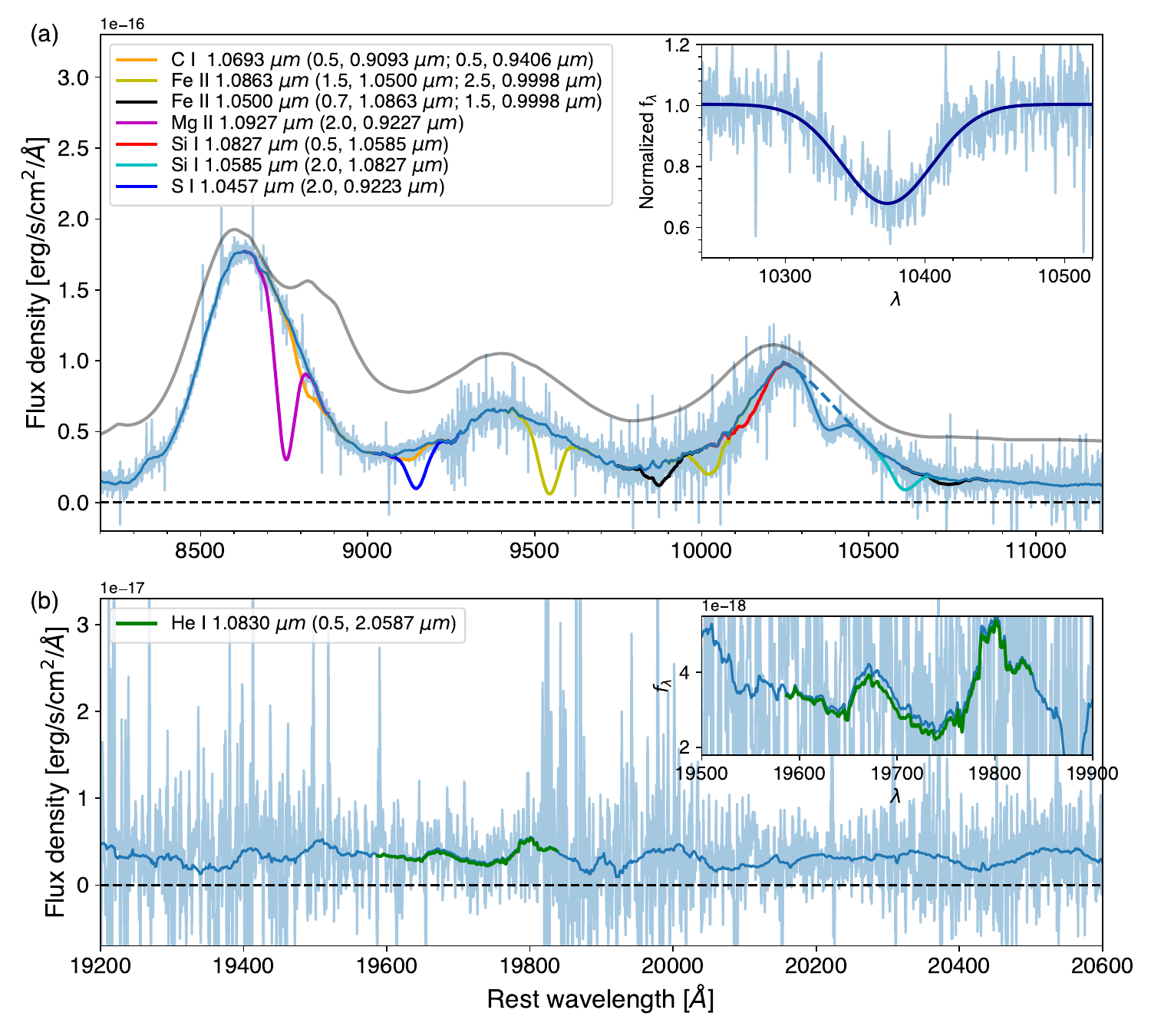}}
\caption{Line identification of the absorption feature in the NIR spectrum of SN\,2022an. The +55.4-day X-Shooter spectrum and its smoothed version are shown in blue, with the original data shown in a lighter shade. The synthetic spectrum of the SCH2p0 model at a phase of +58.9 days relative to the $B$-band maximum from \cite{Blondin2018} is shown in gray for comparison. The absorption feature at around 1.037 $\mu$m is superposed on an emission feature attributed to \ion{S}{2} \citep{Blondin2018}. The blue dashed line shows the equivalent pseudocontinuum, i.e., the estimated line profile of the underlying emission in the absence of the absorption feature. It is constructed by fitting the selected regions at 10,240--10,300 \AA\, and 10,450--10,550 \AA\, with a third-order polynomial. The inset in panel (a) shows the normalized spectrum (blue) of SN\,2022an around the absorption feature along with its best-fit Gaussian profile (dark blue). Assuming that the detected absorption feature arises from a given ion, the expected absorptions from other transitions of that same ion \citep{Marion2009} are shown as colored line segments. The legend labels have the format ``Ion $\lambda
_1 (k_2, \lambda_2; k_3, \lambda_3; ...)$,'' where $\lambda
_1$ is the rest-frame wavelength of the transition of the given ion to account for the observed absorption, $\lambda_{2,3,...}$ are the rest-frame wavelengths of other transitions of the same ion, and $k_{2,3,...}$ are the corresponding ratios of the mock absorptions at $\lambda_{2,3,...}$ relative to the observed absorption at $\lambda_1$. For example, if the observed absorption feature is identified as the \ion{Mg}{2} 1.0927 $\mu m$ line, then \ion{Mg}{2} 0.9227 $\mu m$ line, with twice the strength, would cause the absorption shown by the magenta line. The inset in panel (b) shows a zoomed-in view around the expected \ion{He}{1} 2.058 $\mu m$ line.}
\label{fig:nir_hei_iden}
\end{figure*}

\begin{figure*}
\centerline{\includegraphics[width=18cm]{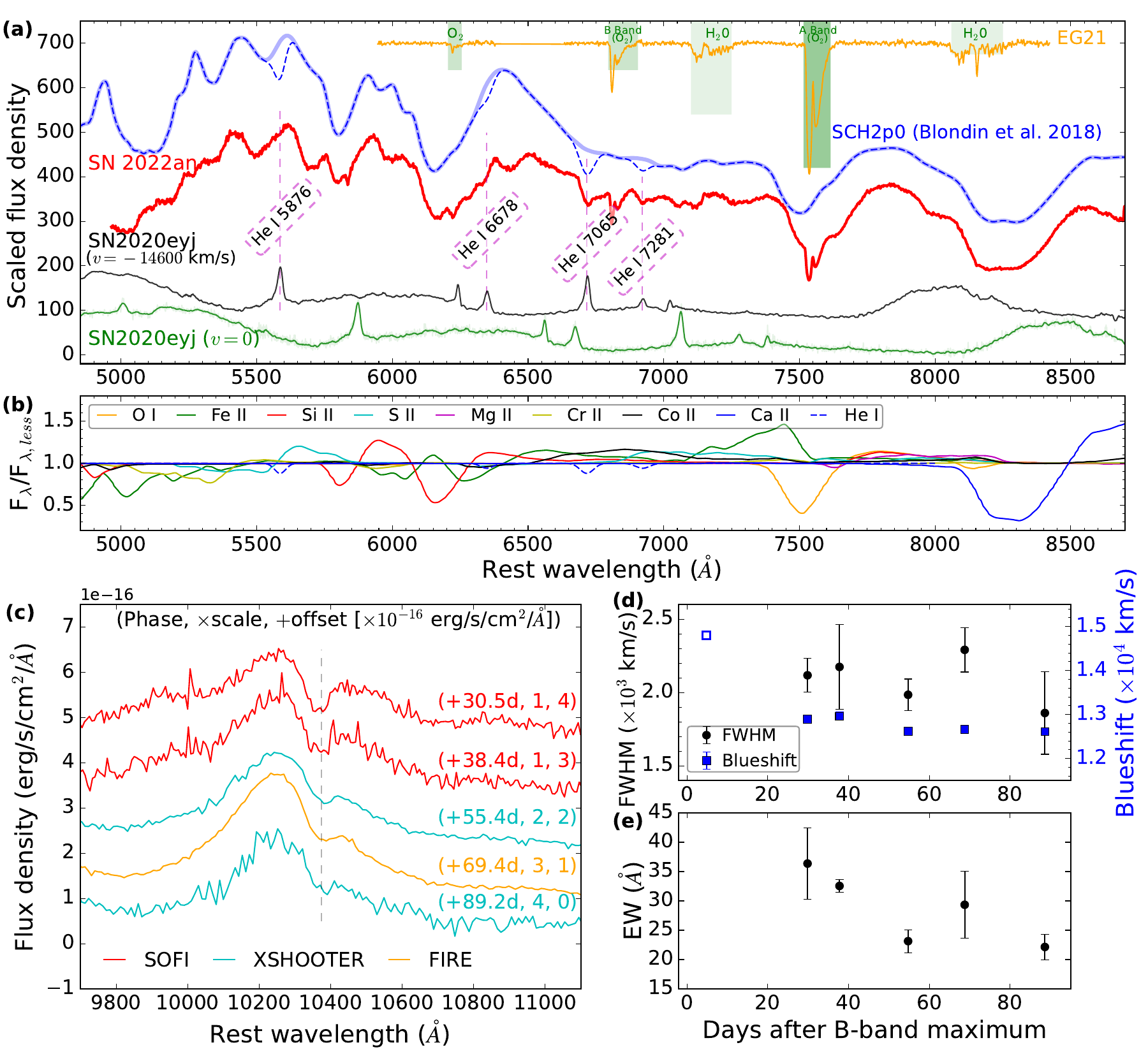}}
\caption{{\bf (a)} \ion{He}{1} absorption features in the optical spectrum of SN 2022an (red line), obtained with SOAR/GHTS at +5.3 days relative to $B$-band maximum. The blue vertical dashed lines denote the identified \ion{He}{1} lines at $\lambda$5876, $\lambda$6678, $\lambda$7065, $\lambda$7281. For comparison, spectra of SN 2020eyj, a SN Ia interacting with helium-rich CSM that exhibits \ion{He}{1} emission \citep{Kool2023_2020eyj}, are shown in green (rest-frame) and black (blueshifted by 14,600\kms). Also shown is the synthetic spectrum from a sub-$M_{\mathrm{CH}}$ explosion model \citep{Blondin2018}. The solid blue line represents the SCH2p0 model at +5.3 days, while the dashed blue line shows the same model after the addition of \ion{He}{1} absorption features (see Section~\ref{sec:helium_iden} for details). A telluric standard star (EG21) observed on the same night is plotted to indicate the location of atmospheric absorption bands.
{\bf (b)} Ion contributions to the synthetic spectrum shown in panel (a). The spectra of individual ions, all except \ion{He}{1} extracted from \cite{Blondin2018}, are represented by the ratio of the full spectrum (F$_\lambda$) to the spectrum excluding bound–bound transitions of the corresponding ion (F$_{\lambda, \rm{less}}$). Only ions that alter the flux by $>$10\% within the wavelength range 5000–8500 \AA\ are displayed. The \ion{He}{1} mock spectrum is indicated by the dashed line. This comparison illustrates how line blending complicates the identification of individual ions. 
{\bf (c)}  Zoomed-in NIR spectra of SN\,2022an centered on the \ion{He}{1} 1.083\,$\mu m$ absorption feature (marked by vertical dashed line). For clarity, the spectra have been scaled and vertically offset; the applied scaling factors, offsets, and phases relative to $B$-band maximum are indicated in parentheses. 
{\bf (d)}{\bf (e)} Measured properties of the \ion{He}{1} 1.083\,$\mu m$ line. The FWHM and the blueshifted velocity of the absorption line are presented in panel (d), while the equivalent width (EW) is presented in panel (e). The velocity measured from the optical helium lines is added as an unfilled symbol.}
\label{fig:HeI_iden}
\end{figure*}

Although SN 2022an closely resembles 91bg-like SNe Ia in general, it exhibits several atypical spectral features. The most significant is a persistent absorption trough around 1.037 $\mu m$ in the NIR spectra (see the highlighted region in the zoomed-in view of Figure~\ref{fig:specs}), a feature absent from other SNe Ia at similar phases. We attribute this feature to the blueshifted \ion{He}{1} 1.083 $\mu m$ line. This identification is supported by the following two arguments: (1) alternative identifications involving common SN~Ia ions are inconsistent with the absence of their accompanying comparable or stronger transitions (Section~\ref{sec:heI_nir}; see also Figure~\ref{fig:nir_hei_iden}); (2) several absorption lines identified in early-time optical spectra are consistent with arising from the same helium-rich material (Section~\ref{sec:heI_optical}; see also Figure~\ref{fig:HeI_iden}).

\subsection{Identification of the Near-infrared absorption}
\label{sec:heI_nir}

Figure~\ref{fig:nir_hei_iden} presents the relevant section of the SN\,2022an spectrum used for line identification. The telluric-corrected X-Shooter spectrum, obtained 54.8 days after $B$-band maximum, exhibits a well-defined absorption trough around 1.037\,$\mu$m. This absorption is superimposed on an \ion{S}{2} emission feature \citep{Blondin2018} and lies slightly redward of the \ion{S}{2} line center. The inset panel displays the normalized profile, which is well described by a Gaussian with a full width at half-maximum (FWHM) of $\sim70$\,\AA.

Given that the observed absorption in SN\,2022an has a blueshifted velocity of $v<20,000$\kms, the rest-frame wavelength of the responsible transition must fall between 1.037 and 1.106 $\mu$m. In addition to \ion{He}{1} 1.083\,$\mu$m, other candidate lines and their respective blueshifted velocities include 
\ion{He}{1} 1.083\,$\mu$m ($v=12740\,\rm km\,s^{-1}$), 
\ion{C}{1} 1.0693\,$\mu$m ($v=9{,}070$\kms), 
\ion{Mg}{2} 1.0927\,$\mu$m ($v=15{,}300$\kms), 
\ion{Fe}{2} 1.0500\,$\mu$m ($v=3{,}720$\kms) and 1.0863\,$\mu$m ($v=13{,}620$\kms), 
\ion{S}{1} 1.0457\,$\mu$m ($v=2{,}505$\kms), 
and \ion{Si}{1} 1.0827\,$\mu$m ($v=12{,}670$\kms) and 1.0585\,$\mu$m ($v=6{,}100$\kms). These are among the most prominent lines expected in supernova NIR spectra \citep{Valenti2008_2007gr, Marion2009, Shahbandeh2022}.

We first test these candidate ions observationally by checking for other transitions of the same species under LTE assumptions, while deferring questions of physical feasibility. The observed narrow absorption sits atop a broad emission pseudocontinuum, so any comparable narrow feature would stand out clearly. Except for He I, each ion capable of producing the observed trough has a corresponding transition of comparable or greater strength \citep{Marion2009}; the nondetection of these lines therefore argues against their identification.
For example, attributing the 1.037\,$\mu$m trough to \ion{Mg}{2} 1.0927\,$\mu$m would require the companion \ion{Mg}{2} 0.9227\,$\mu$m line to be present. Under LTE conditions at 5000--10000~K, the latter is expected to be 2--3 times stronger \citep{Marion2009}. NLTE model spectra \citep{Blondin2017, Blondin2018} independently confirm that the 0.9227\,$\mu$m line is of comparable or greater strength whenever the 1.0927\,$\mu$m line appears. The absence of this corresponding absorption (Figure~\ref{fig:nir_hei_iden}) therefore argues against the \ion{Mg}{2} identification. We applied the same test to the remaining candidate ions, simulating the expected strengths of companion lines using LTE ratios over the same temperature range and adopting conservative values equal to or smaller than the theoretical predictions.

The \ion{He}{1} 1.083\,$\mu$m line provides a viable explanation for the observed absorption in SN\,2022an. For \ion{He}{1}, the high excitation energy of the 1.083\,$\mu$m transition requires nonthermal processes. The NLTE calculations of \cite{Callan2025} predict a ratio of $\mathrm{EW(He\,I\, 2.058\,\mu m)}/\mathrm{EW(He\,I\,1.083\,\mu m)} \approx 0.4$ (see panel (b) of Figure~\ref{fig:hei_model_spec} and Section~\ref{sec:model_spec}). The expected line profile for the \ion{He}{1} 2.058\,$\mu$m transition, assuming an EW equal to half that of the 1.083\,$\mu$m line, is shown by the green curve in the bottom panel of Figure~\ref{fig:nir_hei_iden}. No clear absorption can be identified at the expected position of the \ion{He}{1} 2.058\,$\mu$m line given the signal-to-noise ratio of the observed spectrum. The supernova emission in the \ion{He}{1} 2.058\,$\mu$m region is weak, and the spectrum is highly affected by strong telluric absorption; the nondetection of a clear absorption feature cannot rule out the existence of \ion{He}{1} 2.058\,$\mu$m absorption consistent with the predicted EW \citet{Callan2025}.


The inferred blueshift of 12,740\kms~places the absorbing helium outside the silicon layer, whose blueshifted velocity is 9,400\kms~in the +5.3-day optical spectrum. Radiative-transfer calculations of helium-shell double-detonation models likewise produce comparable blueshifts for the \ion{He}{1} 1.083\,$\mu$m line (Figure~\ref{fig:hei_model_spec}).

In conclusion, none of the alternative ions can reproduce the observed 1.037\,$\mu$m absorption feature without simultaneously predicting additional lines of comparable or greater strength that are clearly absent from the data. In contrast, the \ion{He}{1} identification is fully consistent with the observed spectral morphology, expected line ratios, and blueshifted velocities. We therefore favor \ion{He}{1} 1.083\,$\mu$m as the most plausible origin of the absorption feature in SN\,2022an.

\subsection{Evidence of Optical \ion{He}{1} Absorption}
\label{sec:heI_optical}

To further test the helium interpretation, we examine the optical spectra to search for corresponding helium absorption. No absorption features are clearly detected in the optical spectra at epochs similar to those of the NIR spectra. Instead, absorption features that could originate from helium are seen only at earlier epochs. In panel (a) of Figure~\ref{fig:HeI_iden}, we show the +5.3\,day spectrum of SN\,2022an (red). For comparison, we include the synthetic spectrum from \citet{Blondin2018}, which reproduces SN~1999by well (blue); its individual ion contributions are shown in panel (b). Redward of 6000\,\AA, the overall agreement between SN\,2022an and the model is good, with \ion{Si}{2} $\lambda$6355, \ion{O}{1} $\lambda$7774, and the \ion{Ca}{2} NIR triplet dominating both spectra.

Two absorption dips appear around 6720 and 6930\,\AA\ in the SN\,2022an spectrum, well outside telluric regions. These features could be explained by \ion{He}{1} $\lambda$7065 and $\lambda$7281 at a blueshifted velocity around $\sim 14{,}800$\kms. If indeed from \ion{He}{1}, other accompanying optical \ion{He}{1} lines, such as \ion{He}{1} $\lambda$5876 and \ion{He}{1} $\lambda$6678, are also expected. These lines are observed, for example, in the Ia-CSM event SN~2020eyj \citep{Kool2023_2020eyj}, shown in rest-frame (orange) and blueshifted (black) in Figure~\ref{fig:HeI_iden}.  
 Both \ion{He}{1} $\lambda$5876 and \ion{He}{1} $\lambda$6678 lie within the observed spectral range but are not obviously detected in SN\,2022an due to complex line blending. To illustrate how these transitions may appear when blended with other ions, we add \ion{He}{1} absorption to the synthetic spectrum with an illustrative strength ratio:  
\[
f_{\lambda 5876} : f_{\lambda 6678} : f_{\lambda 7065} : f_{\lambda 7281} = 2:1:2:1.
\]

The mock contributions are shown as dashed lines in panel (b), and the resulting composite spectrum as a blue dashed curve in panel (a). The predicted and observed spectra exhibit similar detectability patterns: \ion{He}{1} $\lambda7065$ and $\lambda7281$ appear as distinct absorption dips, while $\lambda5876$ and $\lambda6678$ are blended with broad neighboring features. Even where these lines are unresolved, the spectral morphology of SN\,2022an is consistent with an additional \ion{He}{1} contribution. 

The \ion{He}{1} velocity measured from the optical spectrum at $+5.3$~days exceeds that in the NIR spectra at $+30.5$~days by $\sim$2000~km\,s$^{-1}$, consistent with the absorption layer receding into slower-moving ejecta as the supernova expands.

To summarize, the optical spectra of SN\,2022an at earlier phases exhibit features consistent with the presence of \ion{He}{1} absorption, which corroborate the identification of the \ion{He}{1} 1.083 $\mu$m line in Section~\ref{sec:heI_nir}. 

\subsection{Properties of the observed \ion{He}{1} absorptions}
\label{sec:heI_property}

Panel (c) of Figure~\ref{fig:HeI_iden} shows the observed NIR spectra of SN\,2022an around the region of the detected \ion{He}{1} 1.083 $\mu$m line. The absorption feature is persistent and appears in all five spectra, spanning epochs from around 1 month to 3 months after the $B$-band peak. 

Panel (d) of Figure~\ref{fig:HeI_iden} presents the blueshifted velocities derived from the absorption minima of the \ion{He}{1} lines (blue symbols), with filled points corresponding to the 1.083\,$\mu$m line and the open point to the optical lines. The velocities decline from $\sim 14{,}800$ to $\sim 13{,}000$\kms~over the first month after maximum light, consistent with the absorbing helium receding into a deeper layer due to a reduced heating rate of the outer layer by nonthermal processes.

The FWHM of the \ion{He}{1} 1.083\,$\mu$m line, measured from Gaussian fits, is shown in panel (d) (black points). The line width remains nearly constant at approximately 2000\kms. The high blueshifted velocity and a relatively narrow absorption feature imply that the line-forming region is restricted to a narrow range of the outer ejecta. The EW of the 1.083\,$\mu$m line decreases by roughly a factor of 2 over 2 months (panel (e)), consistent with a declining optical depth. This decline could result from the combined effects of decreasing density and weaker excitation as the ejecta expand and as radioactive $^{56}$Co decays.

\begin{figure*}
\centerline{\includegraphics[width=18cm]{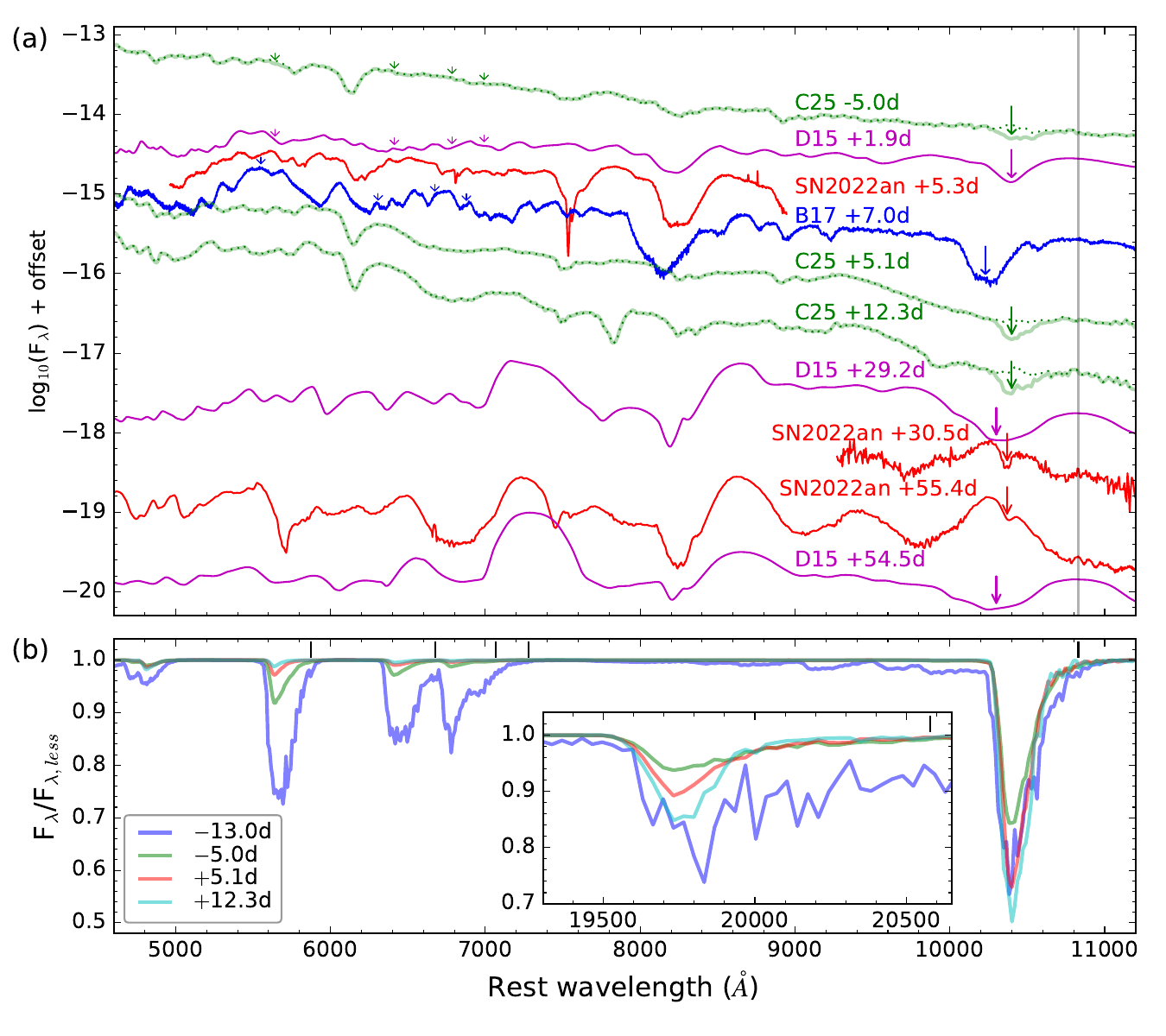}}
\caption{{\bf (a)} Spectra of SN\,2022an in red compared with model spectra from \cite{Dessart2015} (D15) in magenta, \cite{Boyle2017}(B17) in blue, and \cite{Callan2025}(C25) in green. For C25, both the full model spectra and those computed without \ion{He}{1} line opacity are plotted with solid and dotted lines, respectively. The phase of each spectrum, relative to the $B$-band peak in days, is labeled. The vertical black line indicates the rest-frame wavelength of the \ion{He}{1} 1.083 $\mu m$ line, while the corresponding observed absorption in SN\,2022an and the synthetic features in the model spectra are marked by arrows. Absorption features corresponding to the \ion{He}{1} $\lambda$5876, $\lambda$6678, $\lambda$7065, and $\lambda$7281 lines, which have the same blueshifted velocity as the 1.083 $\mu$m line, are indicated with short arrows in the model spectra for the earliest epochs of each category. {\bf (b)} Normalized version of the C25 model spectra, i.e., the full model spectra divided by the corresponding spectra computed without the He line opacity. Different colors indicate different phases. The wavelength range around \ion{He}{1} 2.058 $\mu m$ is shown in the inset panel. The rest wavelengths of the \ion{He}{1} lines are indicated by vertical black lines. }
\label{fig:hei_model_spec}
\end{figure*}

\subsection{Comparison with Model Predictions}
\label{sec:model_spec}
Helium is an essential ingredient in the helium-shell double-detonation scenario \citep{Nomoto1982a, Nomoto1982b}. In this framework, a surface helium layer on a sub-$M_{\mathrm{CH}}$ WD detonates, triggering a secondary carbon detonation in the core. Numerical simulations of such explosions generally predict that a significant fraction of unburned helium remains in the outermost ejecta \citep{Woosley1994, Fink2010, Shen2014}. Early radiative-transfer calculations, however, neglected nonthermal excitation and ionization, and therefore produced spectra that lacked \ion{He}{1} features \citep{Shen2010, Waldman2011}. \citet{Dessart2015} carried out time-dependent NLTE radiative-transfer simulations of the helium-shell detonation model COp45HEp2 of \citet{Waldman2011}, demonstrating that nonthermal processes naturally yield unambiguous \ion{He}{1} absorption features in both the optical and NIR at epochs around maximum light. Their models were tailored to faint, rapidly evolving transients such as .Ia explosions and Ca-rich events, in which the ejecta originate only from the helium-shell detonation itself.

\citet{Boyle2017} investigated the visibility of unburned helium in more contemporary double-detonation models designed to explain normal SNe\,Ia. Using an analytical approximation for the \ion{He}{1} level populations and the explosion models of \citet{Fink2010}, they predicted strong high-velocity absorption from \ion{He}{1}~$1.083\,\mu$m and \ion{He}{1}~$2.058\,\mu$m in a low-luminosity model containing 0.077 \msun\, of helium. Their approach assumed that \ion{He}{2} dominates the ionization balance, an assumption that may not hold in all regimes and therefore affects the predicted strength and temporal evolution of the \ion{He}{1} lines.

A more complete treatment was presented by \citet{Collins2023}, who computed full NLTE ionization and excitation, including nonthermal electrons for the M2a double-detonation model of \citet{Gronow2020}. Their synthetic spectra exhibit a prominent high-velocity ($\sim 19{,}000$\kms) \ion{He}{1}~$1.083\,\mu$m absorption feature. In these models, the \ion{He}{1} absorption is strongest during the first few days after explosion and fades rapidly, becoming negligible by roughly 2 weeks postexplosion.

Different realizations of the double-detonation scenario produce widely varying amounts and velocity distributions of unburned helium, and line-of-sight effects further increase the diversity of predicted spectral features. \citet{Callan2025} explored another set of NLTE radiative-transfer calculations for the M08\_03 double-detonation model of \citet{Gronow2021}, a sub-$M_{\mathrm{CH}}$  C--O WD with a $\sim$0.8\,\msun\ core and a $\sim$0.03\,\msun\ helium shell, using a 1D ejecta model constructed from the south polar region of the 3D explosion (total mass $1.01\,\msun$, $^{56}$Ni mass $0.47\,\msun$, and unburned He mass $\sim$0.04\,\msun). In this model, the surviving helium resides at significantly lower velocities, with the distribution peaking near $\sim$13{,}000\,\kms.

In Figure~\ref{fig:hei_model_spec}, we compare the $+5.3$, $+30.5$, and $+55.4$\,days spectra of SN\,2022an with representative model spectra from \citet{Dessart2015}, \citet{Boyle2017}, and \citet{Callan2025}. All of the models exhibit a clear \ion{He}{1}~$1.083\,\mu$m absorption, marked by vertical arrows. In contrast, the associated optical \ion{He}{1} transitions are difficult to identify in the model spectra, as they are weak and blended with broader neighboring features; their expected wavelengths are indicated for reference.

To better illustrate the \ion{He}{1} lines predicted by the models of \citet{Callan2025}, we show normalized spectra in panel~(b), using the synthetic spectra without \ion{He}{1} opacity as a pseudocontinuum. This reveals optical \ion{He}{1} features that are otherwise buried in the blended line forest seen in panel~(a). The optical \ion{He}{1} absorption is discernible only at very early times, whereas the NIR \ion{He}{1}~$1.083\,\mu$m and $2.058\,\mu$m lines strengthen with time in the days surrounding maximum light.  For a more quantitative analysis, we measure the EWs of the \ion{He}{1}~$1.083\,\mu$m line in the models as
$\mathrm{EW_{1.083\,\mu m}^{-5.0d} = 73\,\AA}$,
$\mathrm{EW_{1.083\,\mu m}^{+5.1d} = 89\,\AA}$, and
$\mathrm{EW_{1.083\,\mu m}^{+12.3d} = 104\,\AA}$.
For the \ion{He}{1}~$2.058\,\mu$m line, we get
$\mathrm{EW_{2.058\,\mu m}^{-5.0d} = 28\,\AA}$,
$\mathrm{EW_{2.058\,\mu m}^{+5.1d} = 34\,\AA}$, and
$\mathrm{EW_{2.058\,\mu m}^{+12.3d} = 41\,\AA}$. These measurements give a ratio of $\mathrm{EW_{2.058\,\mu m}}/\mathrm{EW_{1.083\,\mu m}}\approx0.4$.
Unfortunately, the available model spectra do not extend to the later epochs at which we observe SN\,2022an, preventing a direct one-to-one comparison at those times.

The \ion{He}{1}~$1.083\,\mu$m absorption in the \citet{Callan2025} models exhibits a blueshift similar to that observed in SN\,2022an. However, the model line is significantly broader, with an FWHM of $\sim 6{,}000$\kms, around 3 times the width of the corresponding feature in SN\,2022an. This highlights the diversity among double-detonation predictions, and suggests that the structure and kinematics of the helium layer may vary substantially across different realizations of the explosion scenario.

\begin{figure*}[t]
\centering  \includegraphics[width=\textwidth]{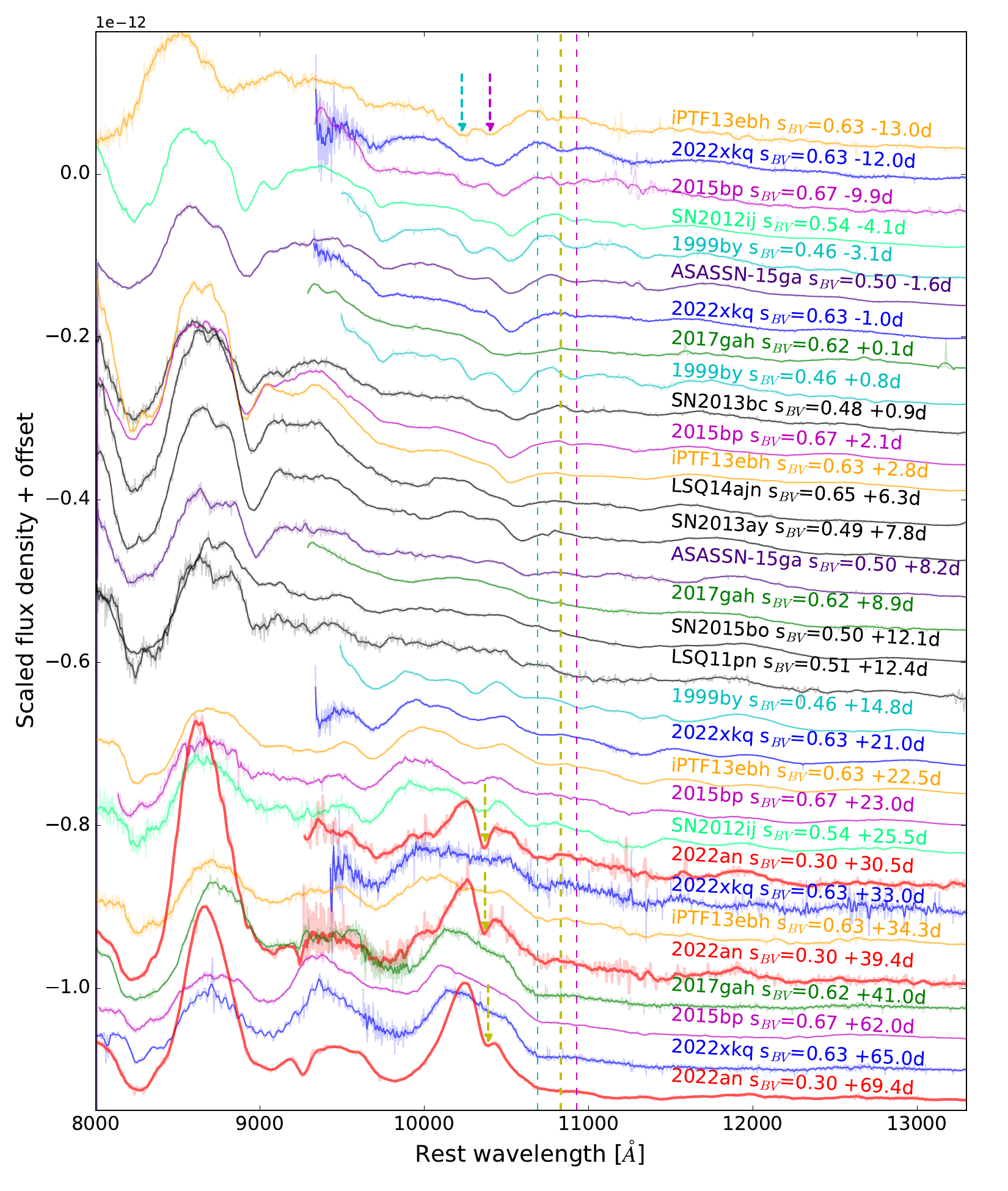} \\
\caption{NIR spectra of SN\,2022an compared with those of fast-declining SNe Ia with $s_{BV}<0.7$. The vertical dashed lines show the rest-frame wavelengths of \ion{C}{1} 1.069\,$\mu m$, \ion{He}{1} 1.083\,$\mu m$, and \ion{Mg}{2} 1.0927\,$\mu m$ in cyan, yellow, and magenta, respectively. The magenta arrow indicates the absorption feature commonly attributed to the \ion{Mg}{2} 1.0927\,$\mu m$ line in the literature. The cyan arrow indicates an absorption with ambiguous identifications, potentially corresponding to either \ion{C}{1} 1.069\,$\mu m$ or \ion{He}{1} 1.083\,$\mu m$. The yellow arrows mark the feature identified as \ion{He}{1} 1.083\,$\mu m$ absorption in SN\,2022an. The spectra are sorted by phase relative to the $B$-band peak, which is labeled after each object name together with the color-stretch parameter $s_{BV}$. The comparison objects include iPTF13ebh \citep{Hsiao2015}, SN\,2022xkq \citep{Pearson2024}, SN\,2015bp \citep{Wyatt2021}, SN2012ij \citep{Li2022_2012ij}, SN\,1999by \citep{Hoflich2002, Garnavich2004}, ASASSN-15ga \citep{Lu2023_CSP_SNIa_NIR}, SN\,2017gah \citep{Smartt2015, Chen2022_pmpyeasy}, SN\,2013bc \citep{Lu2023_CSP_SNIa_NIR}, LSQ14ajn \citep{Lu2023_CSP_SNIa_NIR}, SN\,2013ay \citep{Lu2023_CSP_SNIa_NIR}, SN\,2015bo \citep{Lu2023_CSP_SNIa_NIR}, LSQ11pn \citep{Lu2023_CSP_SNIa_NIR}. 
}
\label{fig:nir_spec_sequence_compare}
\end{figure*}

\section{Discussion}
\label{sec:discussion}


\subsection{Origin of the Helium}

Several possible origins for the helium must be considered. In the helium-shell double-detonation scenario, a significant fraction of unburned helium remaining in the outermost ejecta has long been predicted \citep{Woosley1994, Fink2010, Shen2014}. Such helium around the progenitor could be accreted from eithera degenerate or nondegenerate helium-star companion. It could also result from the merger of the progenitor with a low-mass helium WD \citep{Dan2014, Dan2015}. 

An alternative possibility is helium stripped from a nondegenerate helium-star companion. Hydrodynamic simulations show that a small fraction of the companion envelope becomes unbound, typically of order $10^{-3}$ to a few $\times10^{-2}$ \msun\, depending on the binary separation and stellar structure \citep[e.g.,][]{WangB2009, Pan2010, LiuZW2013}. The stripped material is mixed into the $\mathrm{^{56}Ni}$-rich ejecta, with a velocity distribution strongly peaked at low speeds: Most of the unbound helium resides at $\lesssim(2$--$5)\times10^{3}$\kms, with only a weak high-velocity tail. Such material is therefore expected to produce narrow nebular emission \citep{Botyanszki2018, Dessart2020} rather than high-velocity absorption. The helium absorption detected in SN\,2022an, by contrast, appears at much higher velocities of $\sim(1.3$--$1.5)\times10^{4}$\kms~and persists for several months, strongly disfavoring this scenario.

We also consider whether the helium could originate from a very low-mass helium WD, such as an AM CVn star analogous to GP Com ($M \approx 0.01\,\msun$) that orbits the progenitor at the time of explosion. We focus on a simple kinematic feasibility test: Can such a low-mass companion be accelerated to the high velocities observed in SN\,2022an? Treating this as an inelastic collision between the SN ejecta ($M_{\rm ej} \lesssim 1.4\,\msun$) and the helium WD, the companion velocity is $ v_{\rm c} = \frac{M_{\rm int}}{M_{\rm int} + M_{\rm c}} \, v_{\rm ej}$, where $M_{\rm int}$ is the intercepted ejecta mass and $M_{\rm c}$ is the companion mass. Using the helium WD mass--radius relation from \citet{Nelemans2001},
\begin{equation}
R \approx 0.0106 - 0.0064\,\ln(M_{\rm WD}) + 0.0015\,M_{\rm WD}^2,
\end{equation}
and assuming the companion fills its Roche lobe, we derive a companion star radius $R_{\rm c} \approx 0.04\,\rsun$ and an orbital semi-major axis $a \approx 0.44\,\rsun$. This configuration yields $v \lesssim 0.2\, v_{ej}$. Therefore, even a very low-mass AM CVn companion cannot be accelerated to the high helium velocities observed in SN\,2022an, disfavoring this scenario.

Thus, the helium velocity provides the most direct discriminant among the proposed origin scenarios: Stripped or dynamically accelerated helium is confined to low velocities, whereas SN\,2022an exhibits helium at $\sim1.3-1.5\times10^4$\kms.

Overall, the most natural explanation for the helium in SN\,2022an is unburned helium that was already present on the WD at the time of explosion, residing in the outermost layers of the progenitor and subsequently ejected at high velocities during the detonation. In this picture, the detected helium is an intrinsic component of the progenitor system rather than material stripped from or dynamically accelerated by a companion star.

\subsection{Physical Implications of the Helium Absorption}

The detection of persistent \ion{He}{1} absorption in SN\,2022an provides strong evidence for the presence of helium embedded in the outer ejecta. The nearly constant line width (FWHM $\sim2000$\kms) and the slow decline of the EW over $\sim60$~days imply that the absorbing helium occupies a well-defined velocity layer and remains at least partially optically thick well into the postmaximum phase. Such long-lived absorption requires a sustained population of metastable \ion{He}{1} levels, which must be maintained by nonthermal excitation from fast electrons produced by the radioactive decay of $^{56}$Ni and $^{56}$Co. This, in turn, indicates that a nonnegligible fraction of radioactive material or its decay products must penetrate into the high-velocity helium-rich layers.

At the same time, establishing a direct quantitative connection between the helium mass, its velocity distribution, and the resulting spectroscopic signatures remains challenging. The appearance of \ion{He}{1} lines depends sensitively on the spatial distribution of $^{56}$Ni, the degree of mixing, the local radiation field, and NLTE effects. Even within a given explosion model, the amount of unburned helium and its observable velocity range can vary significantly with viewing angle. Consequently, reproducing the observed helium features in SN\,2022an will require dedicated multidimensional hydrodynamic and radiative-transfer calculations tailored to individual progenitor configurations.

\subsection{Implications for progenitor and explosion models of 91bg-like SNe Ia and other WLR-sequence SNe Ia}

The observed helium absorption in SN\,2022an demands a helium-rich outer layer that underwent highly incomplete burning during the explosion. In sub-$M_{\mathrm{CH}}$ progenitors, the mass of the accreted helium shell is expected to increase toward lower WD masses, as a larger helium mass is required to reach ignition conditions on lower-mass, larger-radius WDs \citep{Shen2009, Fink2010, WoosleyKasen2011}. For 91bg-like SNe~Ia, helium-shell double-detonation models occupy a particularly favorable region of parameter space for producing observable helium signatures. These events are best matched by low-mass sub-Chandrasekhar WDs ($M_{\rm WD}\sim0.85$--$0.95$ \msun), which require comparatively massive helium shells to reach ignition conditions \citep{Fink2010, Sim2010}. In this low-mass regime, helium burning is expected to be highly incomplete due to lower shell densities and stronger detonation curvature, leaving a substantial fraction of unburned helium at high velocities \citep{Shen2014, Gronow2020}. Consequently, 91bg-like SNe~Ia are expected to be the subclass most likely to exhibit detectable \ion{He}{1} absorption, while such features may remain weak or absent in higher-luminosity SNe~Ia with thinner helium shells and more complete burning. In this picture, SN\,2022an represents an extreme but physically informative realization of a sub-$M_{\mathrm{CH}}$ explosion with an unusually prominent helium signature.

Double-degenerate merger channels are not mutually exclusive with helium-shell double-detonation models; instead, they can provide a natural physical pathway to such explosions. In particular, mergers between a C--O WD and a low-mass helium WD lead to the deposition of a substantial helium layer onto the primary, which may subsequently ignite under degenerate conditions and trigger a secondary detonation in the C--O core \citep{Woosley1986, Bildsten2007, Fink2010, Shen2010,  Dan2014, Dan2015}. Such C--O+He WD mergers preferentially occur at long delay times and yield lower total ejecta masses and reduced $^{56}$Ni production, naturally aligning with the fast-evolving, subluminous properties and old host environments of 91bg-like SNe~Ia. By contrast, mergers involving two C--O WDs generally lack a substantial helium reservoir and are therefore less likely to produce observable helium-rich outer ejecta.

The rarity of detected helium in literature SNe\,Ia may reflect both observational limitations and intrinsic diversity. Figure~\ref{fig:nir_spec_sequence_compare} shows a collection of NIR spectra for all $s_{BV}<0.7$ WLR-sequence SNe Ia with publicly available NIR spectra. 
All the observed NIR spectra in Figure~\ref{fig:nir_spec_sequence_compare} correspond to transitional SNe Ia except for SN\,2022an and SN\,1999by. For 91bg-like SNe Ia, the number of observed NIR spectra is too small. The nondetection of similar \ion{He}{1} absorption in SN\,1999by may be due in part to line-of-sight effects. If the helium has an aspherical distribution at explosion, for example, in a disklike structure in the postmerger remnant formed through a double WD merger \citep{Dan2014}, where helium excess has a preference on the equatorial plane. Only observed from the edge-on direction, the background radiation powered by the radioactive decay of $^{56}$Ni and $^{56}$Co could be absorbed. SN\,2022an may therefore represent either an extreme realization of a broader helium-bearing population of 91bg-like SNe~Ia or the first case in which favorable geometry and data quality have revealed an otherwise common but elusive feature.
For transitional SNe Ia, core-normal SNe Ia, and 91T-like SNe, if they come from helium-shell double detonations or violent double WD mergers, the absence of helium absorption features could simply be due to a lack of enough helium mass to produce 2022an-like absorption.

\subsection{Limitations and Future Prospects}
\label{sec:limit}
The primary limitation of our analysis is the lack of early-time NIR spectra, when the development of the \ion{He}{1} 1.083\,$\mu$m feature could be directly traced and compared with model predictions for its initial formation. A complete temporal sequence combining optical and NIR spectroscopy from the premaximum phase onward is essential for constraining the excitation mechanism, ionization balance, and helium mass in the outer ejecta. Future facilities optimized for rapid transient follow-up, such as SoXS \citep{Schipani2018}, will provide the sensitivity and cadence needed to systematically search for helium in faint, fast-declining SNe\,Ia and conduct detailed studies. A larger sample will clarify whether SN\,2022an represents a rare outlier or a key example of a broader population of helium-bearing sub-$M_\mathrm{{Ch}}$ explosions. On the theoretical side, systematic NLTE radiative-transfer calculations spanning the low-mass sub-$M_{\mathrm{Ch}}$ regime are needed to determine under which conditions helium absorption becomes detectable.

\acknowledgments
We thank Yuri Beletsky for the Magellan observation. We thank Doron Kushnir for useful discussions. We thank Aoife Boyle for sending the model spectra. We thank Wynn V. Jacobson-Galán for help in acquiring the raw data of the SOAR spectrum.

P.C. acknowledges support from the Zhejiang Provincial Top-Level Research Support Program.

This work is supported by the National Natural Science Foundation of China (grant No. 12133005). This research uses data obtained through the Telescope Access Program (TAP), which has been funded by the TAP member institutes. 

A.G.Y.'s research is supported by ISF, IMOS and BSF grants, as well as the André Deloro Institute for Space and Optics Research, the Center for Experimental Physics, a WIS-MIT Sagol grant, the Norman E. Alexander Family M Foundation ULTRASAT Data Center Fund, and Yeda-Sela;  A.G.Y. is the incumbent of the Arlyn Imberman Professorial Chair. 

K.M. acknowledges funding from Horizon Europe ERC grant no. 101125877.

T.-W.C. acknowledges financial support from the Yushan Fellow Program of the Ministry of Education, Taiwan (MOE-111-YSFMS-0008-001-P1), and from the National Science and Technology Council, Taiwan (NSTC grant No. 114-2112-M-008-021-MY3).

L.G. acknowledges financial support from CSIC, MCIN and AEI 10.13039/501100011033 under projects PID2023-151307NB-I00, PIE 20215AT016, and CEX2020-001058-M.

T.E.M.B. is funded by Horizon Europe ERC grant No. 101125877.

C.L. is supported by DoE award No. DE-SC0025599.

T.P. acknowledges the financial support from the Slovenian Research Agency (grants P1-0031, I0-0033, J1-2460, and N1-0344) and the Hungarian project NKFIH SNN-147362 grant.

This work makes use of observations from the Las Cumbres Observatory Global Telescope Network. Based on observations collected at the European Southern Observatory under ESO programmes 108.220C.006, 108.220C.010, 108.220C.011, 108.220C.012, 108.220C.016, 108.23MS.001 and 108.23MS.002. Based in part on observations obtained at the Southern Astrophysical Research (SOAR) telescope, which is a joint project of the Minist\'{e}rio da Ci\^{e}ncia, Tecnologia e Inova\c{c}\~{o}es (MCTI/LNA) do Brasil, the US National Science Foundation’s NOIRLab, the University of North Carolina at Chapel Hill (UNC), and Michigan State University (MSU). This work made use of the Heidelberg Supernova Model Archive (HESMA; \url{https://hesma.h-its.org}). This research made use of the NASA/IPAC Extragalactic Database (NED), which is funded by the National Aeronautics and Space Administration and operated by the California Institute of Technology.

\bibliography{ref}

@ARTICLE{Ogando2008,
       author = {{Ogando}, Ricardo L.~C. and {Maia}, Marcio A.~G. and {Pellegrini}, Paulo S. and {da Costa}, Luiz N.},
        title = "{Line Strengths of Early-Type Galaxies}",
      journal = {\aj},
         year = 2008,
        month = jun,
       volume = {135},
       number = {6},
        pages = {2424-2445},
          doi = {10.1088/0004-6256/135/6/2424},
archivePrefix = {arXiv},
       eprint = {0803.3477},
 primaryClass = {astro-ph},
       adsurl = {https://ui.adsabs.harvard.edu/abs/2008AJ....135.2424O}
}

@ARTICLE{Wegner2003,
       author = {{Wegner}, G. and {Bernardi}, M. and {Willmer}, C.~N.~A. and {da Costa}, L.~N. and {Alonso}, M.~V. and {Pellegrini}, P.~S. and {Maia}, M.~A.~G. and {Chaves}, O.~L. and {Rit{\'e}}, C.},
        title = "{Redshift-Distance Survey of Early-Type Galaxies: Spectroscopic Data}",
      journal = {\aj},
         year = 2003,
        month = nov,
       volume = {126},
       number = {5},
        pages = {2268-2280},
          doi = {10.1086/378959},
archivePrefix = {arXiv},
       eprint = {astro-ph/0308357},
 primaryClass = {astro-ph},
       adsurl = {https://ui.adsabs.harvard.edu/abs/2003AJ....126.2268W}
}

@ARTICLE{Tonry2000,
       author = {{Tonry}, John L. and {Blakeslee}, John P. and {Ajhar}, Edward A. and {Dressler}, Alan},
        title = "{The Surface Brightness Fluctuation Survey of Galaxy Distances. II. Local and Large-Scale Flows}",
      journal = {\apj},
         year = 2000,
        month = feb,
       volume = {530},
       number = {2},
        pages = {625-651},
          doi = {10.1086/308409},
archivePrefix = {arXiv},
       eprint = {astro-ph/9907062},
 primaryClass = {astro-ph},
       adsurl = {https://ui.adsabs.harvard.edu/abs/2000ApJ...530..625T}
}

@ARTICLE{Mieske2003,
       author = {{Mieske}, S. and {Hilker}, M.},
        title = "{Distance to the Centaurus cluster and its subcomponents  from surface brightness fluctuations}",
      journal = {\aap},
         year = 2003,
        month = nov,
       volume = {410},
        pages = {445-459},
          doi = {10.1051/0004-6361:20031296},
archivePrefix = {arXiv},
       eprint = {astro-ph/0309680},
 primaryClass = {astro-ph},
       adsurl = {https://ui.adsabs.harvard.edu/abs/2003A&A...410..445M}
}

@ARTICLE{Schlafly2011,
       author = {{Schlafly}, Edward F. and {Finkbeiner}, Douglas P.},
        title = "{Measuring Reddening with Sloan Digital Sky Survey Stellar Spectra and Recalibrating SFD}",
      journal = {\apj},
         year = 2011,
        month = aug,
       volume = {737},
       number = {2},
          eid = {103},
        pages = {103},
          doi = {10.1088/0004-637X/737/2/103},
archivePrefix = {arXiv},
       eprint = {1012.4804},
 primaryClass = {astro-ph.GA},
       adsurl = {https://ui.adsabs.harvard.edu/abs/2011ApJ...737..103S}
}

@ARTICLE{Fitzpatrick1999,
       author = {{Fitzpatrick}, Edward L.},
        title = "{Correcting for the Effects of Interstellar Extinction}",
      journal = {\pasp},
         year = 1999,
        month = jan,
       volume = {111},
       number = {755},
        pages = {63-75},
          doi = {10.1086/316293},
archivePrefix = {arXiv},
       eprint = {astro-ph/9809387},
 primaryClass = {astro-ph},
       adsurl = {https://ui.adsabs.harvard.edu/abs/1999PASP..111...63F}
}

@ARTICLE{Branch2006,
       author = {{Branch}, David and {Dang}, Leeann Chau and {Hall}, Nicholas and {Ketchum}, Wesley and {Melakayil}, Mercy and {Parrent}, Jerod and {Troxel}, M.~A. and {Casebeer}, D. and {Jeffery}, David J. and {Baron}, E.},
        title = "{Comparative Direct Analysis of Type Ia Supernova Spectra. II. Maximum Light}",
      journal = {\pasp},
         year = 2006,
        month = apr,
       volume = {118},
       number = {842},
        pages = {560-571},
          doi = {10.1086/502778},
archivePrefix = {arXiv},
       eprint = {astro-ph/0601048},
 primaryClass = {astro-ph},
       adsurl = {https://ui.adsabs.harvard.edu/abs/2006PASP..118..560B}
}

@INCOLLECTION{Phillips2017_WLR,
       author = {{Phillips}, Mark M. and {Burns}, Christopher R.},
        title = "{The Peak Luminosity - Decline Rate Relationship for Type Ia Supernovae}",
    booktitle = {Handbook of Supernovae},
         year = 2017,
       editor = {{Alsabti}, Athem W. and {Murdin}, Paul},
        pages = {2543},
          doi = {10.1007/978-3-319-21846-5_100},
       adsurl = {https://ui.adsabs.harvard.edu/abs/2017hsn..book.2543P}
}

@ARTICLE{Phillips2026,
       author = {{Phillips}, M.~M. and {Uddin}, Syed A. and {Burns}, Christopher R. and {Suntzeff}, Nicholas B. and {Ashall}, C. and {Baron}, E. and {Galbany}, L. and {Hoeflich}, P. and {Hsiao}, E.~Y. and {Morrell}, Nidia and {Persson}, S.~E. and {Stritzinger}, Maximilian and {Contreras}, Carlos and {Freedman}, Wendy L. and {Krisciunas}, Kevin and {Kumar}, S. and {Lu}, J. and {Piro}, Anthony L. and {Shahbandeh}, M.},
        title = "{Carnegie Supernova Project: Fast-declining Type Ia Supernovae as Cosmological Distance Indicators}",
      journal = {\apj},
         year = 2026,
        month = feb,
       volume = {998},
       number = {1},
          eid = {101},
        pages = {101},
          doi = {10.3847/1538-4357/ae2fef},
archivePrefix = {arXiv},
       eprint = {2509.07093},
 primaryClass = {astro-ph.CO},
       adsurl = {https://ui.adsabs.harvard.edu/abs/2026ApJ...998..101P}
}

@ARTICLE{Burns2011,
       author = {{Burns}, Christopher R. and {Stritzinger}, Maximilian and {Phillips}, M.~M. and {Kattner}, ShiAnne and {Persson}, S.~E. and {Madore}, Barry F. and {Freedman}, Wendy L. and {Boldt}, Luis and {Campillay}, Abdo and {Contreras}, Carlos and {Folatelli}, Gaston and {Gonzalez}, Sergio and {Krzeminski}, Wojtek and {Morrell}, Nidia and {Salgado}, Francisco and {Suntzeff}, Nicholas B.},
        title = "{The Carnegie Supernova Project: Light-curve Fitting with SNooPy}",
      journal = {\aj},
         year = 2011,
        month = jan,
       volume = {141},
       number = {1},
          eid = {19},
        pages = {19},
          doi = {10.1088/0004-6256/141/1/19},
archivePrefix = {arXiv},
       eprint = {1010.4040},
 primaryClass = {astro-ph.CO},
       adsurl = {https://ui.adsabs.harvard.edu/abs/2011AJ....141...19B}
}

@ARTICLE{Burns2014,
       author = {{Burns}, Christopher R. and {Stritzinger}, Maximilian and {Phillips}, M.~M. and {Hsiao}, E.~Y. and {Contreras}, Carlos and {Persson}, S.~E. and {Folatelli}, Gaston and {Boldt}, Luis and {Campillay}, Abdo and {Castell{\'o}n}, Sergio and {Freedman}, Wendy L. and {Madore}, Barry F. and {Morrell}, Nidia and {Salgado}, Francisco and {Suntzeff}, Nicholas B.},
        title = "{The Carnegie Supernova Project: Intrinsic Colors of Type Ia Supernovae}",
      journal = {\apj},
         year = 2014,
        month = jul,
       volume = {789},
       number = {1},
          eid = {32},
        pages = {32},
          doi = {10.1088/0004-637X/789/1/32},
archivePrefix = {arXiv},
       eprint = {1405.3934},
 primaryClass = {astro-ph.CO},
       adsurl = {https://ui.adsabs.harvard.edu/abs/2014ApJ...789...32B}
}

@ARTICLE{Burns2018,
       author = {{Burns}, Christopher R. and {Parent}, Emilie and {Phillips}, M.~M. and {Stritzinger}, Maximilian and {Krisciunas}, Kevin and {Suntzeff}, Nicholas B. and {Hsiao}, Eric Y. and {Contreras}, Carlos and {Anais}, Jorge and {Boldt}, Luis and {Busta}, Luis and {Campillay}, Abdo and {Castell{\'o}n}, Sergio and {Folatelli}, Gast{\'o}n and {Freedman}, Wendy L. and {Gonz{\'a}lez}, Consuelo and {Hamuy}, Mario and {Heoflich}, Peter and {Krzeminski}, Wojtek and {Madore}, Barry F. and {Morrell}, Nidia and {Persson}, S.~E. and {Roth}, Miguel and {Salgado}, Francisco and {Ser{\'o}n}, Jacqueline and {Torres}, Sim{\'o}n},
        title = "{The Carnegie Supernova Project: Absolute Calibration and the Hubble Constant}",
      journal = {\apj},
         year = 2018,
        month = dec,
       volume = {869},
       number = {1},
          eid = {56},
        pages = {56},
          doi = {10.3847/1538-4357/aae51c},
archivePrefix = {arXiv},
       eprint = {1809.06381},
 primaryClass = {astro-ph.CO},
       adsurl = {https://ui.adsabs.harvard.edu/abs/2018ApJ...869...56B}
}

@ARTICLE{Graur2024,
       author = {{Graur}, O.},
        title = "{Underluminous 1991bg-like Type Ia supernovae are standardizable candles}",
      journal = {\mnras},
         year = 2024,
        month = jun,
       volume = {530},
       number = {4},
        pages = {4950-4960},
          doi = {10.1093/mnras/stae949},
archivePrefix = {arXiv},
       eprint = {2311.16245},
 primaryClass = {astro-ph.HE},
       adsurl = {https://ui.adsabs.harvard.edu/abs/2024MNRAS.530.4950G}
}

@ARTICLE{Blondin2018,
       author = {{Blondin}, St{\'e}phane and {Dessart}, Luc and {Hillier}, D. John},
        title = "{The detonation of a sub-Chandrasekhar-mass white dwarf at the origin of the low-luminosity Type Ia supernova 1999by}",
      journal = {\mnras},
         year = 2018,
        month = mar,
       volume = {474},
       number = {3},
        pages = {3931-3953},
          doi = {10.1093/mnras/stx3058},
archivePrefix = {arXiv},
       eprint = {1711.09107},
 primaryClass = {astro-ph.SR},
       adsurl = {https://ui.adsabs.harvard.edu/abs/2018MNRAS.474.3931B}
}

@ARTICLE{Garnavich2004_1999by,
       author = {{Garnavich}, Peter M. and {Bonanos}, Alceste Z. and {Krisciunas}, Kevin and {Jha}, Saurabh and {Kirshner}, Robert P. and {Schlegel}, Eric M. and {Challis}, Peter and {Macri}, Lucas M. and {Hatano}, Kazuhito and {Branch}, David and {Bothun}, Gregory D. and {Freedman}, Wendy L.},
        title = "{The Luminosity of SN 1999by in NGC 2841 and the Nature of ``Peculiar'' Type Ia Supernovae}",
      journal = {\apj},
         year = 2004,
        month = oct,
       volume = {613},
       number = {2},
        pages = {1120-1132},
          doi = {10.1086/422986},
archivePrefix = {arXiv},
       eprint = {astro-ph/0105490},
 primaryClass = {astro-ph},
       adsurl = {https://ui.adsabs.harvard.edu/abs/2004ApJ...613.1120G}
}

@ARTICLE{Hoflich2002_1999by,
       author = {{H{\"o}flich}, Peter and {Gerardy}, Christopher L. and {Fesen}, Robert A. and {Sakai}, Shoko},
        title = "{Infrared Spectra of the Subluminous Type Ia Supernova SN 1999by}",
      journal = {\apj},
         year = 2002,
        month = apr,
       volume = {568},
       number = {2},
        pages = {791-806},
          doi = {10.1086/339063},
archivePrefix = {arXiv},
       eprint = {astro-ph/0112126},
 primaryClass = {astro-ph},
       adsurl = {https://ui.adsabs.harvard.edu/abs/2002ApJ...568..791H}
}

@ARTICLE{Krisciunas2017_CSP_dr3,
       author = {{Krisciunas}, Kevin and {Contreras}, Carlos and {Burns}, Christopher R. and {Phillips}, M.~M. and {Stritzinger}, Maximilian D. and {Morrell}, Nidia and {Hamuy}, Mario and {Anais}, Jorge and {Boldt}, Luis and {Busta}, Luis and {Campillay}, Abdo and {Castell{\'o}n}, Sergio and {Folatelli}, Gast{\'o}n and {Freedman}, Wendy L. and {Gonz{\'a}lez}, Consuelo and {Hsiao}, Eric Y. and {Krzeminski}, Wojtek and {Persson}, Sven Eric and {Roth}, Miguel and {Salgado}, Francisco and {Ser{\'o}n}, Jacqueline and {Suntzeff}, Nicholas B. and {Torres}, Sim{\'o}n and {Filippenko}, Alexei V. and {Li}, Weidong and {Madore}, Barry F. and {DePoy}, D.~L. and {Marshall}, Jennifer L. and {Rheault}, Jean-Philippe and {Villanueva}, Steven},
        title = "{The Carnegie Supernova Project. I. Third Photometry Data Release of Low-redshift Type Ia Supernovae and Other White Dwarf Explosions}",
      journal = {\aj},
         year = 2017,
        month = nov,
       volume = {154},
       number = {5},
          eid = {211},
        pages = {211},
          doi = {10.3847/1538-3881/aa8df0},
archivePrefix = {arXiv},
       eprint = {1709.05146},
 primaryClass = {astro-ph.IM},
       adsurl = {https://ui.adsabs.harvard.edu/abs/2017AJ....154..211K}
}

@ARTICLE{Pereira2013_2011fe,
       author = {{Pereira}, R. and {Thomas}, R.~C. and {Aldering}, G. and {Antilogus}, P. and {Baltay}, C. and {Benitez-Herrera}, S. and {Bongard}, S. and {Buton}, C. and {Canto}, A. and {Cellier-Holzem}, F. and {Chen}, J. and {Childress}, M. and {Chotard}, N. and {Copin}, Y. and {Fakhouri}, H.~K. and {Fink}, M. and {Fouchez}, D. and {Gangler}, E. and {Guy}, J. and {Hillebrandt}, W. and {Hsiao}, E.~Y. and {Kerschhaggl}, M. and {Kowalski}, M. and {Kromer}, M. and {Nordin}, J. and {Nugent}, P. and {Paech}, K. and {Pain}, R. and {P{\'e}contal}, E. and {Perlmutter}, S. and {Rabinowitz}, D. and {Rigault}, M. and {Runge}, K. and {Saunders}, C. and {Smadja}, G. and {Tao}, C. and {Taubenberger}, S. and {Tilquin}, A. and {Wu}, C.},
        title = "{Spectrophotometric time series of SN 2011fe from the Nearby Supernova Factory}",
      journal = {\aap},
         year = 2013,
        month = jun,
       volume = {554},
          eid = {A27},
        pages = {A27},
          doi = {10.1051/0004-6361/201221008},
archivePrefix = {arXiv},
       eprint = {1302.1292},
 primaryClass = {astro-ph.CO},
       adsurl = {https://ui.adsabs.harvard.edu/abs/2013A&A...554A..27P}
}

@ARTICLE{Wyatt2021_2015bp,
       author = {{Wyatt}, S.~D. and {Sand}, D.~J. and {Hsiao}, E.~Y. and {Burns}, C.~R. and {Valenti}, S. and {Bostroem}, K.~A. and {Lundquist}, M. and {Galbany}, L. and {Lu}, J. and {Ashall}, C. and {Diamond}, T.~R. and {Filippenko}, A.~V. and {Graham}, M.~L. and {Hoeflich}, P. and {Kirshner}, R.~P. and {Krisciunas}, K. and {Marion}, G.~H. and {Morrell}, N. and {Persson}, S.~E. and {Phillips}, M.~M. and {Stritzinger}, M.~D. and {Suntzeff}, N.~B. and {Taddia}, F.},
        title = "{Strong Near-infrared Carbon Absorption in the Transitional Type Ia SN 2015bp}",
      journal = {\apj},
         year = 2021,
        month = jun,
       volume = {914},
       number = {1},
          eid = {57},
        pages = {57},
          doi = {10.3847/1538-4357/abf7c3},
archivePrefix = {arXiv},
       eprint = {2012.02858},
 primaryClass = {astro-ph.HE},
       adsurl = {https://ui.adsabs.harvard.edu/abs/2021ApJ...914...57W}
}

@ARTICLE{Filippenko1992b_1991bg,
       author = {{Filippenko}, Alexei V. and {Richmond}, Michael W. and {Branch}, David and {Gaskell}, Martin and {Herbst}, William and {Ford}, Charles H. and {Treffers}, Richard R. and {Matheson}, Thomas and {Ho}, Luis C. and {Dey}, Arjun and {Sargent}, Wallace L.~W. and {Small}, Todd A. and {van Breugel}, Wil J.~M.},
        title = "{The Subluminous, Spectroscopically Peculiar Type 1a Supernova 1991bg in the Elliptical Galaxy NGC 4374}",
      journal = {\aj},
         year = 1992,
        month = oct,
       volume = {104},
        pages = {1543},
          doi = {10.1086/116339},
       adsurl = {https://ui.adsabs.harvard.edu/abs/1992AJ....104.1543F}
}

@ARTICLE{Leibundgut1993_1991bg,
       author = {{Leibundgut}, Bruno and {Kirshner}, Robert P. and {Phillips}, Mark M. and {Wells}, Lisa A. and {Suntzeff}, N.~B. and {Hamuy}, Mario and {Schommer}, R.~A. and {Walker}, A.~R. and {Gonzalez}, L. and {Ugarte}, P. and {Williams}, R.~E. and {Williger}, G. and {Gomez}, Mercedes and {Marzke}, Ronald and {Schmidt}, Brian P. and {Whitney}, Barbara and {Caldwell}, Nelson and {Peters}, J. and {Chaffee}, F.~H. and {Foltz}, Craig B. and {Rehner}, D. and {Siciliano}, L. and {Barnes}, T.~G. and {Cheng}, K.-P. and {Hintzen}, P.~M.~N. and {Kim}, Y.-C. and {Maza}, Jose and {Parker}, J. Wm. and {Porter}, A.~C. and {Schmidtke}, P.~C. and {Sonneborn}, George},
        title = "{SN 1991bg: A Type IA Supernova With a Difference}",
      journal = {\aj},
         year = 1993,
        month = jan,
       volume = {105},
        pages = {301},
          doi = {10.1086/116427},
       adsurl = {https://ui.adsabs.harvard.edu/abs/1993AJ....105..301L}
}

@ARTICLE{Filippenko1992a_91T,
       author = {{Filippenko}, Alexei V. and {Richmond}, Michael W. and {Matheson}, Thomas and {Shields}, Joseph C. and {Burbidge}, E. Margaret and {Cohen}, Ross D. and {Dickinson}, Mark and {Malkan}, Matthew A. and {Nelson}, Brant and {Pietz}, Jochen and {Schlegel}, David and {Schmeer}, Patrick and {Spinrad}, Hyron and {Steidel}, Charles C. and {Tran}, Hien D. and {Wren}, William},
        title = "{The Peculiar Type IA SN 1991T: Detonation of a White Dwarf?}",
      journal = {\apjl},
         year = 1992,
        month = jan,
       volume = {384},
        pages = {L15},
          doi = {10.1086/186252},
       adsurl = {https://ui.adsabs.harvard.edu/abs/1992ApJ...384L..15F}
}

@ARTICLE{Phillips1992_91T,
       author = {{Phillips}, M.~M. and {Wells}, Lisa A. and {Suntzeff}, Nicholas B. and {Hamuy}, Mario and {Leibundgut}, Bruno and {Kirshner}, Robert P. and {Foltz}, Craig B.},
        title = "{SN 1991T: Further Evidence of the Heterogeneous Nature of Type IA Supernovae}",
      journal = {\aj},
         year = 1992,
        month = may,
       volume = {103},
        pages = {1632},
          doi = {10.1086/116177},
       adsurl = {https://ui.adsabs.harvard.edu/abs/1992AJ....103.1632P}
}

@ARTICLE{Taubenberger2008_2005bl,
       author = {{Taubenberger}, S. and {Hachinger}, S. and {Pignata}, G. and {Mazzali}, P.~A. and {Contreras}, C. and {Valenti}, S. and {Pastorello}, A. and {Elias-Rosa}, N. and {B{\"a}rnbantner}, O. and {Barwig}, H. and {Benetti}, S. and {Dolci}, M. and {Fliri}, J. and {Folatelli}, G. and {Freedman}, W.~L. and {Gonzalez}, S. and {Hamuy}, M. and {Krzeminski}, W. and {Morrell}, N. and {Navasardyan}, H. and {Persson}, S.~E. and {Phillips}, M.~M. and {Ries}, C. and {Roth}, M. and {Suntzeff}, N.~B. and {Turatto}, M. and {Hillebrandt}, W.},
        title = "{The underluminous Type Ia supernova 2005bl and the class of objects similar to SN 1991bg}",
      journal = {\mnras},
         year = 2008,
        month = mar,
       volume = {385},
       number = {1},
        pages = {75-96},
          doi = {10.1111/j.1365-2966.2008.12843.x},
archivePrefix = {arXiv},
       eprint = {0711.4548},
 primaryClass = {astro-ph},
       adsurl = {https://ui.adsabs.harvard.edu/abs/2008MNRAS.385...75T}
}

@ARTICLE{Hachinger2009_2005bl,
       author = {{Hachinger}, Stephan and {Mazzali}, Paolo A. and {Taubenberger}, Stefan and {Pakmor}, R{\"u}diger and {Hillebrandt}, Wolfgang},
        title = "{Spectral analysis of the 91bg-like Type Ia SN 2005bl: low luminosity, low velocities, incomplete burning}",
      journal = {\mnras},
         year = 2009,
        month = nov,
       volume = {399},
       number = {3},
        pages = {1238-1254},
          doi = {10.1111/j.1365-2966.2009.15403.x},
archivePrefix = {arXiv},
       eprint = {0907.2542},
 primaryClass = {astro-ph.HE},
       adsurl = {https://ui.adsabs.harvard.edu/abs/2009MNRAS.399.1238H}
}

@ARTICLE{Folatelli2010,
       author = {{Folatelli}, Gast{\'o}n and {Phillips}, M.~M. and {Burns}, Christopher R. and {Contreras}, Carlos and {Hamuy}, Mario and {Freedman}, W.~L. and {Persson}, S.~E. and {Stritzinger}, Maximilian and {Suntzeff}, Nicholas B. and {Krisciunas}, Kevin and {Boldt}, Luis and {Gonz{\'a}lez}, Sergio and {Krzeminski}, Wojtek and {Morrell}, Nidia and {Roth}, Miguel and {Salgado}, Francisco and {Madore}, Barry F. and {Murphy}, David and {Wyatt}, Pamela and {Li}, Weidong and {Filippenko}, Alexei V. and {Miller}, Nicole},
        title = "{The Carnegie Supernova Project: Analysis of the First Sample of Low-Redshift Type-Ia Supernovae}",
      journal = {\aj},
         year = 2010,
        month = jan,
       volume = {139},
       number = {1},
        pages = {120-144},
          doi = {10.1088/0004-6256/139/1/120},
archivePrefix = {arXiv},
       eprint = {0910.3317},
 primaryClass = {astro-ph.CO},
       adsurl = {https://ui.adsabs.harvard.edu/abs/2010AJ....139..120F}
}

@ARTICLE{Folatelli2013_CSPspec_dr1,
       author = {{Folatelli}, Gast{\'o}n and {Morrell}, Nidia and {Phillips}, Mark M. and {Hsiao}, Eric and {Campillay}, Abdo and {Contreras}, Carlos and {Castell{\'o}n}, Sergio and {Hamuy}, Mario and {Krzeminski}, Wojtek and {Roth}, Miguel and {Stritzinger}, Maximilian and {Burns}, Christopher R. and {Freedman}, Wendy L. and {Madore}, Barry F. and {Murphy}, David and {Persson}, S.~E. and {Prieto}, Jos{\'e} L. and {Suntzeff}, Nicholas B. and {Krisciunas}, Kevin and {Anderson}, Joseph P. and {F{\"o}rster}, Francisco and {Maza}, Jos{\'e} and {Pignata}, Giuliano and {Rojas}, P. Andrea and {Boldt}, Luis and {Salgado}, Francisco and {Wyatt}, Pamela and {Olivares E.}, Felipe and {Gal-Yam}, Avishay and {Sako}, Masao},
        title = "{Spectroscopy of Type Ia Supernovae by the Carnegie Supernova Project}",
      journal = {\apj},
         year = 2013,
        month = aug,
       volume = {773},
       number = {1},
          eid = {53},
        pages = {53},
          doi = {10.1088/0004-637X/773/1/53},
archivePrefix = {arXiv},
       eprint = {1305.6997},
 primaryClass = {astro-ph.CO},
       adsurl = {https://ui.adsabs.harvard.edu/abs/2013ApJ...773...53F}
}

@ARTICLE{Goldstein2018,
       author = {{Goldstein}, Daniel A. and {Kasen}, Daniel},
        title = "{Evidence for Sub-Chandrasekhar Mass Type Ia Supernovae from an Extensive Survey of Radiative Transfer Models}",
      journal = {\apjl},
         year = 2018,
        month = jan,
       volume = {852},
       number = {2},
          eid = {L33},
        pages = {L33},
          doi = {10.3847/2041-8213/aaa409},
archivePrefix = {arXiv},
       eprint = {1801.00789},
 primaryClass = {astro-ph.HE},
       adsurl = {https://ui.adsabs.harvard.edu/abs/2018ApJ...852L..33G}
}

@ARTICLE{Kool2023_2020eyj,
       author = {{Kool}, Erik C. and {Johansson}, Joel and {Sollerman}, Jesper and {Mold{\'o}n}, Javier and {Moriya}, Takashi J. and {Mattila}, Seppo and {Schulze}, Steve and {Chomiuk}, Laura and {P{\'e}rez-Torres}, Miguel and {Harris}, Chelsea and {Lundqvist}, Peter and {Graham}, Matthew and {Yang}, Sheng and {Perley}, Daniel A. and {Strotjohann}, Nora Linn and {Fremling}, Christoffer and {Gal-Yam}, Avishay and {Lezmy}, Jeremy and {Maguire}, Kate and {Omand}, Conor and {Smith}, Mathew and {Andreoni}, Igor and {Bellm}, Eric C. and {Bloom}, Joshua S. and {De}, Kishalay and {Groom}, Steven L. and {Kasliwal}, Mansi M. and {Masci}, Frank J. and {Medford}, Michael S. and {Park}, Sungmin and {Purdum}, Josiah and {Reynolds}, Thomas M. and {Riddle}, Reed and {Robert}, Estelle and {Ryder}, Stuart D. and {Sharma}, Yashvi and {Stern}, Daniel},
        title = "{A radio-detected type Ia supernova with helium-rich circumstellar material}",
      journal = {\nat},
         year = 2023,
        month = may,
       volume = {617},
       number = {7961},
        pages = {477-482},
          doi = {10.1038/s41586-023-05916-w},
archivePrefix = {arXiv},
       eprint = {2210.07725},
 primaryClass = {astro-ph.HE},
       adsurl = {https://ui.adsabs.harvard.edu/abs/2023Natur.617..477K}
}

@ARTICLE{Dessart2015,
       author = {{Dessart}, Luc and {Hillier}, D. John},
        title = "{One-dimensional non-LTE time-dependent radiative transfer of an He-detonation model and the connection to faint and fast-decaying supernovae}",
      journal = {\mnras},
         year = 2015,
        month = feb,
       volume = {447},
       number = {2},
        pages = {1370-1382},
          doi = {10.1093/mnras/stu2520},
archivePrefix = {arXiv},
       eprint = {1411.7397},
 primaryClass = {astro-ph.SR},
       adsurl = {https://ui.adsabs.harvard.edu/abs/2015MNRAS.447.1370D}
}

@ARTICLE{Dessart2020,
       author = {{Dessart}, Luc and {Leonard}, Douglas C. and {Prieto}, Jose L.},
        title = "{Spectral signatures of H-rich material stripped from a non-degenerate companion by a Type Ia supernova}",
      journal = {\aap},
         year = 2020,
        month = jun,
       volume = {638},
          eid = {A80},
        pages = {A80},
          doi = {10.1051/0004-6361/202037854},
archivePrefix = {arXiv},
       eprint = {2004.03986},
 primaryClass = {astro-ph.SR},
       adsurl = {https://ui.adsabs.harvard.edu/abs/2020A&A...638A..80D}
}

@ARTICLE{Boyle2017,
       author = {{Boyle}, Aoife and {Sim}, Stuart A. and {Hachinger}, Stephan and {Kerzendorf}, Wolfgang},
        title = "{Helium in double-detonation models of type Ia supernovae}",
      journal = {\aap},
         year = 2017,
        month = mar,
       volume = {599},
          eid = {A46},
        pages = {A46},
          doi = {10.1051/0004-6361/201629712},
archivePrefix = {arXiv},
       eprint = {1611.05938},
 primaryClass = {astro-ph.HE},
       adsurl = {https://ui.adsabs.harvard.edu/abs/2017A&A...599A..46B}
}

@ARTICLE{Gronow2020,
       author = {{Gronow}, Sabrina and {Collins}, Christine and {Ohlmann}, Sebastian T. and {Pakmor}, R{\"u}diger and {Kromer}, Markus and {Seitenzahl}, Ivo R. and {Sim}, Stuart A. and {R{\"o}pke}, Friedrich K.},
        title = "{SNe Ia from double detonations: Impact of core-shell mixing on the carbon ignition mechanism}",
      journal = {\aap},
         year = 2020,
        month = mar,
       volume = {635},
          eid = {A169},
        pages = {A169},
          doi = {10.1051/0004-6361/201936494},
archivePrefix = {arXiv},
       eprint = {2002.00981},
 primaryClass = {astro-ph.SR},
       adsurl = {https://ui.adsabs.harvard.edu/abs/2020A&A...635A.169G}
}

@ARTICLE{Gronow2021,
       author = {{Gronow}, Sabrina and {Collins}, Christine E. and {Sim}, Stuart A. and {R{\"o}pke}, Friedrich K.},
        title = "{Double detonations of sub-M$_{Ch}$ CO white dwarfs: variations in Type Ia supernovae due to different core and He shell masses}",
      journal = {\aap},
         year = 2021,
        month = may,
       volume = {649},
          eid = {A155},
        pages = {A155},
          doi = {10.1051/0004-6361/202039954},
archivePrefix = {arXiv},
       eprint = {2102.06719},
 primaryClass = {astro-ph.SR},
       adsurl = {https://ui.adsabs.harvard.edu/abs/2021A&A...649A.155G}
}

@ARTICLE{Callan2025,
       author = {{Callan}, F.~P. and {Collins}, C.~E. and {Sim}, S.~A. and {Shingles}, L.~J. and {Pakmor}, R. and {Srivastav}, S. and {Pollin}, J.~M. and {Gronow}, S. and {R{\"o}pke}, F.~K. and {Seitenzahl}, I.~R.},
        title = "{Exploring the range of impacts of helium in the spectra of double detonation models for Type Ia supernovae}",
      journal = {\mnras},
         year = 2025,
        month = may,
       volume = {539},
       number = {2},
        pages = {1404-1413},
          doi = {10.1093/mnras/staf539},
archivePrefix = {arXiv},
       eprint = {2408.03048},
 primaryClass = {astro-ph.HE},
       adsurl = {https://ui.adsabs.harvard.edu/abs/2025MNRAS.539.1404C}
}

@ARTICLE{WangB2009,
       author = {{Wang}, B. and {Meng}, X. and {Chen}, X. and {Han}, Z.},
        title = "{The helium star donor channel for the progenitors of Type Ia supernovae}",
      journal = {\mnras},
         year = 2009,
        month = may,
       volume = {395},
       number = {2},
        pages = {847-854},
          doi = {10.1111/j.1365-2966.2009.14545.x},
archivePrefix = {arXiv},
       eprint = {0901.3496},
 primaryClass = {astro-ph.SR},
       adsurl = {https://ui.adsabs.harvard.edu/abs/2009MNRAS.395..847W}
}

@ARTICLE{Pan2010,
       author = {{Pan}, Kuo-Chuan and {Ricker}, Paul M. and {Taam}, Ronald E.},
        title = "{Impact of Type Ia Supernova Ejecta on a Helium-star Binary Companion}",
      journal = {\apj},
         year = 2010,
        month = may,
       volume = {715},
       number = {1},
        pages = {78-85},
          doi = {10.1088/0004-637X/715/1/78},
archivePrefix = {arXiv},
       eprint = {1004.0288},
 primaryClass = {astro-ph.HE},
       adsurl = {https://ui.adsabs.harvard.edu/abs/2010ApJ...715...78P}
}

@ARTICLE{Brown2013_lcogt,
       author = {{Brown}, T.~M. and {Baliber}, N. and {Bianco}, F.~B. and {Bowman}, M. and {Burleson}, B. and {Conway}, P. and {Crellin}, M. and {Depagne}, {\'E}. and {De Vera}, J. and {Dilday}, B. and {Dragomir}, D. and {Dubberley}, M. and {Eastman}, J.~D. and {Elphick}, M. and {Falarski}, M. and {Foale}, S. and {Ford}, M. and {Fulton}, B.~J. and {Garza}, J. and {Gomez}, E.~L. and {Graham}, M. and {Greene}, R. and {Haldeman}, B. and {Hawkins}, E. and {Haworth}, B. and {Haynes}, R. and {Hidas}, M. and {Hjelstrom}, A.~E. and {Howell}, D.~A. and {Hygelund}, J. and {Lister}, T.~A. and {Lobdill}, R. and {Martinez}, J. and {Mullins}, D.~S. and {Norbury}, M. and {Parrent}, J. and {Paulson}, R. and {Petry}, D.~L. and {Pickles}, A. and {Posner}, V. and {Rosing}, W.~E. and {Ross}, R. and {Sand}, D.~J. and {Saunders}, E.~S. and {Shobbrook}, J. and {Shporer}, A. and {Street}, R.~A. and {Thomas}, D. and {Tsapras}, Y. and {Tufts}, J.~R. and {Valenti}, S. and {Vander Horst}, K. and {Walker}, Z. and {White}, G. and {Willis}, M.},
        title = "{Las Cumbres Observatory Global Telescope Network}",
      journal = {\pasp},
         year = 2013,
        month = sep,
       volume = {125},
       number = {931},
        pages = {1031},
          doi = {10.1086/673168},
archivePrefix = {arXiv},
       eprint = {1305.2437},
 primaryClass = {astro-ph.IM},
       adsurl = {https://ui.adsabs.harvard.edu/abs/2013PASP..125.1031B}
}

@software{McCully2018_lcogt_Banzai,
       author = {{McCully}, Curtis and {Turner}, Monica and {Volgenau}, N and {Harbeck}, Daniel and {Valenti}, Stefano and {Riba}, Austin and {Bachelet}, Etienne and {Snyder}, Ira W. and {Kurczynski}, Brodie and {Norbury}, Martin and {Street}, Rachel},
        title = "{LCOGT/banzai: Initial Release}",
         year = 2018,
        month = jun,
          eid = {10.5281/zenodo.1257560},
          doi = {10.5281/zenodo.1257560},
      version = {0.9.4},
    publisher = {Zenodo},
       adsurl = {https://ui.adsabs.harvard.edu/abs/2018zndo...1257560M}
}

@ARTICLE{Tonry2018_refcat2,
       author = {{Tonry}, J.~L. and {Denneau}, L. and {Flewelling}, H. and {Heinze}, A.~N. and {Onken}, C.~A. and {Smartt}, S.~J. and {Stalder}, B. and {Weiland}, H.~J. and {Wolf}, C.},
        title = "{The ATLAS All-Sky Stellar Reference Catalog}",
      journal = {\apj},
         year = 2018,
        month = nov,
       volume = {867},
       number = {2},
          eid = {105},
        pages = {105},
          doi = {10.3847/1538-4357/aae386},
archivePrefix = {arXiv},
       eprint = {1809.09157},
 primaryClass = {astro-ph.IM},
       adsurl = {https://ui.adsabs.harvard.edu/abs/2018ApJ...867..105T}
}

@ARTICLE{Tonry2012,
       author = {{Tonry}, J.~L. and {Stubbs}, C.~W. and {Lykke}, K.~R. and {Doherty}, P. and {Shivvers}, I.~S. and {Burgett}, W.~S. and {Chambers}, K.~C. and {Hodapp}, K.~W. and {Kaiser}, N. and {Kudritzki}, R. -P. and {Magnier}, E.~A. and {Morgan}, J.~S. and {Price}, P.~A. and {Wainscoat}, R.~J.},
        title = "{The Pan-STARRS1 Photometric System}",
      journal = {\apj},
         year = 2012,
        month = may,
       volume = {750},
       number = {2},
          eid = {99},
        pages = {99},
          doi = {10.1088/0004-637X/750/2/99},
archivePrefix = {arXiv},
       eprint = {1203.0297},
 primaryClass = {astro-ph.IM},
       adsurl = {https://ui.adsabs.harvard.edu/abs/2012ApJ...750...99T}
}

@ARTICLE{Chen2022_pmpyeasy,
       author = {{Chen}, Ping and {Dong}, Subo and {Kochanek}, C.~S. and {Stanek}, K.~Z. and {Post}, R.~S. and {Stritzinger}, M.~D. and {Prieto}, J.~L. and {Filippenko}, Alexei V. and {Kollmeier}, Juna A. and {Elias-Rosa}, N. and {Katz}, Boaz and {Tomasella}, Lina and {Bose}, S. and {Ashall}, Chris and {Benetti}, S. and {Bersier}, D. and {Brimacombe}, Joseph and {Brink}, Thomas G. and {Brown}, P. and {Buckley}, David A.~H. and {Cappellaro}, Enrico and {Christie}, Grant W. and {Fraser}, Morgan and {Gromadzki}, Mariusz and {Holoien}, Thomas W. -S. and {Hu}, Shaoming and {Kankare}, Erkki and {Koff}, Robert and {Lundqvist}, P. and {Mattila}, S. and {Milne}, P.~A. and {Morrell}, Nidia and {Mu{\~n}oz}, J.~A. and {Mutel}, Robert and {Natusch}, Tim and {Nicolas}, Joel and {Pastorello}, A. and {Prentice}, Simon and {Roth}, Tyler and {Shappee}, B.~J. and {Stone}, Geoffrey and {Thompson}, Todd A. and {Villanueva}, Steven and {Zheng}, WeiKang},
        title = "{The First Data Release of CNIa0.02-A Complete Nearby (Redshift <0.02) Sample of Type Ia Supernova Light Curves}",
      journal = {\apjs},
         year = 2022,
        month = apr,
       volume = {259},
       number = {2},
          eid = {53},
        pages = {53},
          doi = {10.3847/1538-4365/ac50b7},
archivePrefix = {arXiv},
       eprint = {2011.02461},
 primaryClass = {astro-ph.HE},
       adsurl = {https://ui.adsabs.harvard.edu/abs/2022ApJS..259...53C}
}

@ARTICLE{Chen2023,
       author = {{Chen}, Ping and {Dong}, Subo and {Ashall}, Chris and {Benetti}, S. and {Bersier}, D. and {Bose}, S. and {Brimacombe}, Joseph and {Brink}, Thomas G. and {Buckley}, David A.~H. and {Cappellaro}, Enrico and {Christie}, Grant W. and {Elias-Rosa}, N. and {Filippenko}, Alexei V. and {Gromadzki}, Mariusz and {Holoien}, Thomas W.-S. and {Hu}, Shaoming and {Kochanek}, C.~S. and {Koff}, Robert and {Kollmeier}, Juna A. and {Lundqvist}, P. and {Mattila}, S. and {Milne}, Peter A. and {Mu{\~n}oz}, J.~A. and {Mutel}, Robert and {Natusch}, Tim and {Nicolas}, Joel and {Pastorello}, A. and {Prentice}, Simon and {Prieto}, J.~L. and {Roth}, Tyler and {Shappee}, B.~J. and {Stone}, Geoffrey and {Stanek}, K.~Z. and {Stritzinger}, M.~D. and {Thompson}, Todd A. and {Tomasella}, Lina and {Villanueva}, Steven},
        title = "{A Linear Relation between the Color Stretch s $_{ BV }$ and the Rising Color Slope \{s\}\_\{0\}(\{*) \}(B-V) of Type Ia Supernovae}",
      journal = {\apj},
         year = 2023,
        month = apr,
       volume = {946},
       number = {2},
          eid = {101},
        pages = {101},
          doi = {10.3847/1538-4357/acc404},
archivePrefix = {arXiv},
       eprint = {2112.13364},
 primaryClass = {astro-ph.HE},
       adsurl = {https://ui.adsabs.harvard.edu/abs/2023ApJ...946..101C}
}

@ARTICLE{Chen2024,
       author = {{Chen}, Ping and {Gal-Yam}, Avishay and {Sollerman}, Jesper and {Schulze}, Steve and {Post}, Richard S. and {Liu}, Chang and {Ofek}, Eran O. and {Das}, Kaustav K. and {Fremling}, Christoffer and {Horesh}, Assaf and {Katz}, Boaz and {Kushnir}, Doron and {Kasliwal}, Mansi M. and {Kulkarni}, Shri R. and {Liu}, Dezi and {Liu}, Xiangkun and {Miller}, Adam A. and {Rose}, Kovi and {Waxman}, Eli and {Yang}, Sheng and {Yao}, Yuhan and {Zackay}, Barak and {Bellm}, Eric C. and {Dekany}, Richard and {Drake}, Andrew J. and {Fang}, Yuan and {Fynbo}, Johan P.~U. and {Groom}, Steven L. and {Helou}, George and {Irani}, Ido and {Jegou du Laz}, Theophile and {Liu}, Xiaowei and {Mazzali}, Paolo A. and {Neill}, James D. and {Qin}, Yu-Jing and {Riddle}, Reed L. and {Sharon}, Amir and {Strotjohann}, Nora L. and {Wold}, Avery and {Yan}, Lin},
        title = "{A 12.4-day periodicity in a close binary system after a supernova}",
      journal = {\nat},
         year = 2024,
        month = jan,
       volume = {625},
       number = {7994},
        pages = {253-258},
          doi = {10.1038/s41586-023-06787-x},
archivePrefix = {arXiv},
       eprint = {2310.07784},
 primaryClass = {astro-ph.HE},
       adsurl = {https://ui.adsabs.harvard.edu/abs/2024Natur.625..253C}
}

@ARTICLE{Buzzoni1984_efosc,
       author = {{Buzzoni}, B. and {Delabre}, B. and {Dekker}, H. and {Dodorico}, S. and {Enard}, D. and {Focardi}, P. and {Gustafsson}, B. and {Nees}, W. and {Paureau}, J. and {Reiss}, R.},
        title = "{The ESO Faint Object Spectrograph and Camera / EFOSC}",
      journal = {The Messenger},
         year = 1984,
        month = dec,
       volume = {38},
        pages = {9},
       adsurl = {https://ui.adsabs.harvard.edu/abs/1984Msngr..38....9B}
}

@ARTICLE{Smartt2015_pessto,
       author = {{Smartt}, S.~J. and {Valenti}, S. and {Fraser}, M. and {Inserra}, C. and {Young}, D.~R. and {Sullivan}, M. and {Pastorello}, A. and {Benetti}, S. and {Gal-Yam}, A. and {Knapic}, C. and {Molinaro}, M. and {Smareglia}, R. and {Smith}, K.~W. and {Taubenberger}, S. and {Yaron}, O. and {Anderson}, J.~P. and {Ashall}, C. and {Balland}, C. and {Baltay}, C. and {Barbarino}, C. and {Bauer}, F.~E. and {Baumont}, S. and {Bersier}, D. and {Blagorodnova}, N. and {Bongard}, S. and {Botticella}, M.~T. and {Bufano}, F. and {Bulla}, M. and {Cappellaro}, E. and {Campbell}, H. and {Cellier-Holzem}, F. and {Chen}, T. -W. and {Childress}, M.~J. and {Clocchiatti}, A. and {Contreras}, C. and {Dall'Ora}, M. and {Danziger}, J. and {de Jaeger}, T. and {De Cia}, A. and {Della Valle}, M. and {Dennefeld}, M. and {Elias-Rosa}, N. and {Elman}, N. and {Feindt}, U. and {Fleury}, M. and {Gall}, E. and {Gonzalez-Gaitan}, S. and {Galbany}, L. and {Morales Garoffolo}, A. and {Greggio}, L. and {Guillou}, L.~L. and {Hachinger}, S. and {Hadjiyska}, E. and {Hage}, P.~E. and {Hillebrandt}, W. and {Hodgkin}, S. and {Hsiao}, E.~Y. and {James}, P.~A. and {Jerkstrand}, A. and {Kangas}, T. and {Kankare}, E. and {Kotak}, R. and {Kromer}, M. and {Kuncarayakti}, H. and {Leloudas}, G. and {Lundqvist}, P. and {Lyman}, J.~D. and {Hook}, I.~M. and {Maguire}, K. and {Manulis}, I. and {Margheim}, S.~J. and {Mattila}, S. and {Maund}, J.~R. and {Mazzali}, P.~A. and {McCrum}, M. and {McKinnon}, R. and {Moreno-Raya}, M.~E. and {Nicholl}, M. and {Nugent}, P. and {Pain}, R. and {Pignata}, G. and {Phillips}, M.~M. and {Polshaw}, J. and {Pumo}, M.~L. and {Rabinowitz}, D. and {Reilly}, E. and {Romero-Ca{\~n}izales}, C. and {Scalzo}, R. and {Schmidt}, B. and {Schulze}, S. and {Sim}, S. and {Sollerman}, J. and {Taddia}, F. and {Tartaglia}, L. and {Terreran}, G. and {Tomasella}, L. and {Turatto}, M. and {Walker}, E. and {Walton}, N.~A. and {Wyrzykowski}, L. and {Yuan}, F. and {Zampieri}, L.},
        title = "{PESSTO: survey description and products from the first data release by the Public ESO Spectroscopic Survey of Transient Objects}",
      journal = {\aap},
         year = 2015,
        month = jul,
       volume = {579},
          eid = {A40},
        pages = {A40},
          doi = {10.1051/0004-6361/201425237},
archivePrefix = {arXiv},
       eprint = {1411.0299},
 primaryClass = {astro-ph.SR},
       adsurl = {https://ui.adsabs.harvard.edu/abs/2015A&A...579A..40S}
}

@ARTICLE{Dressler2011_imacs,
       author = {{Dressler}, Alan and {Bigelow}, Bruce and {Hare}, Tyson and {Sutin}, Brian and {Thompson}, Ian and {Burley}, Greg and {Epps}, Harland and {Oemler}, Augustus, Jr. and {Bagish}, Alan and {Birk}, Christoph and {Clardy}, Ken and {Gunnels}, Steve and {Kelson}, Daniel and {Shectman}, Stephen and {Osip}, David},
        title = "{IMACS: The Inamori-Magellan Areal Camera and Spectrograph on Magellan-Baade}",
      journal = {\pasp},
         year = 2011,
        month = mar,
       volume = {123},
       number = {901},
        pages = {288},
          doi = {10.1086/658908},
       adsurl = {https://ui.adsabs.harvard.edu/abs/2011PASP..123..288D}
}

@ARTICLE{Moorwood1998_sofi,
       author = {{Moorwood}, A. and {Cuby}, J. -G. and {Lidman}, C.},
        title = "{SOFI sees first light at the NTT.}",
      journal = {The Messenger},
         year = 1998,
        month = mar,
       volume = {91},
        pages = {9-13},
       adsurl = {https://ui.adsabs.harvard.edu/abs/1998Msngr..91....9M}
}

@ARTICLE{Skrutskie2006_2mass,
       author = {{Skrutskie}, M.~F. and {Cutri}, R.~M. and {Stiening}, R. and {Weinberg}, M.~D. and {Schneider}, S. and {Carpenter}, J.~M. and {Beichman}, C. and {Capps}, R. and {Chester}, T. and {Elias}, J. and {Huchra}, J. and {Liebert}, J. and {Lonsdale}, C. and {Monet}, D.~G. and {Price}, S. and {Seitzer}, P. and {Jarrett}, T. and {Kirkpatrick}, J.~D. and {Gizis}, J.~E. and {Howard}, E. and {Evans}, T. and {Fowler}, J. and {Fullmer}, L. and {Hurt}, R. and {Light}, R. and {Kopan}, E.~L. and {Marsh}, K.~A. and {McCallon}, H.~L. and {Tam}, R. and {Van Dyk}, S. and {Wheelock}, S.},
        title = "{The Two Micron All Sky Survey (2MASS)}",
      journal = {\aj},
         year = 2006,
        month = feb,
       volume = {131},
       number = {2},
        pages = {1163-1183},
          doi = {10.1086/498708},
       adsurl = {https://ui.adsabs.harvard.edu/abs/2006AJ....131.1163S}
}

@ARTICLE{Stanek2022_discovery,
       author = {{Stanek}, K.~Z.},
        title = "{ASAS-SN Transient Discovery Report for 2022-01-05}",
      journal = {Transient Name Server Discovery Report},
         year = 2022,
        month = jan,
       volume = {2022-29},
        pages = {1},
       adsurl = {https://ui.adsabs.harvard.edu/abs/2022TNSTR..29....1S}
}

@ARTICLE{Kochanek2017_asassn,
       author = {{Kochanek}, C.~S. and {Shappee}, B.~J. and {Stanek}, K.~Z. and {Holoien}, T.~W. -S. and {Thompson}, Todd A. and {Prieto}, J.~L. and {Dong}, Subo and {Shields}, J.~V. and {Will}, D. and {Britt}, C. and {Perzanowski}, D. and {Pojma{\'n}ski}, G.},
        title = "{The All-Sky Automated Survey for Supernovae (ASAS-SN) Light Curve Server v1.0}",
      journal = {\pasp},
         year = 2017,
        month = oct,
       volume = {129},
       number = {980},
        pages = {104502},
          doi = {10.1088/1538-3873/aa80d9},
archivePrefix = {arXiv},
       eprint = {1706.07060},
 primaryClass = {astro-ph.SR},
       adsurl = {https://ui.adsabs.harvard.edu/abs/2017PASP..129j4502K}
}

@ARTICLE{Shappee2014_asassn,
       author = {{Shappee}, B.~J. and {Prieto}, J.~L. and {Grupe}, D. and {Kochanek}, C.~S. and {Stanek}, K.~Z. and {De Rosa}, G. and {Mathur}, S. and {Zu}, Y. and {Peterson}, B.~M. and {Pogge}, R.~W. and {Komossa}, S. and {Im}, M. and {Jencson}, J. and {Holoien}, T.~W. -S. and {Basu}, U. and {Beacom}, J.~F. and {Szczygie{\l}}, D.~M. and {Brimacombe}, J. and {Adams}, S. and {Campillay}, A. and {Choi}, C. and {Contreras}, C. and {Dietrich}, M. and {Dubberley}, M. and {Elphick}, M. and {Foale}, S. and {Giustini}, M. and {Gonzalez}, C. and {Hawkins}, E. and {Howell}, D.~A. and {Hsiao}, E.~Y. and {Koss}, M. and {Leighly}, K.~M. and {Morrell}, N. and {Mudd}, D. and {Mullins}, D. and {Nugent}, J.~M. and {Parrent}, J. and {Phillips}, M.~M. and {Pojmanski}, G. and {Rosing}, W. and {Ross}, R. and {Sand}, D. and {Terndrup}, D.~M. and {Valenti}, S. and {Walker}, Z. and {Yoon}, Y.},
        title = "{The Man behind the Curtain: X-Rays Drive the UV through NIR Variability in the 2013 Active Galactic Nucleus Outburst in NGC 2617}",
      journal = {\apj},
         year = 2014,
        month = jun,
       volume = {788},
       number = {1},
          eid = {48},
        pages = {48},
          doi = {10.1088/0004-637X/788/1/48},
archivePrefix = {arXiv},
       eprint = {1310.2241},
 primaryClass = {astro-ph.HE},
       adsurl = {https://ui.adsabs.harvard.edu/abs/2014ApJ...788...48S}
}

@ARTICLE{Jacobson-Galan2022_classification_2022an,
       author = {{Jacobson-Gal{\'a}n}, W. and {Margutti}, R. and {DeMarchi}, L. and {Terreran}, G.},
        title = "{Spectroscopic Classification of Transients with SOAR}",
      journal = {Transient Name Server AstroNote},
         year = 2022,
        month = jan,
       volume = {8},
        pages = {1},
       adsurl = {https://ui.adsabs.harvard.edu/abs/2022TNSAN...8....1J}
}

@INPROCEEDINGS{Clemens2004_Goodman,
       author = {{Clemens}, J. Christopher and {Crain}, J. Adam and {Anderson}, Robert},
        title = "{The Goodman spectrograph}",
    booktitle = {Ground-based Instrumentation for Astronomy},
         year = 2004,
       editor = {{Moorwood}, Alan F.~M. and {Iye}, Masanori},
       series = {Society of Photo-Optical Instrumentation Engineers (SPIE) Conference Series},
       volume = {5492},
        month = sep,
        pages = {331-340},
          doi = {10.1117/12.550069},
       adsurl = {https://ui.adsabs.harvard.edu/abs/2004SPIE.5492..331C}
}

@ARTICLE{Vernet2011_xshooter,
       author = {{Vernet}, J. and {Dekker}, H. and {D'Odorico}, S. and {Kaper}, L. and {Kjaergaard}, P. and {Hammer}, F. and {Randich}, S. and {Zerbi}, F. and {Groot}, P.~J. and {Hjorth}, J. and {Guinouard}, I. and {Navarro}, R. and {Adolfse}, T. and {Albers}, P.~W. and {Amans}, J. -P. and {Andersen}, J.~J. and {Andersen}, M.~I. and {Binetruy}, P. and {Bristow}, P. and {Castillo}, R. and {Chemla}, F. and {Christensen}, L. and {Conconi}, P. and {Conzelmann}, R. and {Dam}, J. and {de Caprio}, V. and {de Ugarte Postigo}, A. and {Delabre}, B. and {di Marcantonio}, P. and {Downing}, M. and {Elswijk}, E. and {Finger}, G. and {Fischer}, G. and {Flores}, H. and {Fran{\c{c}}ois}, P. and {Goldoni}, P. and {Guglielmi}, L. and {Haigron}, R. and {Hanenburg}, H. and {Hendriks}, I. and {Horrobin}, M. and {Horville}, D. and {Jessen}, N.~C. and {Kerber}, F. and {Kern}, L. and {Kiekebusch}, M. and {Kleszcz}, P. and {Klougart}, J. and {Kragt}, J. and {Larsen}, H.~H. and {Lizon}, J. -L. and {Lucuix}, C. and {Mainieri}, V. and {Manuputy}, R. and {Martayan}, C. and {Mason}, E. and {Mazzoleni}, R. and {Michaelsen}, N. and {Modigliani}, A. and {Moehler}, S. and {M{\o}ller}, P. and {Norup S{\o}rensen}, A. and {N{\o}rregaard}, P. and {P{\'e}roux}, C. and {Patat}, F. and {Pena}, E. and {Pragt}, J. and {Reinero}, C. and {Rigal}, F. and {Riva}, M. and {Roelfsema}, R. and {Royer}, F. and {Sacco}, G. and {Santin}, P. and {Schoenmaker}, T. and {Spano}, P. and {Sweers}, E. and {Ter Horst}, R. and {Tintori}, M. and {Tromp}, N. and {van Dael}, P. and {van der Vliet}, H. and {Venema}, L. and {Vidali}, M. and {Vinther}, J. and {Vola}, P. and {Winters}, R. and {Wistisen}, D. and {Wulterkens}, G. and {Zacchei}, A.},
        title = "{X-shooter, the new wide band intermediate resolution spectrograph at the ESO Very Large Telescope}",
      journal = {\aap},
         year = 2011,
        month = dec,
       volume = {536},
          eid = {A105},
        pages = {A105},
          doi = {10.1051/0004-6361/201117752},
archivePrefix = {arXiv},
       eprint = {1110.1944},
 primaryClass = {astro-ph.IM},
       adsurl = {https://ui.adsabs.harvard.edu/abs/2011A&A...536A.105V}
}

@ARTICLE{vanDokkum2001,
       author = {{van Dokkum}, Pieter G.},
        title = "{Cosmic-Ray Rejection by Laplacian Edge Detection}",
      journal = {\pasp},
         year = 2001,
        month = nov,
       volume = {113},
       number = {789},
        pages = {1420-1427},
          doi = {10.1086/323894},
archivePrefix = {arXiv},
       eprint = {astro-ph/0108003},
 primaryClass = {astro-ph},
       adsurl = {https://ui.adsabs.harvard.edu/abs/2001PASP..113.1420V}
}

@INPROCEEDINGS{Goldoni2006,
       author = {{Goldoni}, P. and {Royer}, F. and {Fran{\c{c}}ois}, P. and {Horrobin}, M. and {Blanc}, G. and {Vernet}, J. and {Modigliani}, A. and {Larsen}, J.},
        title = "{Data reduction software of the X-shooter spectrograph}",
    booktitle = {Society of Photo-Optical Instrumentation Engineers (SPIE) Conference Series},
         year = 2006,
       editor = {{McLean}, Ian S. and {Iye}, Masanori},
       series = {Society of Photo-Optical Instrumentation Engineers (SPIE) Conference Series},
       volume = {6269},
        month = jun,
          eid = {62692K},
        pages = {62692K},
          doi = {10.1117/12.669986},
       adsurl = {https://ui.adsabs.harvard.edu/abs/2006SPIE.6269E..2KG}
}

@INPROCEEDINGS{Modigliani2010,
       author = {{Modigliani}, Andrea and {Goldoni}, Paolo and {Royer}, Fr{\'e}d{\'e}ric and {Haigron}, Regis and {Guglielmi}, Laurent and {Fran{\c{c}}ois}, Patrick and {Horrobin}, Matthew and {Bristow}, Paul and {Vernet}, Joel and {Moehler}, Sabine and {Kerber}, Florian and {Ballester}, Pascal and {Mason}, Elena and {Christensen}, Lise},
        title = "{The X-shooter pipeline}",
    booktitle = {Observatory Operations: Strategies, Processes, and Systems III},
         year = 2010,
       editor = {{Silva}, David R. and {Peck}, Alison B. and {Soifer}, B. Thomas},
       series = {Society of Photo-Optical Instrumentation Engineers (SPIE) Conference Series},
       volume = {7737},
        month = jul,
          eid = {773728},
        pages = {773728},
          doi = {10.1117/12.857211},
       adsurl = {https://ui.adsabs.harvard.edu/abs/2010SPIE.7737E..28M}
}

@ARTICLE{Freudling2013,
       author = {{Freudling}, W. and {Romaniello}, M. and {Bramich}, D.~M. and {Ballester}, P. and {Forchi}, V. and {Garc{\'\i}a-Dabl{\'o}}, C.~E. and {Moehler}, S. and {Neeser}, M.~J.},
        title = "{Automated data reduction workflows for astronomy. The ESO Reflex environment}",
      journal = {\aap},
         year = 2013,
        month = nov,
       volume = {559},
          eid = {A96},
        pages = {A96},
          doi = {10.1051/0004-6361/201322494},
archivePrefix = {arXiv},
       eprint = {1311.5411},
 primaryClass = {astro-ph.IM},
       adsurl = {https://ui.adsabs.harvard.edu/abs/2013A&A...559A..96F}
}

@ARTICLE{Smette2015,
       author = {{Smette}, A. and {Sana}, H. and {Noll}, S. and {Horst}, H. and {Kausch}, W. and {Kimeswenger}, S. and {Barden}, M. and {Szyszka}, C. and {Jones}, A.~M. and {Gallenne}, A. and {Vinther}, J. and {Ballester}, P. and {Taylor}, J.},
        title = "{Molecfit: A general tool for telluric absorption correction. I. Method and application to ESO instruments}",
      journal = {\aap},
         year = 2015,
        month = apr,
       volume = {576},
          eid = {A77},
        pages = {A77},
          doi = {10.1051/0004-6361/201423932},
archivePrefix = {arXiv},
       eprint = {1501.07239},
 primaryClass = {astro-ph.IM},
       adsurl = {https://ui.adsabs.harvard.edu/abs/2015A&A...576A..77S}
}

@ARTICLE{Simcoe2013_fire,
       author = {{Simcoe}, Robert A. and {Burgasser}, Adam J. and {Schechter}, Paul L. and {Fishner}, Jason and {Bernstein}, Rebecca A. and {Bigelow}, Bruce C. and {Pipher}, Judith L. and {Forrest}, William and {McMurtry}, Craig and {Smith}, Matthew J. and {Bochanski}, John J.},
        title = "{FIRE: A Facility Class Near-Infrared Echelle Spectrometer for the Magellan Telescopes}",
      journal = {\pasp},
         year = 2013,
        month = mar,
       volume = {125},
       number = {925},
        pages = {270},
          doi = {10.1086/670241},
       adsurl = {https://ui.adsabs.harvard.edu/abs/2013PASP..125..270S}
}

@ARTICLE{Yaron2012_wiserep,
       author = {{Yaron}, Ofer and {Gal-Yam}, Avishay},
        title = "{WISeREP{\textemdash}An Interactive Supernova Data Repository}",
      journal = {Publications of the Astronomical Society of the Pacific},
         year = 2012,
        month = Jul,
       volume = {124},
        pages = {668},
          doi = {10.1086/666656},
archivePrefix = {arXiv},
       eprint = {1204.1891},
 primaryClass = {astro-ph.IM},
       adsurl = {https://ui.adsabs.harvard.edu/\#abs/2012PASP..124..668Y}
}

@INPROCEEDINGS{Tody1986_iraf,
       author = {{Tody}, Doug},
        title = "{The IRAF Data Reduction and Analysis System}",
    booktitle = {Instrumentation in astronomy VI},
         year = 1986,
       editor = {{Crawford}, David L.},
       series = {Society of Photo-Optical Instrumentation Engineers (SPIE) Conference Series},
       volume = {627},
        month = jan,
        pages = {733},
          doi = {10.1117/12.968154},
       adsurl = {https://ui.adsabs.harvard.edu/abs/1986SPIE..627..733T}
}

@INPROCEEDINGS{Tody1993_iraf,
       author = {{Tody}, Doug},
        title = "{IRAF in the Nineties}",
    booktitle = {Astronomical Data Analysis Software and Systems II},
         year = 1993,
       editor = {{Hanisch}, R.~J. and {Brissenden}, R.~J.~V. and {Barnes}, J.},
       series = {Astronomical Society of the Pacific Conference Series},
       volume = {52},
        month = jan,
        pages = {173},
       adsurl = {https://ui.adsabs.harvard.edu/abs/1993ASPC...52..173T}
}

@ARTICLE{Dong2018,
       author = {{Dong}, Subo and {Katz}, Boaz and {Kollmeier}, Juna A. and {Kushnir}, Doron and {Elias-Rosa}, N. and {Bose}, Subhash and {Morrell}, Nidia and {Prieto}, J.~L. and {Chen}, Ping and {Kochanek}, C.~S. and {Brandt}, G.~M. and {Holoien}, T.~W.-S. and {Gal-Yam}, Avishay and {Morales-Garoffolo}, Antonia and {Parker}, Stuart and {Phillips}, M.~M. and {Piro}, Anthony L. and {Shappee}, B.~J. and {Simon}, Joshua D. and {Stanek}, K.~Z.},
        title = "{A significantly off-centre $^{56}$Ni distribution for the low-luminosity type Ia supernova SN 2016brx from the 100IAS survey}",
      journal = {\mnras},
         year = 2018,
        month = sep,
       volume = {479},
       number = {1},
        pages = {L70-L75},
          doi = {10.1093/mnrasl/sly098},
archivePrefix = {arXiv},
       eprint = {1805.00010},
 primaryClass = {astro-ph.HE},
       adsurl = {https://ui.adsabs.harvard.edu/abs/2018MNRAS.479L..70D}
}

@ARTICLE{Filippenko1992b_91bg,
       author = {{Filippenko}, Alexei V. and {Richmond}, Michael W. and {Branch}, David and {Gaskell}, Martin and {Herbst}, William and {Ford}, Charles H. and {Treffers}, Richard R. and {Matheson}, Thomas and {Ho}, Luis C. and {Dey}, Arjun and {Sargent}, Wallace L.~W. and {Small}, Todd A. and {van Breugel}, Wil J.~M.},
        title = "{The Subluminous, Spectroscopically Peculiar Type 1a Supernova 1991bg in the Elliptical Galaxy NGC 4374}",
      journal = {\aj},
         year = 1992,
        month = oct,
       volume = {104},
        pages = {1543},
          doi = {10.1086/116339},
       adsurl = {https://ui.adsabs.harvard.edu/abs/1992AJ....104.1543F}
}

@ARTICLE{Leibundgut1993,
       author = {{Leibundgut}, Bruno and {Kirshner}, Robert P. and {Phillips}, Mark M. and {Wells}, Lisa A. and {Suntzeff}, N.~B. and {Hamuy}, Mario and {Schommer}, R.~A. and {Walker}, A.~R. and {Gonzalez}, L. and {Ugarte}, P. and {Williams}, R.~E. and {Williger}, G. and {Gomez}, Mercedes and {Marzke}, Ronald and {Schmidt}, Brian P. and {Whitney}, Barbara and {Caldwell}, Nelson and {Peters}, J. and {Chaffee}, F.~H. and {Foltz}, Craig B. and {Rehner}, D. and {Siciliano}, L. and {Barnes}, T.~G. and {Cheng}, K.-P. and {Hintzen}, P.~M.~N. and {Kim}, Y.-C. and {Maza}, Jose and {Parker}, J. Wm. and {Porter}, A.~C. and {Schmidtke}, P.~C. and {Sonneborn}, George},
        title = "{SN 1991bg: A Type IA Supernova With a Difference}",
      journal = {\aj},
         year = 1993,
        month = jan,
       volume = {105},
        pages = {301},
          doi = {10.1086/116427},
       adsurl = {https://ui.adsabs.harvard.edu/abs/1993AJ....105..301L}
}

@ARTICLE{Mazzali1997,
       author = {{Mazzali}, P.~A. and {Chugai}, N. and {Turatto}, M. and {Lucy}, L.~B. and {Danziger}, I.~J. and {Cappellaro}, E. and {della Valle}, M. and {Benetti}, S.},
        title = "{The properties of the peculiar type IA supernova 1991bg - II. The amount of \^56Ni and the total ejecta mass determined from spectrum synthesis and energetics considerations}",
      journal = {\mnras},
         year = 1997,
        month = jan,
       volume = {284},
       number = {1},
        pages = {151-171},
          doi = {10.1093/mnras/284.1.151},
       adsurl = {https://ui.adsabs.harvard.edu/abs/1997MNRAS.284..151M}
}

@ARTICLE{Howell2001a,
       author = {{Howell}, D. Andrew},
        title = "{The Progenitors of Subluminous Type Ia Supernovae}",
      journal = {\apjl},
         year = 2001,
        month = jun,
       volume = {554},
       number = {2},
        pages = {L193-L196},
          doi = {10.1086/321702},
archivePrefix = {arXiv},
       eprint = {astro-ph/0105246},
 primaryClass = {astro-ph},
       adsurl = {https://ui.adsabs.harvard.edu/abs/2001ApJ...554L.193H}
}

@ARTICLE{Nomoto1982a,
       author = {{Nomoto}, K.},
        title = "{Accreting white dwarf models for type I supernovae. I - Presupernova evolution and triggering mechanisms}",
      journal = {\apj},
         year = 1982,
        month = feb,
       volume = {253},
        pages = {798-810},
          doi = {10.1086/159682},
       adsurl = {https://ui.adsabs.harvard.edu/abs/1982ApJ...253..798N}
}

@ARTICLE{Nomoto1982b,
       author = {{Nomoto}, K.},
        title = "{Accreting white dwarf models for type I supernovae. II. Off-center detonation supernovae.}",
      journal = {\apj},
         year = 1982,
        month = jun,
       volume = {257},
        pages = {780-792},
          doi = {10.1086/160031},
       adsurl = {https://ui.adsabs.harvard.edu/abs/1982ApJ...257..780N}
}

@ARTICLE{Woosley1986,
       author = {{Woosley}, S.~E. and {Taam}, R.~E. and {Weaver}, T.~A.},
        title = "{Models for Type I Supernova. I. Detonations in White Dwarfs}",
      journal = {\apj},
         year = 1986,
        month = feb,
       volume = {301},
        pages = {601},
          doi = {10.1086/163926},
       adsurl = {https://ui.adsabs.harvard.edu/abs/1986ApJ...301..601W}
}

@ARTICLE{Bildsten2007,
       author = {{Bildsten}, Lars and {Shen}, Ken J. and {Weinberg}, Nevin N. and {Nelemans}, Gijs},
        title = "{Faint Thermonuclear Supernovae from AM Canum Venaticorum Binaries}",
      journal = {\apjl},
         year = 2007,
        month = jun,
       volume = {662},
       number = {2},
        pages = {L95-L98},
          doi = {10.1086/519489},
archivePrefix = {arXiv},
       eprint = {astro-ph/0703578},
 primaryClass = {astro-ph},
       adsurl = {https://ui.adsabs.harvard.edu/abs/2007ApJ...662L..95B}
}

@ARTICLE{Shen2009,
       author = {{Shen}, Ken J. and {Bildsten}, Lars},
        title = "{Unstable Helium Shell Burning on Accreting White Dwarfs}",
      journal = {\apj},
         year = 2009,
        month = jul,
       volume = {699},
       number = {2},
        pages = {1365-1373},
          doi = {10.1088/0004-637X/699/2/1365},
archivePrefix = {arXiv},
       eprint = {0903.0654},
 primaryClass = {astro-ph.HE},
       adsurl = {https://ui.adsabs.harvard.edu/abs/2009ApJ...699.1365S}
}

@ARTICLE{Fink2010,
       author = {{Fink}, M. and {R{\"o}pke}, F.~K. and {Hillebrandt}, W. and {Seitenzahl}, I.~R. and {Sim}, S.~A. and {Kromer}, M.},
        title = "{Double-detonation sub-Chandrasekhar supernovae: can minimum helium shell masses detonate the core?}",
      journal = {\aap},
         year = 2010,
        month = may,
       volume = {514},
          eid = {A53},
        pages = {A53},
          doi = {10.1051/0004-6361/200913892},
archivePrefix = {arXiv},
       eprint = {1002.2173},
 primaryClass = {astro-ph.SR},
       adsurl = {https://ui.adsabs.harvard.edu/abs/2010A&A...514A..53F}
}

@ARTICLE{Woosley2011,
       author = {{Woosley}, S.~E. and {Kasen}, Daniel},
        title = "{Sub-Chandrasekhar Mass Models for Supernovae}",
      journal = {\apj},
         year = 2011,
        month = jun,
       volume = {734},
       number = {1},
          eid = {38},
        pages = {38},
          doi = {10.1088/0004-637X/734/1/38},
archivePrefix = {arXiv},
       eprint = {1010.5292},
 primaryClass = {astro-ph.HE},
       adsurl = {https://ui.adsabs.harvard.edu/abs/2011ApJ...734...38W}
}

@ARTICLE{Shen2014,
       author = {{Shen}, Ken J. and {Moore}, Kevin},
        title = "{The Initiation and Propagation of Helium Detonations in White Dwarf Envelopes}",
      journal = {\apj},
         year = 2014,
        month = dec,
       volume = {797},
       number = {1},
          eid = {46},
        pages = {46},
          doi = {10.1088/0004-637X/797/1/46},
archivePrefix = {arXiv},
       eprint = {1409.3568},
 primaryClass = {astro-ph.HE},
       adsurl = {https://ui.adsabs.harvard.edu/abs/2014ApJ...797...46S}
}

@ARTICLE{Woosley1994,
       author = {{Woosley}, S.~E. and {Weaver}, Thomas A.},
        title = "{Sub--Chandrasekhar Mass Models for Type IA Supernovae}",
      journal = {\apj},
         year = 1994,
        month = mar,
       volume = {423},
        pages = {371},
          doi = {10.1086/173813},
       adsurl = {https://ui.adsabs.harvard.edu/abs/1994ApJ...423..371W}
}

@ARTICLE{Hsiao2015,
       author = {{Hsiao}, E.~Y. and {Burns}, C.~R. and {Contreras}, C. and {H{\"o}flich}, P. and {Sand}, D. and {Marion}, G.~H. and {Phillips}, M.~M. and {Stritzinger}, M. and {Gonz{\'a}lez-Gait{\'a}n}, S. and {Mason}, R.~E. and {Folatelli}, G. and {Parent}, E. and {Gall}, C. and {Amanullah}, R. and {Anupama}, G.~C. and {Arcavi}, I. and {Banerjee}, D.~P.~K. and {Beletsky}, Y. and {Blanc}, G.~A. and {Bloom}, J.~S. and {Brown}, P.~J. and {Campillay}, A. and {Cao}, Y. and {De Cia}, A. and {Diamond}, T. and {Freedman}, W.~L. and {Gonzalez}, C. and {Goobar}, A. and {Holmbo}, S. and {Howell}, D.~A. and {Johansson}, J. and {Kasliwal}, M.~M. and {Kirshner}, R.~P. and {Krisciunas}, K. and {Kulkarni}, S.~R. and {Maguire}, K. and {Milne}, P.~A. and {Morrell}, N. and {Nugent}, P.~E. and {Ofek}, E.~O. and {Osip}, D. and {Palunas}, P. and {Perley}, D.~A. and {Persson}, S.~E. and {Piro}, A.~L. and {Rabus}, M. and {Roth}, M. and {Schiefelbein}, J.~M. and {Srivastav}, S. and {Sullivan}, M. and {Suntzeff}, N.~B. and {Surace}, J. and {Wo{\'z}niak}, P.~R. and {Yaron}, O.},
        title = "{Strong near-infrared carbon in the Type Ia supernova iPTF13ebh}",
      journal = {\aap},
         year = 2015,
        month = jun,
       volume = {578},
          eid = {A9},
        pages = {A9},
          doi = {10.1051/0004-6361/201425297},
archivePrefix = {arXiv},
       eprint = {1503.02293},
 primaryClass = {astro-ph.SR},
       adsurl = {https://ui.adsabs.harvard.edu/abs/2015A&A...578A...9H}
}

@ARTICLE{Pearson2024,
       author = {{Pearson}, Jeniveve and {Sand}, David J. and {Lundqvist}, Peter and {Galbany}, Llu{\'\i}s and {Andrews}, Jennifer E. and {Bostroem}, K. Azalee and {Dong}, Yize and {Hoang}, Emily and {Hosseinzadeh}, Griffin and {Janzen}, Daryl and {Jencson}, Jacob E. and {Lundquist}, Michael J. and {Mehta}, Darshana and {Meza Retamal}, Nicol{\'a}s and {Shrestha}, Manisha and {Valenti}, Stefano and {Wyatt}, Samuel and {Anderson}, Joseph P. and {Ashall}, Chris and {Auchettl}, Katie and {Baron}, Eddie and {Blondin}, St{\'e}phane and {Burns}, Christopher R. and {Cai}, Yongzhi and {Chen}, Ting-Wan and {Chomiuk}, Laura and {Coulter}, David A. and {Cross}, Dane and {Davis}, Kyle W. and {de Jaeger}, Thomas and {DerKacy}, James M. and {Desai}, Dhvanil D. and {Dimitriadis}, Georgios and {Do}, Aaron and {Farah}, Joseph R. and {Foley}, Ryan J. and {Gromadzki}, Mariusz and {Guti{\'e}rrez}, Claudia P. and {Haislip}, Joshua and {Gonz{\'a}lez Hern{\'a}ndez}, Jonay I. and {Hinkle}, Jason T. and {Hoogendam}, Willem B. and {Howell}, D. Andrew and {Hoeflich}, Peter and {Hsiao}, Eric and {Huber}, Mark E. and {Jha}, Saurabh W. and {Jim{\'e}nez Palau}, Cristina and {Kilpatrick}, Charles D. and {Kouprianov}, Vladimir and {Kumar}, Sahana and {Kwok}, Lindsey A. and {Larison}, Conor and {LeBaron}, Natalie and {Le Saux}, Xavier and {Lu}, Jing and {McCully}, Curtis and {Mera Evans}, Tycho and {Milne}, Peter and {Modjaz}, Maryam and {Morrell}, Nidia and {M{\"u}ller-Bravo}, Tom{\'a}s E. and {Newsome}, Megan and {Nicholl}, Matt and {Padilla Gonzalez}, Estefania and {Payne}, Anna V. and {Pellegrino}, Craig and {Phan}, Kim and {Pineda-Garc{\'\i}a}, Jonathan and {Piro}, Anthony L. and {Piscarreta}, Lara and {Polin}, Abigail and {Reichart}, Daniel E. and {Rojas-Bravo}, C{\'e}sar and {Ryder}, Stuart D. and {Salmaso}, Irene and {Schwab}, Michaela and {Shahbandeh}, Melissa and {Shappee}, Benjamin J. and {Siebert}, Matthew R. and {Smith}, Nathan and {Strader}, Jay and {Taggart}, Kirsty and {Terreran}, Giacomo and {Tinyanont}, Samaporn and {Tucker}, M.~A. and {Valerin}, Giorgio and {Young}, D.~R.},
        title = "{Strong Carbon Features and a Red Early Color in the Underluminous Type Ia SN 2022xkq}",
      journal = {\apj},
         year = 2024,
        month = jan,
       volume = {960},
       number = {1},
          eid = {29},
        pages = {29},
          doi = {10.3847/1538-4357/ad0153},
archivePrefix = {arXiv},
       eprint = {2309.10054},
 primaryClass = {astro-ph.HE},
       adsurl = {https://ui.adsabs.harvard.edu/abs/2024ApJ...960...29P}
}

@ARTICLE{Wyatt2021,
       author = {{Wyatt}, S.~D. and {Sand}, D.~J. and {Hsiao}, E.~Y. and {Burns}, C.~R. and {Valenti}, S. and {Bostroem}, K.~A. and {Lundquist}, M. and {Galbany}, L. and {Lu}, J. and {Ashall}, C. and {Diamond}, T.~R. and {Filippenko}, A.~V. and {Graham}, M.~L. and {Hoeflich}, P. and {Kirshner}, R.~P. and {Krisciunas}, K. and {Marion}, G.~H. and {Morrell}, N. and {Persson}, S.~E. and {Phillips}, M.~M. and {Stritzinger}, M.~D. and {Suntzeff}, N.~B. and {Taddia}, F.},
        title = "{Strong Near-infrared Carbon Absorption in the Transitional Type Ia SN 2015bp}",
      journal = {\apj},
         year = 2021,
        month = jun,
       volume = {914},
       number = {1},
          eid = {57},
        pages = {57},
          doi = {10.3847/1538-4357/abf7c3},
archivePrefix = {arXiv},
       eprint = {2012.02858},
 primaryClass = {astro-ph.HE},
       adsurl = {https://ui.adsabs.harvard.edu/abs/2021ApJ...914...57W}
}

@ARTICLE{Li2022_2012ij,
       author = {{Li}, Zhitong and {Zhang}, Tianmeng and {Wang}, Xiaofeng and {Sai}, Hanna and {Zhang}, Jujia and {Chen}, Juncheng and {Zhao}, Xulin and {Yan}, Shengyu and {Wang}, Bo and {Phillips}, Mark M. and {Hsiao}, Eric Y. and {Morrell}, Nidia and {Contreras}, Carlos and {Burns}, Christopher R. and {Ashall}, Christopher and {Stritzinger}, Maximilian and {Krisciunas}, Kevin and {Prieto}, Jose and {Zou}, Hu and {Wang}, Jiali and {Ma}, Jun and {Nie}, Jundan and {Xue}, Suijian and {Zhou}, Xu and {Zhou}, Zhimin and {Xiang}, Danfeng and {Xi}, Gaobo},
        title = "{SN 2012ij: A Low-luminosity Type Ia Supernova and Evidence for a Continuous Distribution from a 91bg-like Explosion to Normal Ones}",
      journal = {\apj},
         year = 2022,
        month = mar,
       volume = {927},
       number = {2},
          eid = {142},
        pages = {142},
          doi = {10.3847/1538-4357/ac4e17},
archivePrefix = {arXiv},
       eprint = {2201.06066},
 primaryClass = {astro-ph.HE},
       adsurl = {https://ui.adsabs.harvard.edu/abs/2022ApJ...927..142L}
}

@ARTICLE{Hoflich2002,
       author = {{H{\"o}flich}, Peter and {Gerardy}, Christopher L. and {Fesen}, Robert A. and {Sakai}, Shoko},
        title = "{Infrared Spectra of the Subluminous Type Ia Supernova SN 1999by}",
      journal = {\apj},
         year = 2002,
        month = apr,
       volume = {568},
       number = {2},
        pages = {791-806},
          doi = {10.1086/339063},
archivePrefix = {arXiv},
       eprint = {astro-ph/0112126},
 primaryClass = {astro-ph},
       adsurl = {https://ui.adsabs.harvard.edu/abs/2002ApJ...568..791H}
}

@ARTICLE{Garnavich2004,
       author = {{Garnavich}, Peter M. and {Bonanos}, Alceste Z. and {Krisciunas}, Kevin and {Jha}, Saurabh and {Kirshner}, Robert P. and {Schlegel}, Eric M. and {Challis}, Peter and {Macri}, Lucas M. and {Hatano}, Kazuhito and {Branch}, David and {Bothun}, Gregory D. and {Freedman}, Wendy L.},
        title = "{The Luminosity of SN 1999by in NGC 2841 and the Nature of ``Peculiar'' Type Ia Supernovae}",
      journal = {\apj},
         year = 2004,
        month = oct,
       volume = {613},
       number = {2},
        pages = {1120-1132},
          doi = {10.1086/422986},
archivePrefix = {arXiv},
       eprint = {astro-ph/0105490},
 primaryClass = {astro-ph},
       adsurl = {https://ui.adsabs.harvard.edu/abs/2004ApJ...613.1120G}
}

@ARTICLE{Smartt2015,
       author = {{Smartt}, S.~J. and {Valenti}, S. and {Fraser}, M. and {Inserra}, C. and {Young}, D.~R. and {Sullivan}, M. and {Pastorello}, A. and {Benetti}, S. and {Gal-Yam}, A. and {Knapic}, C. and {Molinaro}, M. and {Smareglia}, R. and {Smith}, K.~W. and {Taubenberger}, S. and {Yaron}, O. and {Anderson}, J.~P. and {Ashall}, C. and {Balland}, C. and {Baltay}, C. and {Barbarino}, C. and {Bauer}, F.~E. and {Baumont}, S. and {Bersier}, D. and {Blagorodnova}, N. and {Bongard}, S. and {Botticella}, M.~T. and {Bufano}, F. and {Bulla}, M. and {Cappellaro}, E. and {Campbell}, H. and {Cellier-Holzem}, F. and {Chen}, T. -W. and {Childress}, M.~J. and {Clocchiatti}, A. and {Contreras}, C. and {Dall'Ora}, M. and {Danziger}, J. and {de Jaeger}, T. and {De Cia}, A. and {Della Valle}, M. and {Dennefeld}, M. and {Elias-Rosa}, N. and {Elman}, N. and {Feindt}, U. and {Fleury}, M. and {Gall}, E. and {Gonzalez-Gaitan}, S. and {Galbany}, L. and {Morales Garoffolo}, A. and {Greggio}, L. and {Guillou}, L.~L. and {Hachinger}, S. and {Hadjiyska}, E. and {Hage}, P.~E. and {Hillebrandt}, W. and {Hodgkin}, S. and {Hsiao}, E.~Y. and {James}, P.~A. and {Jerkstrand}, A. and {Kangas}, T. and {Kankare}, E. and {Kotak}, R. and {Kromer}, M. and {Kuncarayakti}, H. and {Leloudas}, G. and {Lundqvist}, P. and {Lyman}, J.~D. and {Hook}, I.~M. and {Maguire}, K. and {Manulis}, I. and {Margheim}, S.~J. and {Mattila}, S. and {Maund}, J.~R. and {Mazzali}, P.~A. and {McCrum}, M. and {McKinnon}, R. and {Moreno-Raya}, M.~E. and {Nicholl}, M. and {Nugent}, P. and {Pain}, R. and {Pignata}, G. and {Phillips}, M.~M. and {Polshaw}, J. and {Pumo}, M.~L. and {Rabinowitz}, D. and {Reilly}, E. and {Romero-Ca{\~n}izales}, C. and {Scalzo}, R. and {Schmidt}, B. and {Schulze}, S. and {Sim}, S. and {Sollerman}, J. and {Taddia}, F. and {Tartaglia}, L. and {Terreran}, G. and {Tomasella}, L. and {Turatto}, M. and {Walker}, E. and {Walton}, N.~A. and {Wyrzykowski}, L. and {Yuan}, F. and {Zampieri}, L.},
        title = "{PESSTO: survey description and products from the first data release by the Public ESO Spectroscopic Survey of Transient Objects}",
      journal = {\aap},
         year = 2015,
        month = jul,
       volume = {579},
          eid = {A40},
        pages = {A40},
          doi = {10.1051/0004-6361/201425237},
archivePrefix = {arXiv},
       eprint = {1411.0299},
 primaryClass = {astro-ph.SR},
       adsurl = {https://ui.adsabs.harvard.edu/abs/2015A&A...579A..40S}
}

@ARTICLE{Lu2023_CSP_SNIa_NIR,
       author = {{Lu}, Jing and {Hsiao}, Eric Y. and {Phillips}, Mark M. and {Burns}, Christopher R. and {Ashall}, Chris and {Morrell}, Nidia and {Ng}, Lawrence and {Kumar}, Sahana and {Shahbandeh}, Melissa and {Hoeflich}, Peter and {Baron}, E. and {Uddin}, Syed and {Stritzinger}, Maximilian D. and {Suntzeff}, Nicholas B. and {Baltay}, Charles and {Davis}, Scott and {Diamond}, Tiara R. and {Folatelli}, Gaston and {F{\"o}rster}, Francisco and {Gagn{\'e}}, Jonathan and {Galbany}, Llu{\'\i}s and {Gall}, Christa and {Gonz{\'a}lez-Gait{\'a}n}, Santiago and {Holmbo}, Simon and {Kirshner}, Robert P. and {Krisciunas}, Kevin and {Marion}, G.~H. and {Perlmutter}, Saul and {Pessi}, Priscila J. and {Piro}, Anthony L. and {Rabinowitz}, David and {Ryder}, Stuart D. and {Sand}, David J.},
        title = "{Carnegie Supernova Project. II. Near-infrared Spectral Diversity and Template of Type Ia Supernovae}",
      journal = {\apj},
         year = 2023,
        month = may,
       volume = {948},
       number = {1},
          eid = {27},
        pages = {27},
          doi = {10.3847/1538-4357/acc100},
archivePrefix = {arXiv},
       eprint = {2211.05998},
 primaryClass = {astro-ph.HE},
       adsurl = {https://ui.adsabs.harvard.edu/abs/2023ApJ...948...27L}
}

@ARTICLE{Liu2023_2020jgb,
       author = {{Liu}, Chang and {Miller}, Adam A. and {Polin}, Abigail and {Nugent}, Anya E. and {De}, Kishalay and {Nugent}, Peter E. and {Schulze}, Steve and {Gal-Yam}, Avishay and {Fremling}, Christoffer and {Anand}, Shreya and {Andreoni}, Igor and {Blanchard}, Peter and {Brink}, Thomas G. and {Dhawan}, Suhail and {Filippenko}, Alexei V. and {Maguire}, Kate and {Schweyer}, Tassilo and {Sears}, Huei and {Sharma}, Yashvi and {Graham}, Matthew J. and {Groom}, Steven L. and {Hale}, David and {Kasliwal}, Mansi M. and {Masci}, Frank J. and {Purdum}, Josiah and {Racine}, Benjamin and {Sollerman}, Jesper and {Kulkarni}, Shrinivas R.},
        title = "{SN 2020jgb: A Peculiar Type Ia Supernova Triggered by a Helium-shell Detonation in a Star-forming Galaxy}",
      journal = {\apj},
         year = 2023,
        month = apr,
       volume = {946},
       number = {2},
          eid = {83},
        pages = {83},
          doi = {10.3847/1538-4357/acbb5e},
archivePrefix = {arXiv},
       eprint = {2209.04463},
 primaryClass = {astro-ph.HE},
       adsurl = {https://ui.adsabs.harvard.edu/abs/2023ApJ...946...83L}
}

@ARTICLE{Waldman2011,
       author = {{Waldman}, Roni and {Sauer}, Daniel and {Livne}, Eli and {Perets}, Hagai and {Glasner}, Ami and {Mazzali}, Paolo and {Truran}, James W. and {Gal-Yam}, Avishay},
        title = "{Helium Shell Detonations on Low-mass White Dwarfs as a Possible Explanation for SN 2005E}",
      journal = {\apj},
         year = 2011,
        month = sep,
       volume = {738},
       number = {1},
          eid = {21},
        pages = {21},
          doi = {10.1088/0004-637X/738/1/21},
archivePrefix = {arXiv},
       eprint = {1009.3829},
 primaryClass = {astro-ph.SR},
       adsurl = {https://ui.adsabs.harvard.edu/abs/2011ApJ...738...21W}
}

@ARTICLE{Shen2010,
       author = {{Shen}, Ken J. and {Kasen}, Dan and {Weinberg}, Nevin N. and {Bildsten}, Lars and {Scannapieco}, Evan},
        title = "{Thermonuclear .Ia Supernovae from Helium Shell Detonations: Explosion Models and Observables}",
      journal = {\apj},
         year = 2010,
        month = jun,
       volume = {715},
       number = {2},
        pages = {767-774},
          doi = {10.1088/0004-637X/715/2/767},
archivePrefix = {arXiv},
       eprint = {1002.2258},
 primaryClass = {astro-ph.HE},
       adsurl = {https://ui.adsabs.harvard.edu/abs/2010ApJ...715..767S}
}

@ARTICLE{Kromer2010,
       author = {{Kromer}, M. and {Sim}, S.~A. and {Fink}, M. and {R{\"o}pke}, F.~K. and {Seitenzahl}, I.~R. and {Hillebrandt}, W.},
        title = "{Double-detonation Sub-Chandrasekhar Supernovae: Synthetic Observables for Minimum Helium Shell Mass Models}",
      journal = {\apj},
         year = 2010,
        month = aug,
       volume = {719},
       number = {2},
        pages = {1067-1082},
          doi = {10.1088/0004-637X/719/2/1067},
archivePrefix = {arXiv},
       eprint = {1006.4489},
 primaryClass = {astro-ph.HE},
       adsurl = {https://ui.adsabs.harvard.edu/abs/2010ApJ...719.1067K}
}

@ARTICLE{WoosleyKasen2011,
       author = {{Woosley}, S.~E. and {Kasen}, Daniel},
        title = "{Sub-Chandrasekhar Mass Models for Supernovae}",
      journal = {\apj},
         year = 2011,
        month = jun,
       volume = {734},
       number = {1},
          eid = {38},
        pages = {38},
          doi = {10.1088/0004-637X/734/1/38},
archivePrefix = {arXiv},
       eprint = {1010.5292},
 primaryClass = {astro-ph.HE},
       adsurl = {https://ui.adsabs.harvard.edu/abs/2011ApJ...734...38W}
}

@ARTICLE{Collins2023,
       author = {{Collins}, Christine E. and {Sim}, Stuart A. and {Shingles}, Luke J. and {Gronow}, Sabrina and {R{\"o}pke}, Friedrich K. and {Pakmor}, R{\"u}diger and {Seitenzahl}, Ivo R. and {Kromer}, Markus},
        title = "{Helium as a signature of the double detonation in Type Ia supernovae}",
      journal = {\mnras},
         year = 2023,
        month = sep,
       volume = {524},
       number = {3},
        pages = {4447-4454},
          doi = {10.1093/mnras/stad2170},
archivePrefix = {arXiv},
       eprint = {2307.08660},
 primaryClass = {astro-ph.SR},
       adsurl = {https://ui.adsabs.harvard.edu/abs/2023MNRAS.524.4447C}
}

@INPROCEEDINGS{Schipani2018,
       author = {{Schipani}, P. and {Campana}, S. and {Claudi}, R. and {K{\"a}ufl}, H.~U. and {Accardo}, M. and {Aliverti}, M. and {Baruffolo}, A. and {Ben Ami}, S. and {Biondi}, F. and {Brucalassi}, A. and {Capasso}, G. and {Cosentino}, R. and {D'Alessio}, F. and {D'Avanzo}, P. and {Hershko}, O. and {Gardiol}, D. and {Kuncarayacti}, H. and {Munari}, M. and {Rubin}, A. and {Scuderi}, S. and {Vitali}, F. and {Achr{\'e}n}, J. and {Araiza-Duran}, J. Antonio and {Arcavi}, I. and {Bianco}, A. and {Cappellaro}, E. and {Colapietro}, M. and {Della Valle}, M. and {Diner}, O. and {D'Orsi}, S. and {Fantinel}, D. and {Fynbo}, J. and {Gal-Yam}, A. and {Genoni}, M. and {Hirvonen}, M. and {Kotilainen}, J. and {Kumar}, T. and {Landoni}, M. and {Lehti}, J. and {Li Causi}, G. and {Loreggia}, D. and {Marafatto}, L. and {Mattila}, S. and {Pariani}, G. and {Pignata}, G. and {Rappaport}, M. and {Ricci}, D. and {Riva}, M. and {Salasnich}, B. and {Zanmar Sanchez}, R. and {Smartt}, S. and {Turatto}, M.},
        title = "{SOXS: a wide band spectrograph to follow up transients}",
    booktitle = {Ground-based and Airborne Instrumentation for Astronomy VII},
         year = 2018,
       editor = {{Evans}, Christopher J. and {Simard}, Luc and {Takami}, Hideki},
       series = {Society of Photo-Optical Instrumentation Engineers (SPIE) Conference Series},
       volume = {10702},
        month = jul,
          eid = {107020F},
        pages = {107020F},
          doi = {10.1117/12.2307349},
archivePrefix = {arXiv},
       eprint = {1807.08828},
 primaryClass = {astro-ph.IM},
       adsurl = {https://ui.adsabs.harvard.edu/abs/2018SPIE10702E..0FS}
}

@ARTICLE{Marion2009,
       author = {{Marion}, G.~H. and {H{\"o}flich}, P. and {Gerardy}, C.~L. and {Vacca}, W.~D. and {Wheeler}, J.~C. and {Robinson}, E.~L.},
        title = "{A Catalog of Near-Infrared Spectra from Type Ia Supernovae}",
      journal = {\aj},
         year = 2009,
        month = sep,
       volume = {138},
       number = {3},
        pages = {727-757},
          doi = {10.1088/0004-6256/138/3/727},
archivePrefix = {arXiv},
       eprint = {0906.4085},
 primaryClass = {astro-ph.CO},
       adsurl = {https://ui.adsabs.harvard.edu/abs/2009AJ....138..727M}
}

@ARTICLE{Shahbandeh2022,
       author = {{Shahbandeh}, M. and {Hsiao}, E.~Y. and {Ashall}, C. and {Teffs}, J. and {Hoeflich}, P. and {Morrell}, N. and {Phillips}, M.~M. and {Anderson}, J.~P. and {Baron}, E. and {Burns}, C.~R. and {Contreras}, C. and {Davis}, S. and {Diamond}, T.~R. and {Folatelli}, G. and {Galbany}, L. and {Gall}, C. and {Hachinger}, S. and {Holmbo}, S. and {Karamehmetoglu}, E. and {Kasliwal}, M.~M. and {Kirshner}, R.~P. and {Krisciunas}, K. and {Kumar}, S. and {Lu}, J. and {Marion}, G.~H. and {Mazzali}, P.~A. and {Piro}, A.~L. and {Sand}, D.~J. and {Stritzinger}, M.~D. and {Suntzeff}, N.~B. and {Taddia}, F. and {Uddin}, S.~A.},
        title = "{Carnegie Supernova Project-II: Near-infrared Spectroscopy of Stripped-envelope Core-collapse Supernovae}",
      journal = {\apj},
         year = 2022,
        month = feb,
       volume = {925},
       number = {2},
          eid = {175},
        pages = {175},
          doi = {10.3847/1538-4357/ac4030},
archivePrefix = {arXiv},
       eprint = {2110.12083},
 primaryClass = {astro-ph.HE},
       adsurl = {https://ui.adsabs.harvard.edu/abs/2022ApJ...925..175S}
}

@ARTICLE{Valenti2008_2007gr,
       author = {{Valenti}, S. and {Elias-Rosa}, N. and {Taubenberger}, S. and {Stanishev}, V. and {Agnoletto}, I. and {Sauer}, D. and {Cappellaro}, E. and {Pastorello}, A. and {Benetti}, S. and {Riffeser}, A. and {Hopp}, U. and {Navasardyan}, H. and {Tsvetkov}, D. and {Lorenzi}, V. and {Patat}, F. and {Turatto}, M. and {Barbon}, R. and {Ciroi}, S. and {Di Mille}, F. and {Frandsen}, S. and {Fynbo}, J.~P.~U. and {Laursen}, P. and {Mazzali}, P.~A.},
        title = "{The Carbon-rich Type Ic SN 2007gr: The Photospheric Phase}",
      journal = {\apjl},
         year = 2008,
        month = feb,
       volume = {673},
       number = {2},
        pages = {L155},
          doi = {10.1086/527672},
archivePrefix = {arXiv},
       eprint = {0712.1899},
 primaryClass = {astro-ph},
       adsurl = {https://ui.adsabs.harvard.edu/abs/2008ApJ...673L.155V}
}

@ARTICLE{Botyanszki2018,
       author = {{Boty{\'a}nszki}, J{\'a}nos and {Kasen}, Daniel and {Plewa}, Tomasz},
        title = "{Multidimensional Models of Type Ia Supernova Nebular Spectra: Strong Emission Lines from Stripped Companion Gas Rule Out Classic Single-degenerate Systems}",
      journal = {\apjl},
         year = 2018,
        month = jan,
       volume = {852},
       number = {1},
          eid = {L6},
        pages = {L6},
          doi = {10.3847/2041-8213/aaa07b},
archivePrefix = {arXiv},
       eprint = {1712.03274},
 primaryClass = {astro-ph.SR},
       adsurl = {https://ui.adsabs.harvard.edu/abs/2018ApJ...852L...6B}
}

@ARTICLE{LiuZW2013,

       author = {{Liu}, Zheng-Wei and {Pakmor}, R. and {Seitenzahl}, I.~R. and {Hillebrandt}, W. and {Kromer}, M. and {R{\"o}pke}, F.~K. and {Edelmann}, P. and {Taubenberger}, S. and {Maeda}, K. and {Wang}, B. and {Han}, Z.~W.},

        title = "{The Impact of Type Ia Supernova Explosions on Helium Companions in the Chandrasekhar-mass Explosion Scenario}",

      journal = {\apj},

         year = 2013,

        month = sep,

       volume = {774},

       number = {1},

          eid = {37},

        pages = {37},

          doi = {10.1088/0004-637X/774/1/37},

archivePrefix = {arXiv},

       eprint = {1307.5579},

 primaryClass = {astro-ph.SR},

       adsurl = {https://ui.adsabs.harvard.edu/abs/2013ApJ...774...37L}

}

@ARTICLE{Gall2018,
       author = {{Gall}, C. and {Stritzinger}, M.~D. and {Ashall}, C. and {Baron}, E. and {Burns}, C.~R. and {Hoeflich}, P. and {Hsiao}, E.~Y. and {Mazzali}, P.~A. and {Phillips}, M.~M. and {Filippenko}, A.~V. and {Anderson}, J.~P. and {Benetti}, S. and {Brown}, P.~J. and {Campillay}, A. and {Challis}, P. and {Contreras}, C. and {Elias de la Rosa}, N. and {Folatelli}, G. and {Foley}, R.~J. and {Fraser}, M. and {Holmbo}, S. and {Marion}, G.~H. and {Morrell}, N. and {Pan}, Y.-C. and {Pignata}, G. and {Suntzeff}, N.~B. and {Taddia}, F. and {Torres Robledo}, S. and {Valenti}, S.},
        title = "{Two transitional type Ia supernovae located in the Fornax cluster member NGC 1404: SN 2007on and SN 2011iv}",
      journal = {\aap},
         year = 2018,
        month = mar,
       volume = {611},
          eid = {A58},
        pages = {A58},
          doi = {10.1051/0004-6361/201730886},
archivePrefix = {arXiv},
       eprint = {1707.03823},
 primaryClass = {astro-ph.SR},
       adsurl = {https://ui.adsabs.harvard.edu/abs/2018A&A...611A..58G}
}

@ARTICLE{Wang2009,
       author = {{Wang}, X. and {Li}, W. and {Filippenko}, A.~V. and {Foley}, R.~J. and {Kirshner}, R.~P. and {Modjaz}, M. and {Bloom}, J. and {Brown}, P.~J. and {Carter}, D. and {Friedman}, A.~S. and {Gal-Yam}, A. and {Ganeshalingam}, M. and {Hicken}, M. and {Krisciunas}, K. and {Milne}, P. and {Silverman}, J.~M. and {Suntzeff}, N.~B. and {Wood-Vasey}, W.~M. and {Cenko}, S.~B. and {Challis}, P. and {Fox}, D.~B. and {Kirkman}, D. and {Li}, J.~Z. and {Li}, T.~P. and {Malkan}, M.~A. and {Moore}, M.~R. and {Reitzel}, D.~B. and {Rich}, R.~M. and {Serduke}, F.~J.~D. and {Shang}, R.~C. and {Steele}, T.~N. and {Swift}, B.~J. and {Tao}, C. and {Wong}, D.~S. and {Zhang}, S.~N.},
        title = "{The Golden Standard Type Ia Supernova 2005cf: Observations from the Ultraviolet to the Near-Infrared Wavebands}",
      journal = {\apj},
         year = 2009,
        month = may,
       volume = {697},
       number = {1},
        pages = {380-408},
          doi = {10.1088/0004-637X/697/1/380},
archivePrefix = {arXiv},
       eprint = {0811.1205},
 primaryClass = {astro-ph},
       adsurl = {https://ui.adsabs.harvard.edu/abs/2009ApJ...697..380W}
}

@ARTICLE{Chen2019,
       author = {{Chen}, Ping and {Dong}, Subo and {Katz}, Boaz and {Kochanek}, C.~S. and {Kollmeier}, Juna A. and {Maguire}, K. and {Phillips}, M.~M. and {Prieto}, J.~L. and {Shappee}, B.~J. and {Stritzinger}, M.~D. and {Bose}, Subhash and {Brown}, Peter J. and {Holoien}, T.~W.-S. and {Galbany}, L. and {Milne}, Peter A. and {Morrell}, Nidia and {Piro}, Anthony L. and {Stanek}, K.~Z. and {Thompson}, Todd A. and {Young}, D.~R.},
        title = "{ASASSN-15pz: Revealing Significant Photometric Diversity among 2009dc-like, Peculiar SNe Ia}",
      journal = {\apj},
         year = 2019,
        month = jul,
       volume = {880},
       number = {1},
          eid = {35},
        pages = {35},
          doi = {10.3847/1538-4357/ab2630},
archivePrefix = {arXiv},
       eprint = {1904.03198},
 primaryClass = {astro-ph.HE},
       adsurl = {https://ui.adsabs.harvard.edu/abs/2019ApJ...880...35C}
}

@ARTICLE{Sharon2025,
       author = {{Sharon}, Amir and {Kushnir}, Doron and {Wygoda}, Nahliel},
        title = "{All known Type Ia supernovae models fail to reproduce the observed bolometric luminosity{\textendash}width correlation}",
      journal = {\mnras},
         year = 2025,
        month = jul,
       volume = {540},
       number = {4},
        pages = {3247-3262},
          doi = {10.1093/mnras/staf808},
archivePrefix = {arXiv},
       eprint = {2407.06859},
 primaryClass = {astro-ph.HE},
       adsurl = {https://ui.adsabs.harvard.edu/abs/2025MNRAS.540.3247S}
}

@ARTICLE{Wygoda2019,
       author = {{Wygoda}, Nahliel and {Elbaz}, Yonatan and {Katz}, Boaz},
        title = "{Type Ia supernovae have two physical width-luminosity relations and they favour sub-Chandrasekhar and direct collision models - I. Bolometric}",
      journal = {\mnras},
         year = 2019,
        month = apr,
       volume = {484},
       number = {3},
        pages = {3941-3950},
          doi = {10.1093/mnras/stz145},
archivePrefix = {arXiv},
       eprint = {1711.00969},
 primaryClass = {astro-ph.HE},
       adsurl = {https://ui.adsabs.harvard.edu/abs/2019MNRAS.484.3941W}
}

@ARTICLE{Jeffery1999,
       author = {{Jeffery}, David J.},
        title = "{Radioactive Decay Energy Deposition in Supernovae and the Exponential/Quasi-Exponential Behavior of Late-Time Supernova Light Curves}",
      journal = {arXiv e-prints},
         year = 1999,
        month = jul,
          eid = {astro-ph/9907015},
        pages = {astro-ph/9907015},
          doi = {10.48550/arXiv.astro-ph/9907015},
archivePrefix = {arXiv},
       eprint = {astro-ph/9907015},
 primaryClass = {astro-ph},
       adsurl = {https://ui.adsabs.harvard.edu/abs/1999astro.ph..7015J}
}

@ARTICLE{Katz2013,
       author = {{Katz}, Boaz and {Kushnir}, Doron and {Dong}, Subo},
        title = "{An exact integral relation between the Ni56 mass and the bolometric light curve of a type Ia supernova}",
      journal = {arXiv e-prints},
         year = 2013,
        month = jan,
          eid = {arXiv:1301.6766},
        pages = {arXiv:1301.6766},
          doi = {10.48550/arXiv.1301.6766},
archivePrefix = {arXiv},
       eprint = {1301.6766},
 primaryClass = {astro-ph.HE},
       adsurl = {https://ui.adsabs.harvard.edu/abs/2013arXiv1301.6766K}
}

@INCOLLECTION{Taubenberger2017,
       author = {{Taubenberger}, Stefan},
        title = "{The Extremes of Thermonuclear Supernovae}",
    booktitle = {Handbook of Supernovae},
         year = 2017,
       editor = {{Alsabti}, Athem W. and {Murdin}, Paul},
        pages = {317},
          doi = {10.1007/978-3-319-21846-5\_37},
       adsurl = {https://ui.adsabs.harvard.edu/abs/2017hsn..book..317T}
}
\bibliographystyle{aasjournal}

\appendix
\twocolumngrid
\section{Imaging and Photometry of SN\,2022an}

The multiband photometric observations of SN\,2022an are summarized as follows.

\paragraph{\textbf{LCOGT-1m/Sinistro}}  We observed SN 2022an in $BVgri$ bands using the Sinistro cameras mounted on the 1 m class telescopes of the Las Cumbres Observatory Global Telescope Network (LCOGT; \citealt{Brown2013_lcogt}). The observations were conducted from 2022 January 9 to April 2. The images were reduced using the LCOGT/BANZAI pipeline \citep{McCully2018_lcogt_Banzai}. We downloaded the reduced frames from the Las Cumbres Observatory science archive.

\paragraph{\textbf{NTT/EFOSC2}} We observed two epochs of photometric data in $BV$ bands with the ESO Faint Object Spectrograph and Camera version 2 (EFOSC2; \citealt{Buzzoni1984_efosc}) on the ESO 3.58m New Technology Telescope (NTT). Basic data reduction, including bias subtraction, flat-field correction, and astrometric calibration, was performed using the PESSTO pipeline \citep{Smartt2015_pessto}. We also obtained four epochs of acquisition images in the $V$ band from spectroscopic observations. 

\paragraph{\textbf{Magellan/IMACS}} We obtained three epochs of $gri$ imaging with the Inamori-Magellan Areal Camera and Spectrograph (IMACS; \citealt{Dressler2011_imacs}). The images were reduced using {\tt IRAF} \citep{Tody1986_iraf, Tody1993_iraf}, including bias subtraction and flat-field correction. 

\paragraph{\textbf{NTT/SOFI}} We obtained four epochs of NIR imaging of SN\,2022an in the $JHKs$ bands with the Son OF ISAAC (SOFI; \citealt{Moorwood1998_sofi}) on the NTT. \\ 

We performed point-spread function photometry using the {\tt daophot} task in {\tt IRAF} \citep{Tody1986_iraf, Tody1993_iraf}. The host-galaxy background flux was modeled using an isophotal model and iteratively subtracted from the images (see \citealt{Chen2022_pmpyeasy} for the method). We utilized the ATLAS All-sky Stellar Reference Catalog (ATLAS-REFCAT2; \citealt{Tonry2018_refcat2}) to derive photometric zero-points for the optical photometry. Before being used for photometric calibration of our target, the ATLAS-REFCAT2 magnitudes of the reference stars in the fields were first converted to the Johnson $BV$ and Sloan-$gri$ systems using the transformations given in \cite{Tonry2012}. Reference stars in the field of view from the Two Micron All Sky Survey catalog \citep{Skrutskie2006_2mass} were used to derive the photometric zero-points for the $JHKs$ bands. All photometric procedures were performed using the {\tt pmpyeasy} pipeline \citep{Chen2022_pmpyeasy}. 

The photometric data for SN\,2022an are summarized in Table~\ref{tab:phot_BVgri} for the $BVgri$ bands and in Table~\ref{tab:phot_JHKs} for the $JHK$ bands. The $gri$ magnitudes are reported in the AB system, while the $BVJHK$ magnitudes are based on the Vega system.

We also obtained ASAS-SN $g$-band photometry \citep{Shappee2014_asassn} from the ASAS-SN Sky Patrol \footnote{\url{https://asas-sn.osu.edu/}} \citep{Kochanek2017_asassn}. We adopted the ``Image Subtraction (No reference flux added)'' photometry method, which performs aperture photometry on coadded image-subtracted data for each epoch but does not add the source flux from the reference image to the light curve. The results are presented in Table~\ref{tab:phot_asassn}.


\begin{deluxetable*}{cccccccc}
\tablecaption{Optical photometry of SN 2022an\label{tab:phot_BVgri}}
\tablecolumns{8}
\tablehead{
\colhead{JD} &
\colhead{Phase \tablenotemark{a}} &
\colhead{\hspace{.3cm}$B$}\hspace{.3cm} & 
\colhead{\hspace{.3cm}$V$} \hspace{.3cm} & 
\colhead{\hspace{.3cm}$g$} \hspace{.3cm} & 
\colhead{\hspace{.3cm}$r$} \hspace{.3cm}  & 
\colhead{\hspace{.3cm}$i$} \hspace{.3cm}  &
\colhead{Instrument} \\
\colhead{(-2,459,000)} &
\colhead{(day)} &
\colhead{\hspace{.3cm}(mag)}\hspace{.3cm} & 
\colhead{\hspace{.3cm}(mag)} \hspace{.3cm} & 
\colhead{\hspace{.3cm}(mag)} \hspace{.3cm} & 
\colhead{\hspace{.3cm}(mag)} \hspace{.3cm}  & 
\colhead{\hspace{.3cm}(mag)} \hspace{.3cm}  &
\colhead{}
}
\startdata
588.75 & 7.35 & 17.892\,(0.056) & 16.233\,(0.022) & $\cdot\cdot\cdot$ & 15.812\,(0.022) & 15.695\,(0.023) & Sinistro \\
590.14 & 8.74 & 18.112\,(0.071) & 16.392\,(0.027) & 17.296\,(0.035) & 15.983\,(0.022) & 15.816\,(0.028) & Sinistro \\
590.78 & 9.38 & $\cdot\cdot\cdot$ & 16.505\,(0.035) & $\cdot\cdot\cdot$ & $\cdot\cdot\cdot$ & $\cdot\cdot\cdot$ & EFOSC2 \\
591.75 & 10.35 & 18.433\,(0.059) & 16.664\,(0.024) & 17.532\,(0.040) & 16.114\,(0.028) & 15.908\,(0.030) & Sinistro \\
592.85 & 11.45 & $\cdot\cdot\cdot$ & 16.771\,(0.044) & $\cdot\cdot\cdot$ & $\cdot\cdot\cdot$ & $\cdot\cdot\cdot$ & EFOSC2 \\
593.74 & 12.34 & 18.587\,(0.105) & 16.924\,(0.039) & 17.800\,(0.067) & 16.346\,(0.035) & 16.072\,(0.035) & Sinistro \\
595.81 & 14.41 & 18.723\,(0.102) & 17.056\,(0.040) & 17.925\,(0.059) & 16.528\,(0.031) & 16.352\,(0.034) & Sinistro \\
597.73 & 16.33 & 18.811\,(0.141) & 17.179\,(0.053) & 18.040\,(0.102) & 16.687\,(0.040) & 16.428\,(0.040) & Sinistro \\
602.53 & 21.13 & 19.123\,(0.165) & 17.502\,(0.063) & 18.241\,(0.099) & 17.081\,(0.053) & 16.895\,(0.055) & Sinistro \\
603.75 & 22.35 & $\cdot\cdot\cdot$ & 17.689\,(0.024) & $\cdot\cdot\cdot$ & $\cdot\cdot\cdot$ & $\cdot\cdot\cdot$ & EFOSC2 \\
605.73 & 24.33 & 19.188\,(0.094) & 17.717\,(0.057) & 18.468\,(0.031) & 17.255\,(0.036) & 17.057\,(0.043) & Sinistro \\
609.19 & 27.79 & 19.325\,(0.041) & 17.868\,(0.022) & 18.562\,(0.030) & 17.547\,(0.021) & 17.292\,(0.020) & Sinistro \\
613.70 & 32.30 & 19.489\,(0.051) & 18.061\,(0.029) & 18.746\,(0.048) & 17.737\,(0.051) & 17.484\,(0.035) & Sinistro \\
617.44 & 36.04 & 19.436\,(0.049) & 18.251\,(0.039) & 18.807\,(0.046) & 17.953\,(0.058) & 17.667\,(0.042) & Sinistro \\
618.50 & 37.10 & 19.540\,(0.054) & 18.154\,(0.038) & 18.780\,(0.055) & 17.974\,(0.039) & 17.675\,(0.036) & Sinistro \\
620.77 & 39.37 & $\cdot\cdot\cdot$ & 18.447\,(0.021) & $\cdot\cdot\cdot$ & $\cdot\cdot\cdot$ & $\cdot\cdot\cdot$ & EFOSC2 \\
622.61 & 41.21 & $\cdot\cdot\cdot$ & $\cdot\cdot\cdot$ & 18.935\,(0.073) & 18.227\,(0.066) & 18.090\,(0.136) & Sinistro \\
624.77 & 43.37 & 19.751\,(0.069) & 18.492\,(0.039) & 19.035\,(0.054) & 18.403\,(0.039) & 18.180\,(0.028) & Sinistro \\
626.08 & 44.68 & $\cdot\cdot\cdot$ & $\cdot\cdot\cdot$ & 18.987\,(0.092) & 18.535\,(0.067) & 18.032\,(0.059) & Sinistro \\
629.82 & 48.42 & $\cdot\cdot\cdot$ & $\cdot\cdot\cdot$ & 19.059\,(0.121) & 18.716\,(0.087) & 18.230\,(0.064) & Sinistro \\
632.12 & 50.72 & $\cdot\cdot\cdot$ & $\cdot\cdot\cdot$ & 19.448\,(0.275) & 18.883\,(0.202) & 18.413\,(0.189) & Sinistro \\
636.77 & 55.37 & 20.054\,(0.067) & 18.983\,(0.054) & 19.327\,(0.057) & 19.130\,(0.047) & 18.538\,(0.047) & Sinistro \\
640.45 & 59.05 & $\cdot\cdot\cdot$ & $\cdot\cdot\cdot$ & 19.318\,(0.085) & 19.181\,(0.092) & 18.456\,(0.067) & Sinistro \\
642.64 & 61.24 & 20.310\,(0.143) & 19.191\,(0.154) & 19.534\,(0.050) & 19.482\,(0.036) & 18.681\,(0.025) & Sinistro \\
648.48 & 67.08 & 20.168\,(0.103) & 19.281\,(0.063) & 19.556\,(0.065) & 19.609\,(0.101) & 18.746\,(0.067) & Sinistro \\
650.87 & 69.47 & $\cdot\cdot\cdot$ & $\cdot\cdot\cdot$ & $\cdot\cdot\cdot$ & 19.688\,(0.066) & $\cdot\cdot\cdot$ & IMACS \\
654.64 & 73.24 & 20.288\,(0.174) & 19.495\,(0.125) & 19.610\,(0.060) & 19.903\,(0.075) & 18.953\,(0.045) & Sinistro \\
659.80 & 78.40 & 20.596\,(0.104) & 19.803\,(0.106) & $\cdot\cdot\cdot$ & $\cdot\cdot\cdot$ & $\cdot\cdot\cdot$ & EFOSC2 \\
660.12 & 78.72 & 20.497\,(0.305) & 19.818\,(0.080) & 19.829\,(0.078) & 20.102\,(0.241) & 19.009\,(0.041) & Sinistro \\
671.58 & 90.18 & 20.831\,(0.090) & 20.103\,(0.056) & 20.199\,(0.073) & 20.592\,(0.068) & 19.241\,(0.040) & Sinistro \\
720.58 & 139.18 & $\cdot\cdot\cdot$ & $\cdot\cdot\cdot$ & 21.378\,(0.049) & 22.226\,(0.040) & 20.353\,(0.044) & IMACS \\
722.73 & 141.33 & 21.955\,(0.075) & 21.853\,(0.041) & $\cdot\cdot\cdot$ & $\cdot\cdot\cdot$ & $\cdot\cdot\cdot$ & EFOSC2 \\
817.49 & 236.09 & $\cdot\cdot\cdot$ & $\cdot\cdot\cdot$ & 23.670\,(0.282) & $\cdot\cdot\cdot$ & 23.044\,(0.089) & IMACS \\
\enddata
\tablenotetext{a}{Relative to the estimated epoch of $B-$band peak JD = 2459581.4. It is the same for Table~\ref{tab:phot_JHKs} and \ref{tab:phot_asassn}.}
\end{deluxetable*}

\begin{deluxetable}{ccccc}
\tablecaption{NIR photometry of SN 2022an with NTT/SOFI\label{tab:phot_JHKs}}
\tablecolumns{5}
\tabletypesize{\footnotesize}  
\tablehead{
\colhead{JD} &
\colhead{Phase} &
\colhead{$J$} &
\colhead{$H$} &
\colhead{$K_s$} \\
\colhead{(-2,459,000)} &
\colhead{(day)} &
\colhead{(mag)} &
\colhead{(mag)} &
\colhead{(mag)} 
}
\startdata
592.80 & 11.40 & 15.30\,(0.06) & 14.99\,(0.06) & 15.03\,(0.11) \\
611.78 & 30.38 & 17.77\,(0.13) & 16.68\,(0.14) & 16.97\,(0.15) \\
638.70 & 57.30 & 19.51\,(0.13) & 18.30\,(0.12) & 18.80\,(0.29) \\
649.72 & 68.32 & 20.14\,(0.15) & $\cdot\cdot\cdot$ & $\cdot\cdot\cdot$ \\
660.80 & 79.40 & $\cdot\cdot\cdot$ & 19.57\,(0.27) & $\cdot\cdot\cdot$ \\
\enddata
\end{deluxetable}

\begin{deluxetable}{ccc}
\tablecaption{g-band photometry of SN 2022an from ASAS-SN survey\label{tab:phot_asassn}}
\tablehead{
\colhead{JD} & \colhead{Phase} & \colhead{g}\\
\colhead{(-2,459,000)} & \colhead{(day)} &\colhead{(mag)} 
} 
\startdata
567.790 & -13.610 & $>$16.875 \\
579.833 & -1.567 & 16.439\,(0.066) \\
581.544 & 0.144 & 16.377\,(0.065) \\
583.755 & 2.355 & 16.443\,(0.073) \\
585.765 & 4.365 & 16.788\,(0.086) \\
590.521 & 9.121 & 16.938\,(0.101) \\
592.488 & 11.088 & 16.986\,(0.141) \\
593.484 & 12.084 & $>$17.261 \\
594.832 & 13.432 & 17.768\,(0.188) \\
595.776 & 14.376 & $>$17.204 \\
597.731 & 16.331 & $>$16.693 \\
600.720 & 19.320 & $>$16.867 \\
605.814 & 24.414 & $>$17.537 \\
609.613 & 28.213 & $>$17.796 \\
\enddata
\end{deluxetable}

\section{Spectroscopic Follow-up Observations and Data Reduction}

We performed spectroscopic observations of SN\,2022an with multiple instruments covering the optical and NIR wavelength ranges. A summary of these observations is listed in Table~\ref{tab:spec_log}. Details of the follow-up campaign and data reduction are described below.

\paragraph{\textbf{SOAR/GHTS}}
SN\,2022an was observed on 2022 January 7 with the Goodman High Throughput Spectrograph (GHTS; \citealt{Clemens2004_Goodman}) on the 4.1-meter Southern Astrophysical Research (SOAR) Telescope. The observation was conducted with a 1\farcs0 wide slit and the ``400\_SYZY" grating. The spectrum\footnote{We noticed that the original spectrum submitted to TNS had an inaccurate wavelength solution, as indicated by the mismatch with the expected telluric features.} was first presented in the classification report \citep{Jacobson-Galan2022_classification_2022an}. We obtained the raw data from the Las Cumbres Observatory Science Archive\footnote{\url{https://archive.lco.global/}} and performed the data reduction. We reduced the spectrum using {\tt IRAF} \citep{Tody1986_iraf, Tody1993_iraf}, including bias subtraction, flat-field correction, cosmic-ray removal, wavelength calibration using Hg(Ar) and Ne arc-lamp frames taken immediately before the target observation, and flux calibration using spectroscopic standard star observations taken on the same night as the science object was observed. 

\paragraph{\textbf{NTT/EFOSC2 \& NTT/SOFI}}

We obtained 7 epochs of optical spectroscopy with EFOSC2 and 2 epochs of NIR spectroscopy with SOFI. The observations with EFOSC2 were performed with grism \#11, \#13 and \#16. The SOFI observations were performed with the blue grism, which covers approximately 9300--16500 \AA. The data were reduced using the PESSTO pipeline \citep{Smartt2015_pessto}.

\paragraph{\textbf{VLT/X-Shooter}}

We obtained two epochs of intermediate-resolution spectroscopy with the X-shooter echelle spectrograph \citep{Vernet2011_xshooter} on 2022 February 26 and March 31, through a DDT program (Program ID: 108.23MS, P.I.: P. Chen). All observations were performed in nodding mode and with 1\farcs0/0\farcs9/0\farcs9 wide slits (UVB/VIS/NIR). The observations cover the entire spectral range of X-shooter from 3000 to 24800 \AA. We first removed cosmic rays using {\tt astroscrappy}\footnote{\href{https://github.com/astropy/astroscrappy}{https://github.com/astropy/astroscrappy}}, which is based on the cosmic-ray removal algorithm of \cite{vanDokkum2001}. Then the data were processed with the X-shooter pipeline v3.3.5, and the ESO workflow engine ESOReflex \citep{Goldoni2006, Modigliani2010, Freudling2013}. All data from the three arms were reduced in nodding mode. The nodding mode reduction is critical for NIR data to ensure good sky subtraction. The spectra from the three arms were stitched by averaging the overlap regions. The atmospheric absorption in the VIS and NIR arms was corrected with the software tool {\tt molefit}\citep[v4.2.3;][]{Smette2015}.

\paragraph{\textbf{Magellan/FIRE \& Magellan/IMACS}}

We observed SN\,2022an with both the Folded-port InfraRed Echellette (FIRE; \citealt{Simcoe2013_fire}) spectrograph and the Inamori-Magellan Areal Camera and Spectrograph (IMACS; \citealt{Dressler2011_imacs}) on 2022 March 12. We obtained an additional epoch of IMACS spectroscopy of SN\,2022an on 2022 May 21. Both FIRE and IMACS are mounted on the 6.5m Magellan Baade telescope. The FIRE spectra were taken in long-slit mode, and the data were reduced using the IDL pipeline
{\tt firehose} \citep{Simcoe2013_fire}.   The IMACS spectra were reduced with {\tt IRAF} \citep{Tody1986_iraf, Tody1993_iraf}, including bias subtraction, flat-field correction, cosmic-ray removal, wavelength calibration using arc lamp frames taken immediately after the target observation, and flux calibration using spectroscopic standard star observations taken on the same night as the science object was observed.


\begin{deluxetable}{cccc}
\tablecaption{Summary of Spectroscopic Observations of SN 2022an \label{tab:spec_log}}
\tablecolumns{4}
\tabletypesize{\footnotesize}  
\tablehead{
\colhead{JD} &
\colhead{Phase\tablenotemark{a}} &
\colhead{Wavelength(\AA)} &
\colhead{Telescope/Spectrograph}
}
\startdata 
2459586.7 &   +5.3d &  5000 - 9050 &   SOAR/GHTS \\ 
2459590.8 & +9.4d &  3350 -10000 & NTT/EFOSC2 \\
2459592.9 & +11.5d &  3650 - 9250 & NTT/EFOSC2 \\
2459603.8 & +22.4d &  3350 -10000 & NTT/EFOSC2 \\
2459611.9 & +30.5d &  9300 -16500 & NTT/SOFI \\
2459619.8 & +38.4d &  9300 -16500 & NTT/SOFI \\
2459620.8 & +39.4d &  3350 -10000 & NTT/EFOSC2 \\
2459633.7 & +52.3d &  3650 - 9250 & NTT/EFOSC2 \\
2459636.8 & +55.4d &  3000 -25000 & VLT/XSHOOTER \\
2459648.6 & +67.2d &  3650 - 9250 & NTT/EFOSC2 \\
2459650.8 & +69.4d &  8000 -21500 & Magellan/FIRE \\
2459651.1 & +69.7d &  4700 - 9400 & Magellan/IMACS \\
2459651.6 & +70.2d &  3350 - 7500 & NTT/EFOSC2 \\
2459670.6 & +89.2d &  3000 -25000 & VLT/XSHOOTER \\
2459720.5 & +139.1d &  4150 - 9400 & Magellan/IMACS \\
\enddata
\tablenotetext{a}{Relative to the estimated epoch of $B-$band peak JD = 2459581.4}
\end{deluxetable}

All spectroscopic observations will be made publicly available on WISeREP\footnote{\hyperlink{https://www.wiserep.org}{https://www.wiserep.org}} \citep{Yaron2012_wiserep}.

\end{document}